\documentclass[useAMS,usenatbib]{mnras}
\usepackage{float}
\usepackage{hyperref}
\usepackage{hhline}
\usepackage[dvipsnames]{xcolor}
\usepackage{booktabs}
\usepackage{gensymb}
\usepackage{natbib}
\usepackage{lipsum}
\usepackage{tikz}
\usetikzlibrary{spy}
\usepackage{subcaption}
\usepackage{epstopdf}
\usepackage{etex}

\title{On the efficiency of  techniques for the reduction of impulsive noise in astronomical images}
\author[A. Popowicz et al.]{A. Popowicz$^1$\thanks{E-mail:apopowicz@polsl.pl}, A. R. Kurek$^2$, T. Blachowicz$^3$, V. Orlov$^4$  and B. Smolka$^1$\\
$^1$Silesian University of Technology, Institute of Automatic Control, Poland, 44-100 Gliwice, Akademicka 16\\
$^2$Astronomical Observatory of the Jagiellonian University, Poland, 30-244 Krakow, Orla 171\\
$^3$Silesian University of Technology, Institute of Physics - Center for Science and Education, Poland, 44-100 Gliwice, S. Konarskiego 22B\\
$^4$ Instituto de Astronom{\'i}a,Universidad Nacional Autonoma de M{\'e}xico, Apdo. Postal 70-264, Cd. Universitaria, 04510 M{\'e}xico D.F., M\'{e}xico }
\begin{document}
\date{~}
\pagerange{\pageref{firstpage}--\pageref{lastpage}} \pubyear{2016}
\maketitle
\label{firstpage}

\begin{abstract}
The impulsive noise in astronomical images originates from various sources. It develops as a result of thermal generation in pixels, collision of cosmic rays with image sensor or may be induced by high readout voltage in Electron Multiplying CCD (EMCCD). It is usually efficiently removed by employing the dark frames or by averaging several exposures. Unfortunately, there are some circumstances, when either the observed objects or positions of impulsive pixels evolve and therefore each obtained image has to be filtered independently. In this article we present an overview of impulsive noise filtering methods and compare their efficiency for the purpose of astronomical image enhancement. The employed set of noise templates consists of dark frames obtained from CCD and EMCCD cameras working on ground and in space. The experiments conducted on synthetic and real images, allowed for drawing numerous conclusions about the usefulness of several filtering methods for various: (1) widths of stellar profiles, (2) signal to noise ratios, (3) noise distributions and (4) applied imaging techniques. The results of presented evaluation are especially valuable for selection of the most efficient filtering schema in astronomical image processing pipelines.
\end{abstract}


\section{Introduction}
The astronomical imaging techniques are significantly different from those used in other science fields, like in medicine, biology or physics. The exposure times may range from several milliseconds (Lucky Imaging \cite{LuckyImaging}, adaptive optics \cite{Saha1}, meteor registrations \cite{meteor}) to several minutes (high resolution spectroscopy, narrow band photometry). Moreover, the astronomy, in which the distant and low-intensity targets are observed, puts high demands on imaging devices requiring extremely high sensitivity, together with a very low noise readout (\cite{Janesick1,Janesick2}). Ideally, the image sensor should capture each striking photon and allow for precise registration of even a single photo-induced electron. While the sensor's sensitivity can be enhanced only by modification of internal pixel structure (\cite{ar1,ar2}), the noise can be also mitigated to a certain degree by the application of a proper image processing technique.

There is a vast number of competitive algorithms aiming to retrieve the information from noisy images. Most of them were evaluated extensively however, utilizing unrealistic noise models (like uniform or salt \& pepper, in which faulty pixels receives only maximal or minimal possible intensities), which are far from real noise present in CCD or CMOS imagers. Moreover, the set of benchmark images is very limited and includes usually a set of multimedia samples (faces, landscapes, still life, etc.), which present significantly different scenes than the astronomical data.

In this paper we compare the performance of 12  impulsive noise reduction strategies when applied for the denoising of astronomical data. For this purpose a unique benchmarking tool was created, which consists of noise templates (mainly the dark frames) and high-fidelity astronomical reference images (synthetic and real). As our evaluations are dedicated for astronomical imaging, the methods were assessed by the quality indicators employing the magnitude system.

One of the most significant outcomes of our experiments is a set of guide tables, which allows for selection of the most accurate filtering schema (with its optimized parameters values), depending on the noise distribution and intensity, and on the characteristics of observed scene, (sampling of stellar profile, type of observations: traditional imaging, spectroscopic and speckle imaging).

The paper is organized as follows. In Section 2 we show and explain details of sources of non-stationary impulsive noise specific for the astronomical imaging. In Section 3 we briefly characterize all the filtering algorithms employed in the comparison. In the Section 4 we present our collection of noise templates, obtained from both ground-based and orbital cameras. In Section 5 the details of experiments with synthetic and real astronomical images are given. The Section 6 presents our findings and numerous conclusions drawn from the comparison. Finally, we summarize the paper in Section 7.

\section{Impulsive noise in astronomical images}
The astronomical observations in the visible or infrared rely mostly on CCD sensors (\cite{Janesick1}) which consist of a matrix of pixels, wherein the photo-induced electrons are collected during the image acquisition phase. Then, the matrix is read-out, pixel by pixel, through the output amplifier, which is the main source of so-called Gaussian noise (also called the readout noise). The charge measurement on the output gate is always associated with a certain degree of uncertainty. Popular astronomical cameras have the readout noise RMS (root-mean-square) ranging from 3 to 15 e$^-$. The averaging of $Q$ exposures, which present the same scene, results in reduction of Gaussian noise standard deviation by a factor of $\sqrt{Q}$.

A similar type of disturbances in astronomical images is associated with the charge accumulation and is called Poisson noise or "shot" noise. It originates in the discrete nature of charge creation in sensor's pixels. The number of received photo-electrons is governed by the Poisson distribution and therefore, for sufficient number of electrons, it can be approximated by Gaussian distribution with variance equal to the expected value (\cite{Janesick2}). There are plenty of algorithms developed for Gaussian noise suppression, which utilize e.g. the local pixel neighborhood (\cite{Gaussian1}) or make use of information from larger patches of an image (\cite{Gaussian2}). However, in this paper we do not concentrate on this particular type of image disturbance, since it has been already extensively investigated in the rich literature (\cite{bookimageproc1},\cite{bookimageproc2}).

\subsection{Stationary dark current}
The other noise type, called the impulsive noise, is visible as distinctively bright spots/pixels or smudges in an image. The most prominent source of such intensity spikes is the dark current, which is produced by faults in silicon, like point defects (vacancies and intrinsic impurity atoms) or spatial defects (dislocations and clustered vacancies) (\cite{defects}). While the impurity-based defects and dislocations are created mainly during the CCD fabrication, the vacancies and their clusters are induced by energetic protons hitting the CCD matrix (\cite{proton1,proton2,proton3}). The number of defective pixels increases during the sensor lifetime what can be observed in CCDs working in severe space environment (\cite{proton4,proton5}). The only way to, at least partially, remove the defects from silicon crystalline, is annealing, which is regularly employed e.g. in Hubble Space Telescope (\cite{annealing1,annealing2}).

\begin{figure}
\centering
\includegraphics[width=0.5\textwidth]{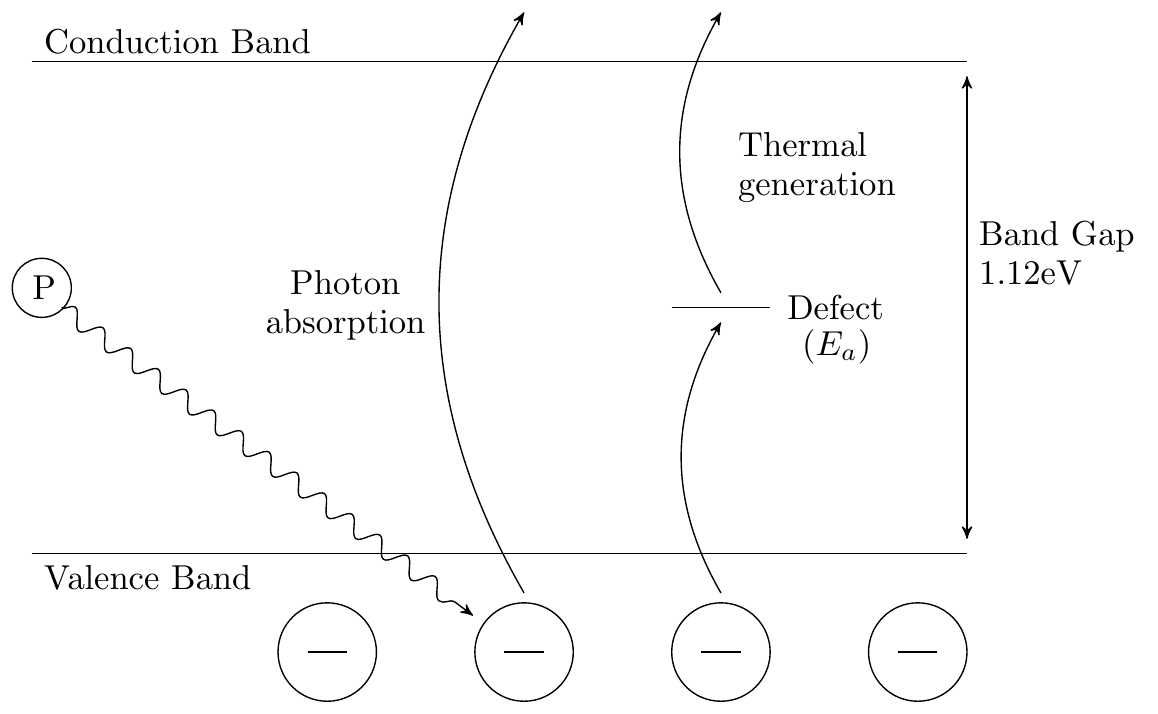}
\caption{Charge generation in silicon: photon absorption (on the left) and thermal generation through defects (on the right).}
\label{ValConDark}
\end{figure}

The silicon defects act as very efficient charge generation centers, which add unwanted bias during the light registration. Such centers have the activation energy $E_a$(amount of energy needed for an electron to release the atom nucleus) within the band gap of silicon, so that they are able to capture the electrons from valence band and transfer them to conduction band, increasing the charge accumulated in a pixel. The both processes: (1) electron creation from photon absorption and (2) thermal generation, are presented in Fig. \ref{ValConDark}. 
\

\begin{figure}
\centering
\begin{subfigure}[b]{\linewidth}
\includegraphics[width=\linewidth]{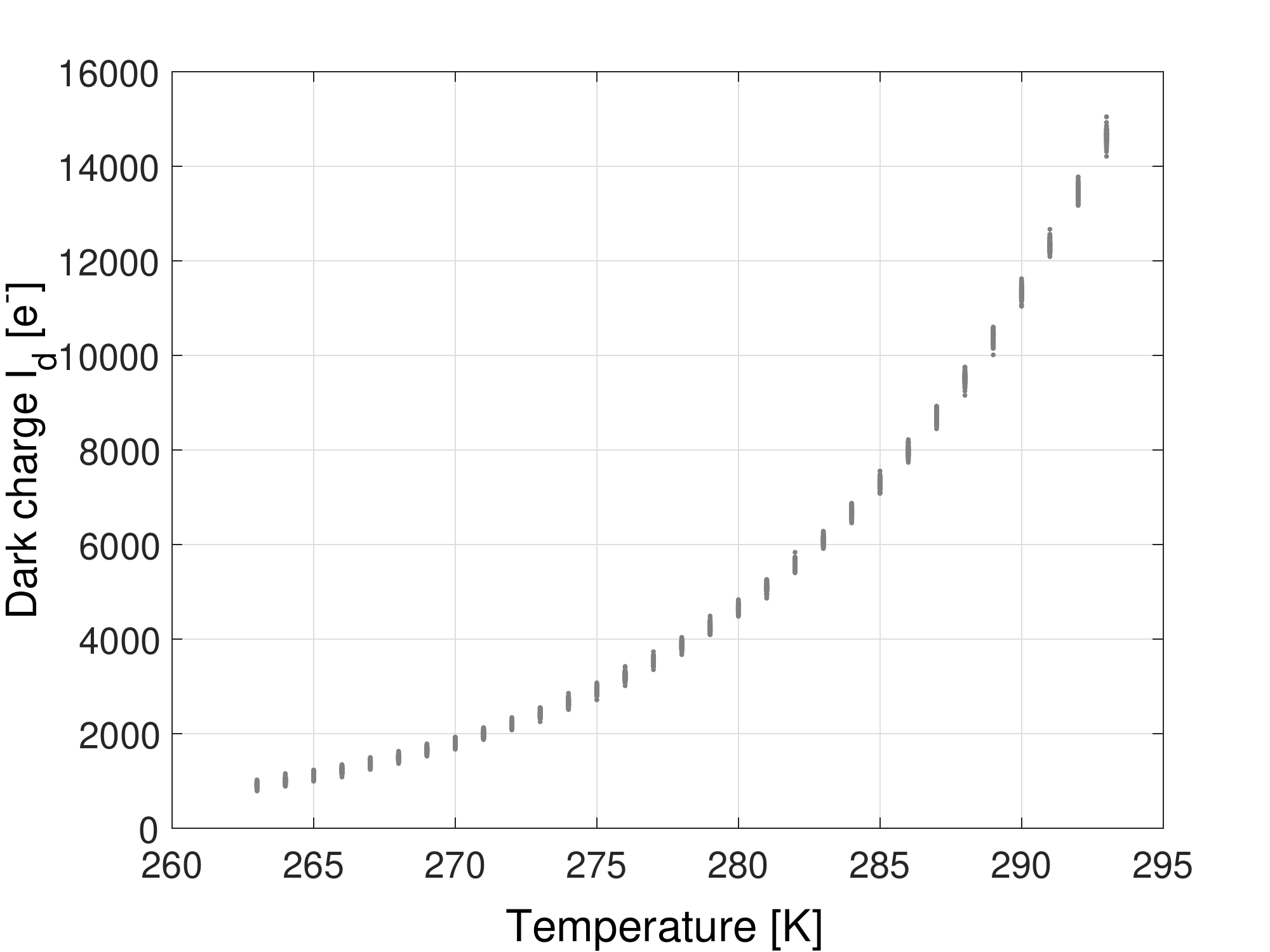}
\end{subfigure}
\begin{subfigure}[b]{\linewidth}
\includegraphics[width=\linewidth]{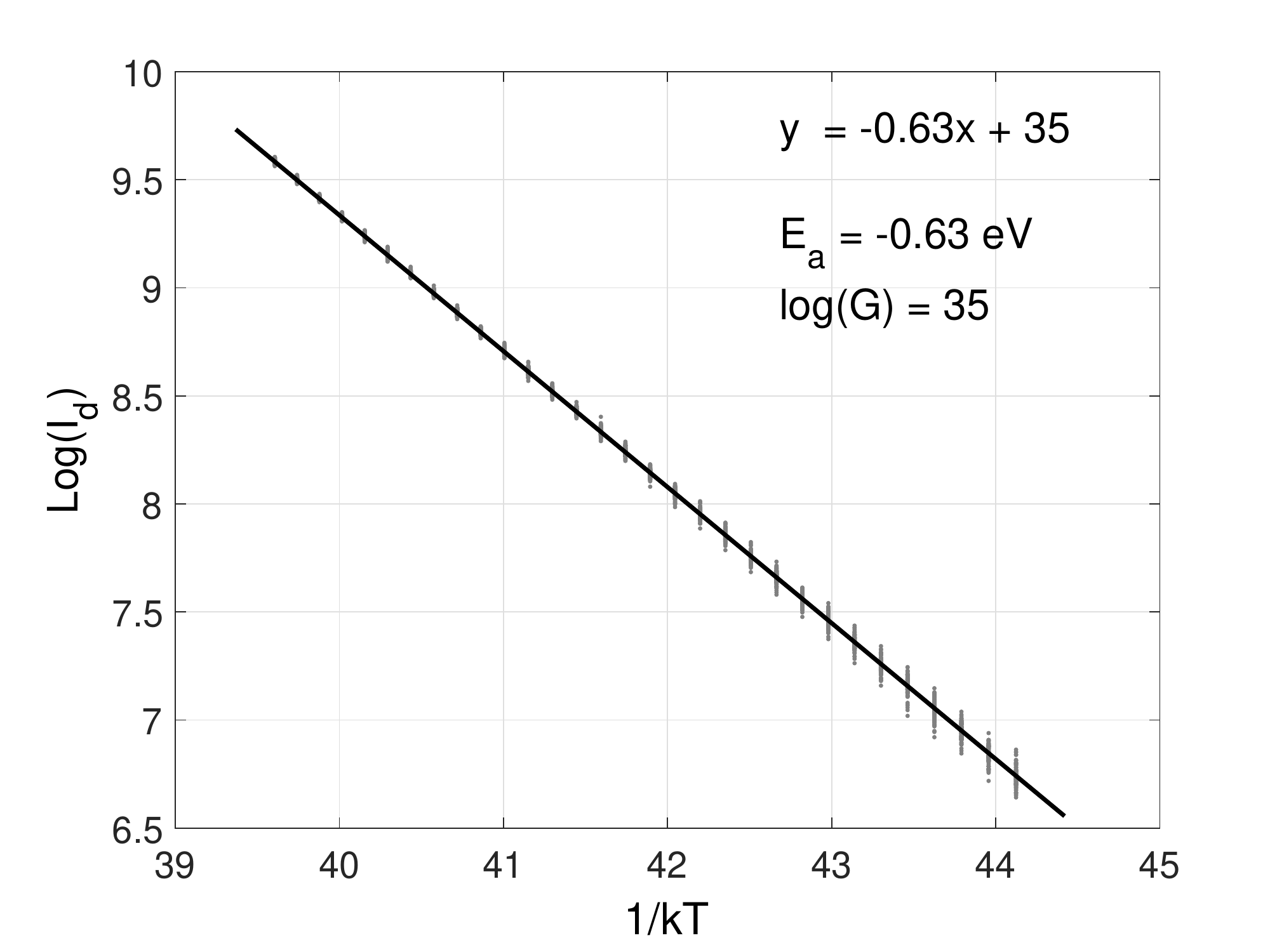}
\end{subfigure}
\caption{The temperature dependency of dark current in an exemplary pixel in KAI 11002M CCD matrix: linear (above) and logarithmic (below) scale. The obtained activation energy $E_a=0.63$ [eV] and log$(G)=35$.}
\label{characteristicsdark}
\end{figure}

The number of thermally generated electrons per time interval depends on the activation energy of defect $E_a$ and on temperature (\cite{WidenTemp}):
\begin{equation}
G_d = G~\textrm{exp}\bigg\{\frac{-E_a}{kT}\bigg\},
\label{dark_equation}
\end{equation}
where $G_d$ is the dark current generation rate, $G$ is a parameter, $k$ is the Boltzmann constant and $T$ is the temperature in Kelvins. A straightforward way to identify both parameters ($E_a$ and $G$) involves a linear approximation of logarithmic dependency of the dark current versus temperature, as exposed in Fig. \ref{characteristicsdark}. The centers (sometimes the called trapping sites) located in the middle of the silicon energetic band gap (i.e. $E_a$ = 0.55 eV, which is half of 1.1 eV silicon band gap) are usually the most efficient generation centers, since for such defects the total probability of thermal transfer from the valence to conduction band using a trapping site, achieves maximum.

\begin{figure}
\centering
\begin{subfigure}[b]{\linewidth}
\includegraphics[width=\linewidth]{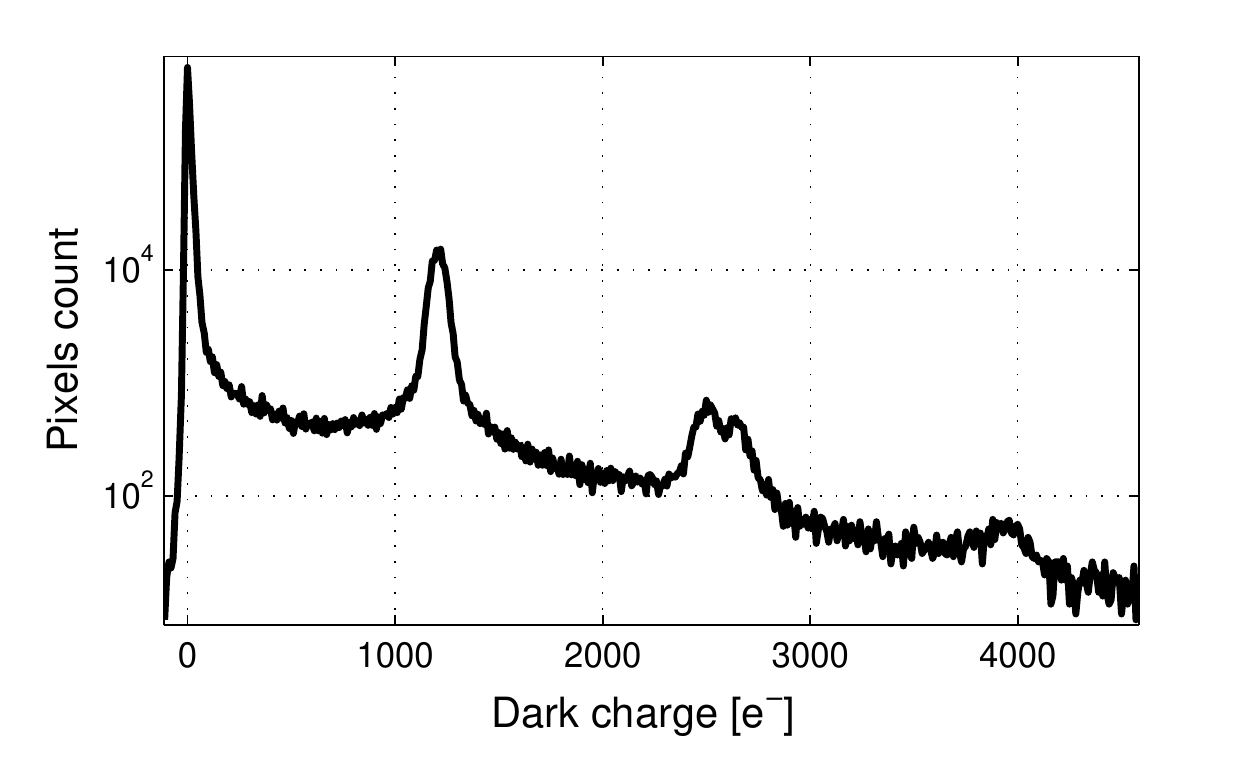}
\caption{KAF-1602ME (ground).}
\end{subfigure}
\begin{subfigure}[b]{\linewidth}
\includegraphics[width=\linewidth]{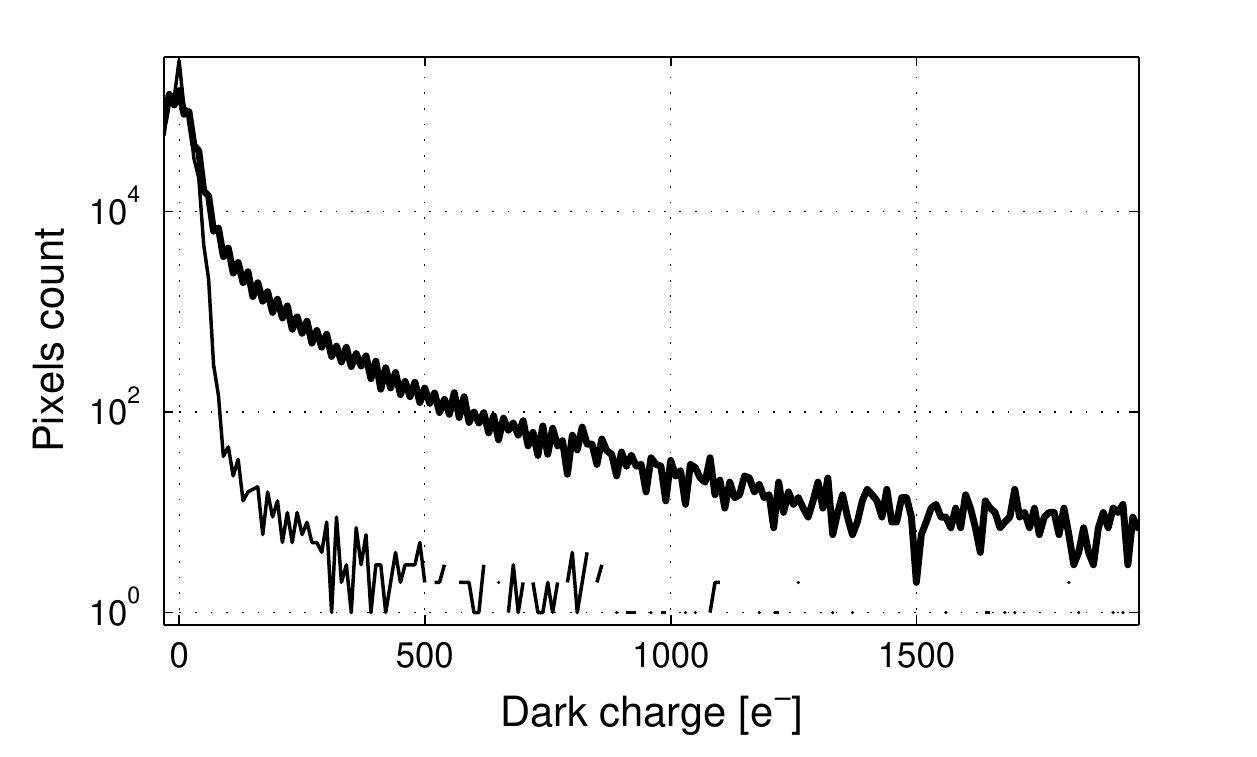}
\caption{KAI-11002M, BRITE (orbital).}
\end{subfigure}
\caption{The dark current distributions in: (a) KAF-1603ME, working on ground and (b) BRITE KAI-11002M, before the launch (thin solid line) and after several months in space (thick solid line). The dark frames were obtained with respective temperatures and exposures times: (a) -5\degree C, 900 s, (b) 25 $^o$C, 1~s.}
\label{distrdarkexamples}
\end{figure}

The distribution of dark current in CCD matrices varies from one sensor to another, as it depends on the type defects. For the fabrication-induced impurities, the quantization of dark current histogram is observed (\cite{darkcurrent1,darkcurrent2,darkcurrent3}), as depicted in Fig. \ref{distrdarkexamples}a. The visible distinct peaks are related to the presence of 1, 2 or more defects of the same type within a pixel structure. On the other hand, the dislocations manifest their presence by a continuous distribution of dark charge, since their generation properties depend on size and location within pixel's electric field (\cite{PopowiczPAK}). The distribution of dark charge of CCDs working on ground is therefore composed of (1) discrete peaks related to the point defects and (2) a continuous background caused by the dislocations. 

A purely continuous distribution may be observed in sensors working in space. It appears due to the dominance of clusters of point defects induced by energetic particles, mainly the protons. In Fig. \ref{distrdarkexamples}b an exemplary dark current histogram (before and after the launch), from one of the BRITE nano-satelites\footnote{~~Based on data collected by the BRITE Constellation satellite mission, designed, built, launched, operated and supported by the Austrian Research Promotion Agency (FFG), the University of Vienna, the Technical University of Graz, the Canadian Space Agency (CSA), the University of Toronto Institute for Aerospace Studies (UTIAS), the Foundation for Polish Science \& Technology (FNiTP MNiSW), and National Science Centre (NCN).}, is presented (\cite{BRITE1}). The BRITEs' cameras are not shielded, due to the lack of space and weights constrains, thus they suffer from radiation damage in orbit. The defects in form of brighter pixels and column offsets emerged after several weeks in space, as a result of growing thermal activity in pixels.

\begin{figure}
\centering
\includegraphics[width=\linewidth]{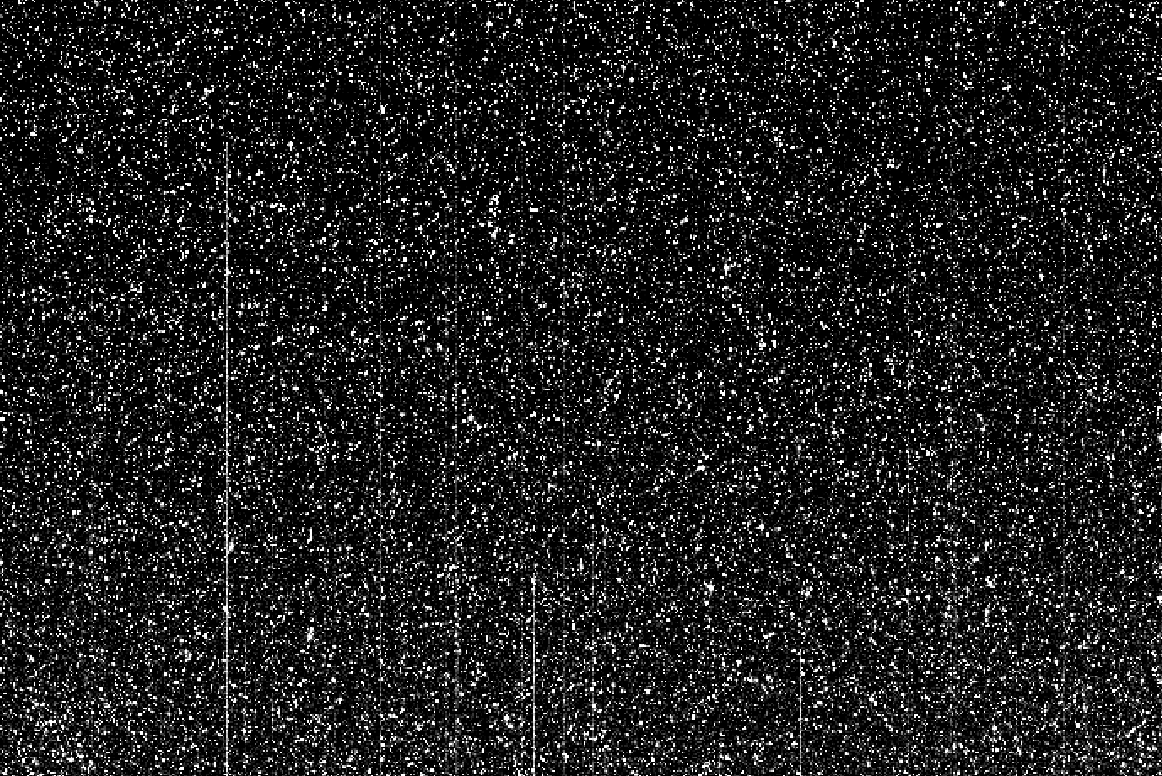}
\caption{Dark frame from KAI 11002M CCD matrix depicting the hot pixels and the column offsets due to the dark current generation.}
\label{darkframe1}
\end{figure}

The dark current is also generated during the short readout period. It is especially visible in interline CCDs, where the pixels are divided into the light-sensitive and charge-transfer parts. Although the transfer part is usually better shielded, some defects may still appear and the associated dark charge, generated during the readout, is spread along column. The overall offset is usually low, since the accumulation time is limited to the readout of a single CCD row. An example of such bias structure is presented in Fig. \ref{darkframe1}.

Nowadays the CCD sensors are optimized to achieve the lowest possible dark current by means of fabrication purity and by utilization of various pixel structures, which may be virtually free from defects (\cite{VeryLowDarkCurrent}). This results in a lower number of hot pixels and reduced average dark current. In the most advanced cameras equipped with extremely strong cooling (down to 70K), the dark current problem is negligible. However, the small observatories with less sophisticated devices, still have to compensate for it. It is usually done by subtraction of a dark frame, which is obtained with the same exposure time and at the same temperature as the astronomical image, but with a sensor protected from any light source. The calibration frames are often stored, since dark current generation rate is considered to be stable for a pixel.

\subsection{Non stationary dark current}
There are some circumstances, when the dark current intensity is not predictable and the correction using dark frame is insufficient. Some of the defects show so-called random telegraph signals (RTS, \cite{RTS3}), which means that the parameter $G$ in (\ref{dark_equation}) fluctuates between many meta-stable discrete values including often a calm state ($G=0$, i.e. no dark current generation). Such problems are observed mainly in space missions, where the sensors are bombarded by energetic protons. They cause a specific type of induced defect (phosphorus-vacancy, P-V pair) in the form of electric dipole, which randomly re-orientates within electric field thus changing its generation properties (\cite{PVpair,PVpair2}).

\begin{figure}
\centering
\begin{subfigure}[b]{\linewidth}
\includegraphics[width=\linewidth,scale=2]{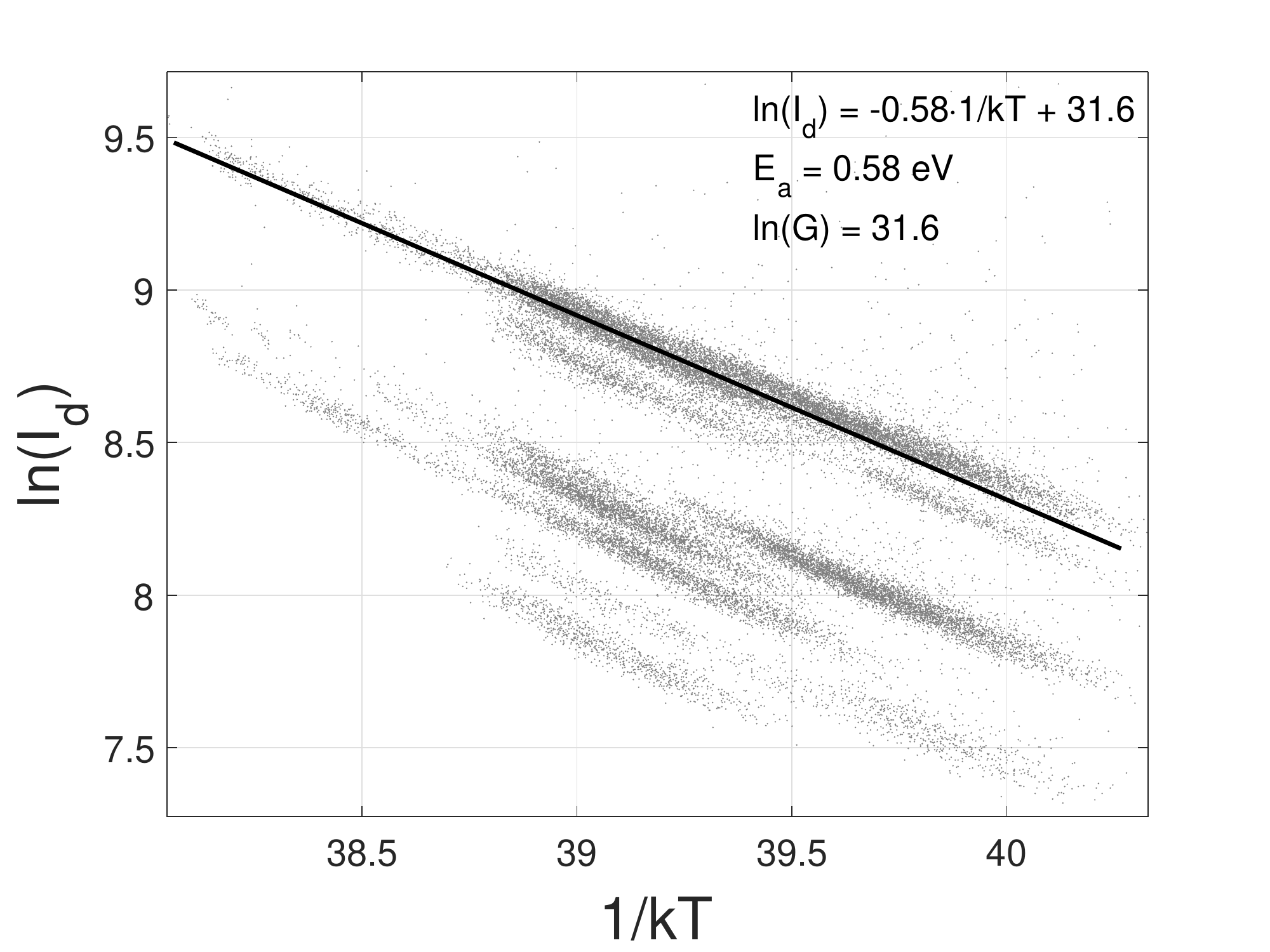}
\caption{Temperature plot.}
\end{subfigure}
\begin{subfigure}[b]{\linewidth}
\includegraphics[width=\linewidth,scale=2]{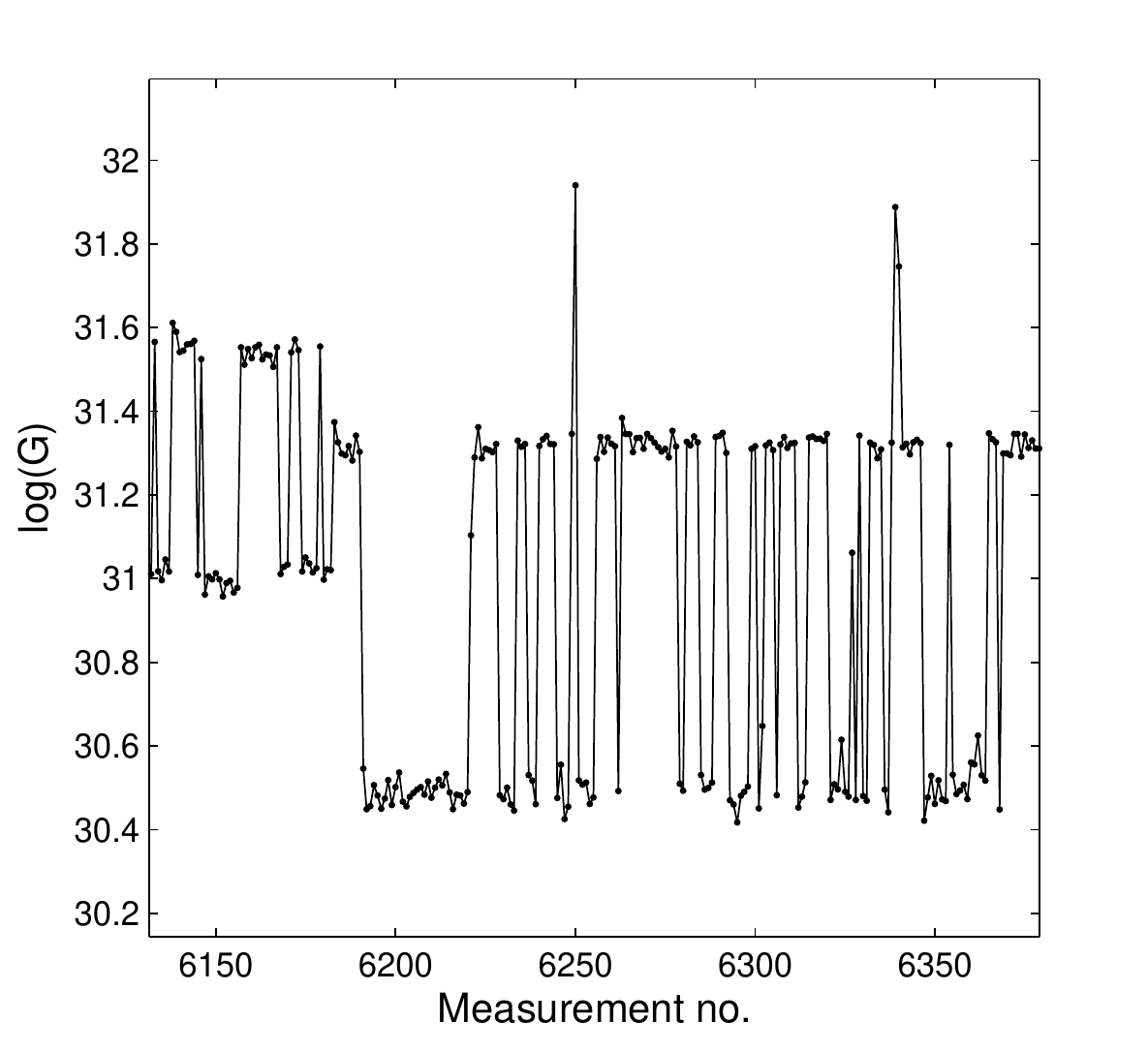}
\caption{Time plot.}
\end{subfigure}
\caption{RTS behavior registered in BRITE satellite: (a) temperature dependencies of dark current ($I$) for exemplary RTS affected pixel, (b) part of time dependencies of $G$ parameter for the same pixel.}
\label{RTSbehaviour1}
\end{figure}

We have registered plenty examples of such a behavior in CCDs installed in BRITE nano-satellites. Not only a growing number of flickering pixels appeared, but also the mentioned column defects exposed RTS behavior (blinking columns). As it can be seen in Fig. \ref{RTSbehaviour1}a, the temperature logarithmic plot shows many shifted linear characteristics, which correspond to various discrete $G$ levels originating from the superposition of switching generation rates in several independent P-V centers. An example of fluctuations of $G$ parameter, which includes 5 meta-stable states, is presented in Fig. \ref{RTSbehaviour1}b.  

\begin{figure}
\centering
\includegraphics[width=0.8\linewidth]{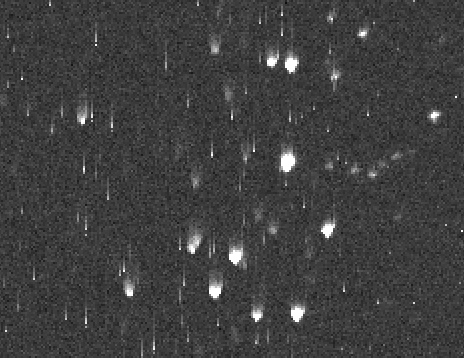}
\caption{Example of a region of charge transfer inefficiency in BRITE's CCD, containing relatively low number of hot pixels and a star cluster Pleiads.}
\label{BRITECTI}
\end{figure}

Shortly after the launch, another impulsive-like problem emerged in BRITE satellites. Some of CCD regions developed charge transfer inefficiency (CTI), which could be observed as characteristic smudges after each brighter pixel (see Fig. \ref{BRITECTI}). This made the precise photometry even more difficult. It is worth to note that the CTI problems appear only in space-based observations, due to the radiation environment. Fortunately, the problem has been successfully mitigated by reduction of frequency of vertical CCD clock. However, there is plenty of valuable data registered before it was fixed, which requires filtering.

In addition to the RTS behavior, for which the standard dark frame subtraction cannot be applied, there are some dark current nonlinearities recently extensively investigated in several works (\cite{nonlinear1,nonlinear2,nonlinear3}). The authors confirm that the generation rate from either a single defect or from dislocations, depends on current amount of charge collected in a pixel. The number of electrons kept in pixel's potential well disturbs the local electric field and gradually decreases the efficiency of thermal activity in defects (it is somewhat similar to the brighter-fatter effect (\cite{brighterfatter})). It results in a lower number of thermally generated electrons in the dark frame than during the light registration. This effect leads to systematic errors introduced by dark frame subtraction, which can be overcome only by extended characterization of CCD's defects using optical methods (\cite{PopowiczOptical}). 

\subsection{Clock induced charge}
The next source of non-predictable impulsive noise in astronomical images is the clock-induced charge (CIC) present in data collected by electron-multiplying CCDs. The EMCCDs are the image sensors utilized in observational techniques which require very low readout noise, so that each photo-induced electron can be counted (\cite{EMCCDs1}). The applications of EMCCDs include Lucky Imaging (\cite{LuckyImaging}), speckle interferometry and adaptive optics, in which the images are registered with very short exposure times (several milliseconds) to retrieve the images less degraded by atmospheric turbulence (\cite{Saha1,Saha2}). Such sensors are also employed when the number of photons received in a pixel is very low, like in spectroscopy and in fast or narrow-band photometry (\cite{EMCCDs2,Optics1}).  The EMCCDs give good premise for future concepts in astronomical imaging, like in hypothetical quantum telescope (\cite{Kurek16}).

\begin{figure}
\centering
\includegraphics[width=\linewidth]{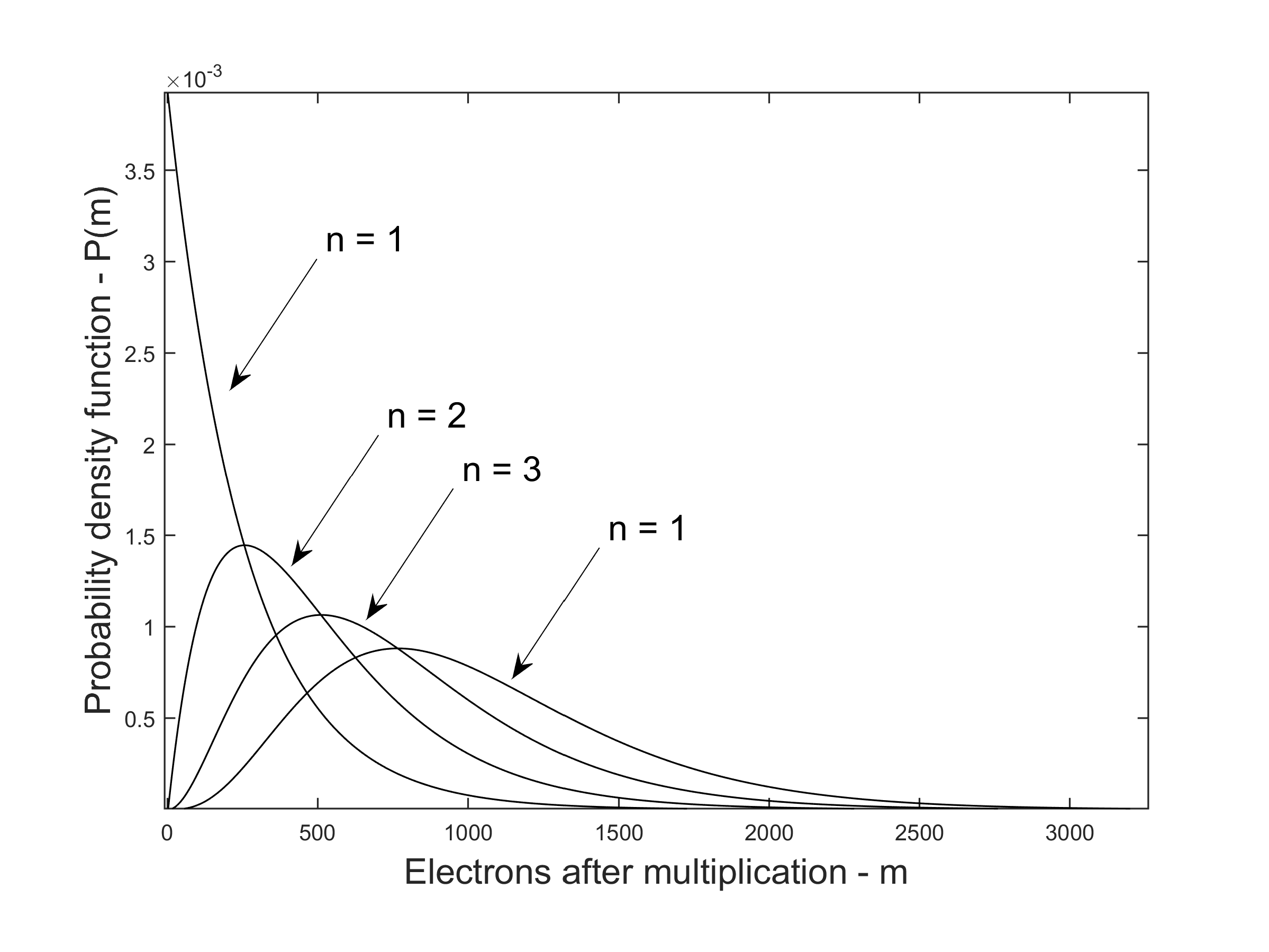}
\caption{Distribution of output electrons in EMCCD, for $n=1, 2, 3$ and $4$ registered photons.}
\label{EMCCD1}
\end{figure}

The idea of electron multiplication is based on the effect of impact ionization, which takes place in horizontal readout register driven by very high voltage (70 V and above). When $n$ electrons enter the output register, the final number of electrons $m$ at the output is governed by the following formula (\cite{EMCCDs1}):
\begin{equation}
P(m) = \frac{(m-n+1)^{n-1}}{(n-1)!\big(g-1+1/n\big)^n}~\textrm{exp}\bigg(-\frac{m-n+1}{g-1+1/n}\bigg),
\label{emccdformula1}
\end{equation}
where $P(m)$ is the probability of receiving $m$ output electrons and $g$ is the average register gain. The corresponding distributions of received counts for various number ($n$) of received photons are presented in Fig. \ref{EMCCD1}.

Unfortunately, very high electric field in readout horizontal register induces additional unwanted charge. The electrons in valence band are swept rapidly during high-voltage switching, thus occasionally some of them gain enough energy to be transferred to the conduction band. Since the chance to generate more than two electrons for a given pixel is negligible (so $n=1$ in eq. \ref{emccdformula1}), the distribution of CIC noise, after the electron multiplication, shows an exponential distribution:
\begin{equation}
P(m) = \frac{1}{g}~\textrm{exp}\bigg(-\frac{m}{g}\bigg).
\end{equation}

The CIC spikes, in contrast to the dark current, appear in random pixels, therefore the phenomenon can not be mitigated by any calibration frame. Due to the skewness of distribution, the averaging of images is not recommended as the averaged frame will be biased. Summarizing, for most of the applications of EMCCDs, i.e. in high resolution and extremely fast imaging, the CIC noise can be calibrated either by image filtering techniques or by some new fabrication technologies.

\begin{figure}
\centering
\includegraphics[width=0.7\linewidth]{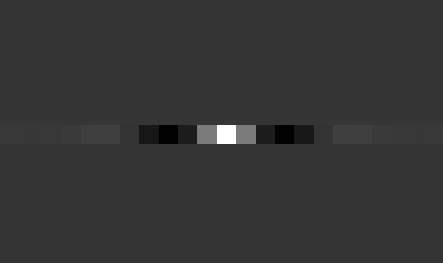}
\caption{Central part of average autocorrelation of dark frames registered by Andor iXon3 EMCCD. The central bright pixel represents the correlation of a pixel with itself. The gray scale was set so the medium gray level corresponds to 0 autocorrelation, while the dimmer and brighter values represent negative and positive autocorrelation, respectively.}
\label{autocorrelation}
\end{figure}

We also found that the electron multiplication process introduces some cross-talk between the pixels in one row. This phenomenon was well observed in the dark frames registered by Andor iXon3 EMCCD during the speckle observations at Mexican National Astronomical Observatory, Sierra San Pedro Martir. The effect of biasing of adjacent pixels in a row was observed in averaged (high signal-to-noise) autocorrelation of dark frames (see Fig. \ref{autocorrelation}). 

The following conclusions can be drawn about the spatial dependency of noise: (1) the two pixels (on both left and right side), closest to the impulse, gain additional charge, (2) the charge of next pixels is slightly lowered, (3) there is only small and diminishing impact on further pixels in a row, (4) there is no evidence of cross-talk between rows. The phenomenon was not observed when EM mode was off. This implies that there must be some mutual influence between electric fields of cells within a row. Summarizing, similarly to the dark current in columns of interline CCD sensors or like for CTI regions in BRITE sensors, the CIC noise in EMCCDs is definitely neither spatially independent nor uncorrelated.

\subsection{Cosmic rays}
\begin{figure}
\centering
\begin{subfigure}[b]{0.49\linewidth}
\includegraphics[height=4cm]{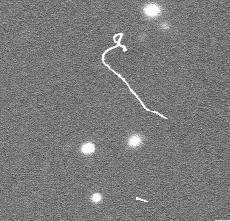}
\caption{}
\end{subfigure}
\begin{subfigure}[b]{0.49\linewidth}
\includegraphics[height=4cm]{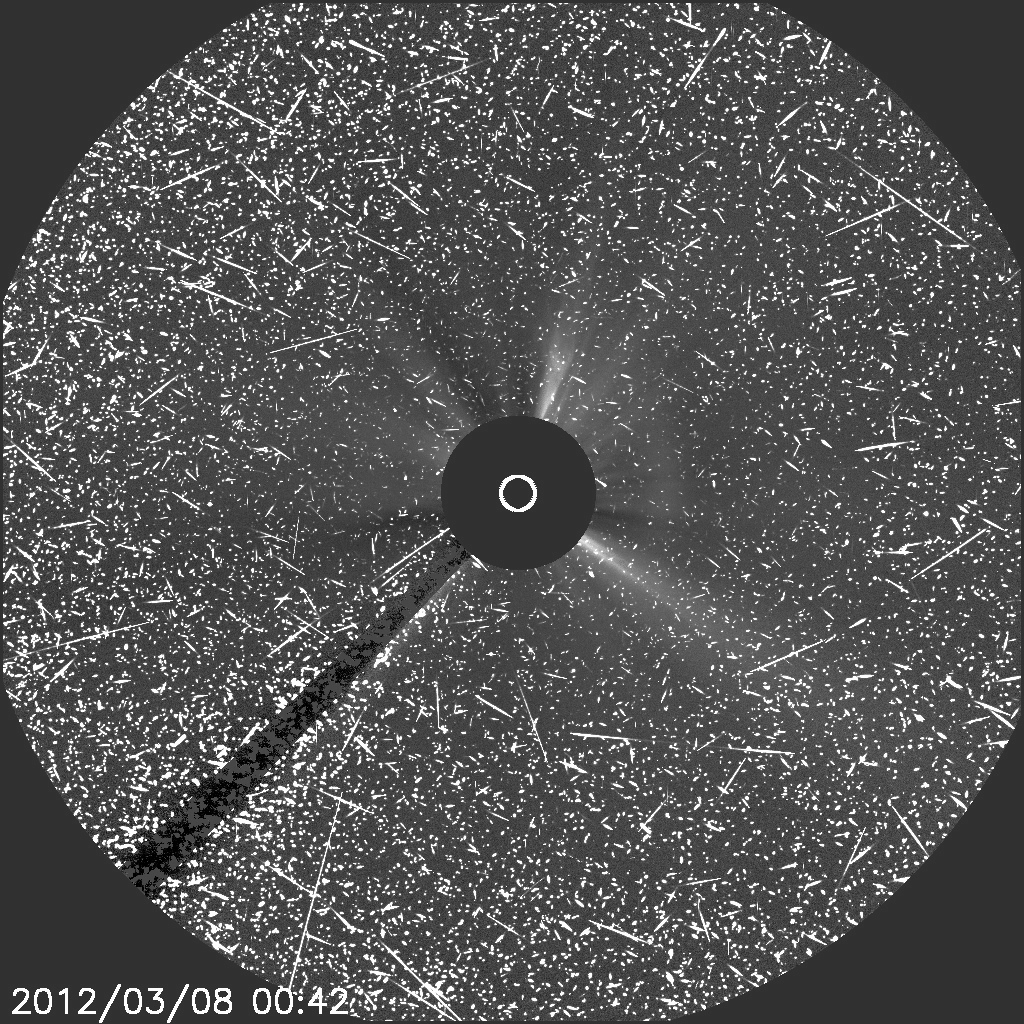}
\caption{}
\end{subfigure}
\caption{Exemplary acquisitions of cosmic ray impacts: (a) a complex cosmic ray event registered by the paper authors (2015), (b) Solar corona as registered by SOHO satellite during an exemplary outburst (2012).}
\label{cosmicsfig}
\end{figure}

The last source of impulses in astronomical images is the flux of cosmic rays (mostly the electrons), which have not enough energy to inflict permanent damage but introduce temporary effects visible in form of smudges in images (see Fig. \ref{cosmicsfig}a). The artifacts are created due to the energy transfer from a particle to CCD electrons in a valence band. As the particles come from different directions and move variously within the CCD internal structure, the cosmic ray impacts show various shapes, often even imitating the astronomical sources. The problem is noticeable in cameras employed in high altitude observatories or operating in space.

Similarly to the Gaussian noise, the cosmic ray impacts can be minimized by image averaging using e.g. a sigma clipping or a median operation. However, this technique is not applicable if the imaged scene changes rapidly like e.g. in the case of SOHO satellite (\cite{SohoPaper}) registering Solar corona phenomena (see Fig. \ref{cosmicsfig}b). Also, due to the required increase of observational time, it is impractical to repeat very long exposures, therefore the cosmic rays cancellation has to be performed separately in each frame.

\section{Methods}
For our comparison study, we have selected several popular, thus frequently cited algorithms, dedicated for impulsive noise removal. The methods usually employ the intensity replacement of a pixel utilizing simple nonlinear operations in a local sliding window (\cite{nonlinearfilters}). A fundamental example of such procedure is median filtering (\textbf{MED}), where a central pixel of a processing window is replaced by the median of its neighboring pixels. The strength of image smoothing depends on the size of operational window (the larger the window, the stronger the smoothing). Although such a simple filter efficiently removes impulses, the outcomes are noticeably blurred. Therefore, more sophisticated filters, presented below, try to preserve the image details by interpolating only the pixels previously detected as faulty. This way the unaffected parts of objects are not disturbed by blurring. In each algorithm, we indicate up to two tunable parameters, which will be further optimized during the experiments.

Alpha-Trimmed Mean (\textbf{ATM})~\cite{ATM} is a special case of the order-statistics filter (\cite{Bovik1983,bookorderstatistics}). The $\alpha$-trimmed mean of pixels intensities in a window $W$ is obtained by (1) sorting the pixels, (2) removing a fixed fraction of pixels at the upper and lower ends of such sorted set, and (3) computing the average $\eta$ of remaining ones:
\begin{equation}
\eta = \frac{1}{N-2 \alpha}\sum_{i=\alpha+1}^{N-\alpha}x_i,
\end{equation}
where $\alpha$ is the number of pixels removed from the sorted set and $W$ has radius $r$, thus it includes $N=(2r+1) \times (2r+1)$ pixels.

Another approach to impulsive noise reduction, called the Switching Median Filter with Boundary Discriminative Noise Detection (\textbf{BDND}), is presented in (\cite{BDND}). The first stage of this filtering schema involves localization of corrupted pixels. First, the pixel is compared with its neighbors in a small, local window $W$, whose size is tuned by radius parameter $r_1$ . The pixels in window $W$, excluding the central one, are sorted according to their intensities into a vector, in which the differences between each consecutive pair are calculated. The maximum of such difference,s in each half of the vector, are found (i.e. in two sets: $(1:N/2)$ and $(N/2+1:N)$), thus the corresponding intensity levels of a lower and upper limit are defined. If the intensity of central window pixel is below or above the intensity of previously defined boundaries, then the same procedure is performed for a larger window, whose size is defined by radius $r_2$. If the pixel again falls out of specified intensity range, it is classified as faulty and replaced by the median of its neighboring pixels classified as uncorrupted.

Central-Weighted Median (\textbf{CWM}~\cite{CWM}) is a median-based filtering method, where the intensity of a central pixel in window is replaced by the weighted median of all the pixels intensities in a local window. In this filter, the central pixel has higher weight, that is obtained by its repetition ($c$ times) of its intensity before the median calculation. The results of filtering may be adjusted by changing the window radius $r$.

Directional-Weighted Median (\textbf{DWM})~\cite{DWM}) is a filter, which includes both detection and filtering phases. This approach utilizes the statistics of pixels intensities aligned in four main directions (up-down, left-right and two diagonal), that significantly improves the filter's performance in terms of edge preservation. The algorithm searches for the direction, which includes the pixels of similar intensities, thus it detects possible edges. If the pixel differs significantly from the intensities in any direction (a detection threshold $T$ is employed), then it is considered as impulsive. The filter output is calculated as a weighted median of pixels in a sliding window (like in CWM), where the weights are higher ($w$=2) for the previously detected best direction (i.e. the one with the lowest gradient) than for all remaining directions ($w$=1, simple non-weighted median). The algorithm is repeated iteratively decreasing the detection threshold ($T_{k+1} = T_k\cdot0.8,k=1,2,...$, the initial value $T_1$ is the first tunable parameter of the algorithm), therefore the main parameter responsible for smoothing strength is the number of iterations ($i$). 

Iterative Truncated Mean (\textbf{ITM}) (\cite{ITM}) is another simple approach to impulsive noise reduction. Initially, a mean of intensities in a local window is calculated. Then, a dynamic threshold is obtained, utilizing a mean of the following absolute differences:
\begin{equation}
\tau = \frac{1}{N}\sum_{i=1}^{N}|x_i-\mu|,
\end{equation}
where $x_i$ is the intensity of $i$-th pixel in local window $W$ and $\mu$ is the mean intensity of the pixels in W. In each step a following truncation procedure is performed: if the intensity of any window pixel is below or above the defined interval $\mu\pm\tau$, then it is replaced by the corresponding boundary value: $\mu+\tau$ or $\mu-\tau$. Eventually, the filtered intensity of a central pixel is the mean of such a truncated set. The two parameters used in the algorithm are: the radius of window ($r$) and the number of iterations ($i$).

The Laplacian Edge Detection Filter (\textbf{LED})~\cite{LAcosmic} is nowadays a standard method of filtering-out the cosmic rays and it is also an important and well known part of the IRAF package (\cite{IRAFold, IRAFnew} - a standard tool for astronomical image processing). The algorithm utilizes the convolution of sub-sampled image with the Laplacian of Gaussian to highlight sharp edges associated with either impulsive pixels or cosmic-rays occurrences. Each pixel in such filtered image is compared with the expected noise to properly identify the impulses. The star-like objects (roundish) are distinguished by the analysis of their symmetry. The procedure is repeated iteratively and in each iteration the flagged pixels are replaced by the median of their undisturbed neighbors. For the case of this study, the CCD characteristics-based parameters (gain and readout noise) were treated as constants, leaving only the remaining two parameters, $\sigma_{\tiny{\textrm{lim}}}$ and $f_{\tiny{\textrm{lim}}}$, which are responsible respectively, for limiting noise level and for maximal allowable gradient of registered star-like objects. The limiting noise level $\sigma_{\tiny{\textrm{lim}}}$ determines how many times the standard deviation of the total noise (i.e. Poisson+readout) should be multiplied to define the limit.

Lower-Upper-Middle filter (\textbf{LUM})~\cite{LUM} belongs to the family of rank-order-based filters. The pixels in a local window are sorted and the algorithm checks, if the central pixel belongs either to the lower (L), upper (U) or middle set (M). The size of lower and upper sets is tunable by $k$ parameter ($k=1,..,(N-1)/2$, where $N$ is number of pixels in a local window). If the pixel is within L or U set, then it is replaced by the median intensity of M set. Otherwise the pixel is not classified as impulsive and remains unchanged. The parameter $k$ adjusts the level of smoothing (e.g. for $k=(N-1)/2$ the filter reduces to simple median filter).

Peak and Valley filter (\textbf{PAV})~\cite{PAV} is another example of a simple and fast smoothing method. It eliminates all the ”peaks” and ”valleys” within the grayscale image. The algorithm compares the pixel intensity with the intensities of pixels in local operational window. If the central pixel is the brightest or the dimmest one, then it is replaced respectively by the highest or lowest intensity of the pixels in local neighborhood, (i.e. the most extreme values are being replaced by the second-most extreme ones). This filter is capable to efficiently remove isolated impulses while preserving undisturbed image regions. 

The modification of Peak and Valley filter is presented in \cite{rpav} and is called recursive PAV filter (\textbf{RPAV}). The pixel is detected as impulsive, the same way it is done in the original concept, however the intensity estimation is performed using a recursive maximum-minimum method presented by \cite{rPAVref}. In this algorithm, the gray scale value is estimated as an average of the minimum and maximum intensity obtained over pixels in specified subsets within the sliding window, (the interested reader is referred to the detailed formula in the original work). The algorithm has two parameters: the number of iterations ($i$) and the radius of sliding window ($r$).

Progressive Switching Median (\textbf{PSM}~\cite{PSM}) is another filtering schema, which utilizes the median operations. Initially each pixel intensity is compared with the median of neighboring pixels in a local window. If the difference is larger than predefined threshold $T$, then such a pixel is considered as noisy. The map of impulsive pixels is created prior to the filtering phase. Next, each faulty pixel is replaced by the median of the pixels intensities in a local window, but excluding those pixels, which were previously identified as impulsive. The algorithm is iterated as long, as there are no corrupted pixels left. The window radius ($r$) and the threshold ($T$) control the smoothing strength.

Another nonlinear filter is the Tri-State Median Filter (\textbf{TSM}~\cite{TSM}). In this algorithm, initially the image is processed by a simple median filter (working in $3\times3$ window) and by CWM filter. Then, the pixel's intensity is compared with the outcomes of both filters. If the deviation from the median filter output is lower than a specified threshold ($T$) then the pixel remains unchanged. If the difference from the CWM output is larger than $T$, the output of a median filter is used for replacement. In the other case, the output of CWM filter is used. This approach utilizes the fact, that the more robust approach (median filtering) is required for strongly outlying pixels, while the CWM should be employed if the pixel is only slightly brighter than its neighborhood.

Due to the significant number of various methods and, especially, their multiple parameters, we aggregated all the important descriptions in Tab. \ref{tab_param}, providing the methods abbreviations, references and their citation report according to Thomphon Reuters Web of Science as checked in May 2016. In Tab. \ref{tab_param2} each method's parameter was described and the employed range of values was also exposed. The range was obtained by either our experiments or by using the literature guidelines and was utilized during the comparison of methods accuracy. It allowed for determining the optimal parameter configurations, so that we were sure that each methods performed its best.

\begin{table*}
\centering
\caption{Filters overview.}
\begin{tabular}{l |c | c | c }  
\hline
Abbreviation & Full name & Referece& Citations \\ \hline\vspace{0.5mm}
ATM & Alpha-Trimmed Mean filter & \cite{ATM}&179 \\\vspace{0.5mm}
BDND & Boundary Discriminative Noise Detection filter & \cite{BDND}&148 \\\vspace{0.5mm}
CWM & Center-Weighted Median filter & \cite{CWM}&502 \\\vspace{0.5mm}
DWM & Directional-Weighted Median filter & \cite{DWM}&110 \\\vspace{0.5mm}
ITM & Iterative Truncated Arithmetic Mean filter & \cite{ITM}&10\\\vspace{0.5mm}
LED & Laplacian Edge Detection filter & \cite{LAcosmic}& 556\\\vspace{0.5mm}
LUM & Lower-Upper-Middle filter & \cite{LUM}&100 \\\vspace{0.5mm}
MED & Median filter & \cite{nonlinearfilters}&- \\\vspace{0.5mm}
PAV & Peak and Valley filter & \cite{PAV}&93 \\\vspace{0.5mm}
PSM & Progressive Switching Median filter & \cite{PSM}&375 \\\vspace{0.5mm}
RPAV & Recursive Peak and Valley filter & \cite{rpav}&49 \\\vspace{0.5mm}
TSM & Tri-State Median filter & \cite{TSM}&288 \\\vspace{0.5mm}

\end{tabular}
\label{tab_param}
\end{table*}

\begin{table*}
\centering
\caption{The parameters of utilized filters with their corresponding range of values used in the experiment. The  local windows utilized in the algorithms are square with a side width of $2r+1$.}
\begin{tabular}{l |c | c | r }  
\hline
Method & Parameter & Description & Values range (step)\\ \hline\vspace{0.5mm}
ATM & $\alpha$ & number of rejected pixels in a window&1$\sim \frac{(2r+1)^2-1}{2}~(1)$ \\
&$r$ & window radius & $1\sim5~(1)$ \\ \hline
BDND & $r_1$ & local, small window radius& $1\sim5~(1)$ \\
 & $r_2$ & global, large window radius& $r_1+1\sim10~(1)$ \\ \hline
CWM & $r$ & window radius & $1\sim5~(1)$ \\
 & $c$ & number of central pixel repetitions & $1\sim\frac{(2r+1)^2-1}{2}~(1)$ \\ \hline
DWM & $T$ &  initial threshold & $100\sim100000~[e^-]~(100)$ \\
 & $i$ & number of algorithm iterations & $1\sim5~(1)$ \\ \hline
ITM &  $r$ & window radius & $1\sim5~(1)$ \\
 & $i$ & number of algorithm iterations & $1\sim5~(1)$ \\ \hline
LED &  $f_{\tiny{\textrm{lim}}}$ & maximal object contrast & $0.5\sim20~(0.25)$ \\
 & $\sigma_{\tiny{\textrm{lim}}}$ & noise limit & $0.5\sim20~(0.25)$ \\ \hline
LUM &  $r$ & window radius & $1\sim5~(1)$ \\
 & $k$ & number of pixels in L and U set & $1\sim\frac{(2r+1)^2-1}{2}~(1)$ \\ \hline
MED &  $r$ & window radius & $1\sim5~(1)$ \\ \hline
PAV &  $r$ & window radius & $1\sim5~(1)$ \\ \hline
PSM & $r$ & window radius & $1\sim5~(1)$ \\
& $T$ &  initial threshold & $20\sim200~[e^-]~(20)$ \\\hline
RPAV &  $r$ & window radius & $1\sim5~(1)$ \\
 & $i$ & number of iterations &$1\sim5~(1)$ \\ \hline
TSM & $c$ & number of central pixel repetitions in CWM filter & $1\sim9~(1)$ \\
& $T$ &  initial threshold & $10\sim100000~[e^-]~(20)$ \\\hline

\end{tabular}
\label{tab_param2}
\end{table*}

\section{Noise templates}
Based on our data, collected for several years, we decided to create a benchmark set of impulsive noise templates, which would be useful for current and future evaluations of filtering methods. The proposed collection includes the dark frames from: (1) widely-used astronomical cameras corrupted by either the dark current or by cosmic ray impacts, (2) irradiated sensors, working in space environment and (3) electron multiplying CCD camera. Therefore, all the previously mentioned types of non-stationary impulsive noise were comprehensively combined in our database.

As the impulsive noise problem is usually related to the dark current generation in pixels, our benchmark set includes the dark frames acquired by off-the-shelf astronomical cameras (e.g. Santa Barbara Group SBIG) and by much more sophisticated equipment (Andor). Most of utilized instruments house the frame transfer or the interline CCDs with a Pelter cooling installed, enabling a thermal stabilization down to about 30-40\degree C below the ambient temperature. The cooling is usually sufficient to perform most of standard photometric series with negligible dark current, however if longer exposures are required (like in the spectroscopy of dim objects or in imaging in short wavelength bands) the impulsive noise bias has to be compensated. 

To include the impulsive noise templates from irradiated sensors working in space environment, we obtained the dark frame from one of the BRITE nanosatellite - BRITE Toronto (BTr, red filter satellite, lunched 19 Jun. 2014). Those nano-telescopes, due to the lack of space, could not be equipped with strong shielding and thus they have been heavily damaged by proton radiation. As a result, many hot pixels appeared and the charge transfer inefficiency developed, which introduced characteristic streaks starting from each hot pixel. The nano-telescopes do not include the mechanical shutter, therefore it was difficult to retrieve the frames with only dark current visible. We decided to take the images during the satellite fast movements in re-stabilization phase, so that the light-induced charge was negligible. It is worth to add, that the BRITE satellites usually work within 10-40\degree C temperature regime which further enhances the dark current generation in pixels. 

The CIC-induced noisy frames were obtained from Andor Luca S EMCCD camera. This camera has been recently successfully utilized, e.g. for the speckle interferometry of double stars (\cite{LucaSpackle1,LucaSpackle2}). The instrument is equipped with Texas Instruments EMCCD and a Pelter cooling -20\degree C. Since the electron multiplication is used mainly in the extremely fast acquisition modes (like in the Lucky Imaging or the wavefront sensing) and the CIC noise appears during the readout phase, we did not require long exposure times and retrieved the noise templates from a single bias frame (0 ms, EM gain set to the maximum). 

To include the noise templates with a significant number of cosmic ray impacts, the extremely long exposures were acquired using well-calibrated Andor iKon full-frame camera, strongly cooled down (i.e. to -90\degree C). We registered 9 exposures of 3000 seconds, and then the images were combined  using the median operation to obtain the master dark template, which showed only a stationary dark current pattern. In the next step, we summed the frames, compensated by master-dark subtraction, to expose all the cosmic rays events registered during our 9-hour recording. As the RTS behavior is extremely rare for the CCDs working on ground, all the spikes in such co-added frame must have originated in collisions of particles with the sensor. We carefully investigated such a resulting frame and found that indeed, nearly all the impulses were strongly clustered, which is typical for cosmic-ray impacts.

\begin{figure*}
\centering
\begin{subfigure}[b]{0.48\linewidth}
\includegraphics[width=\textwidth]{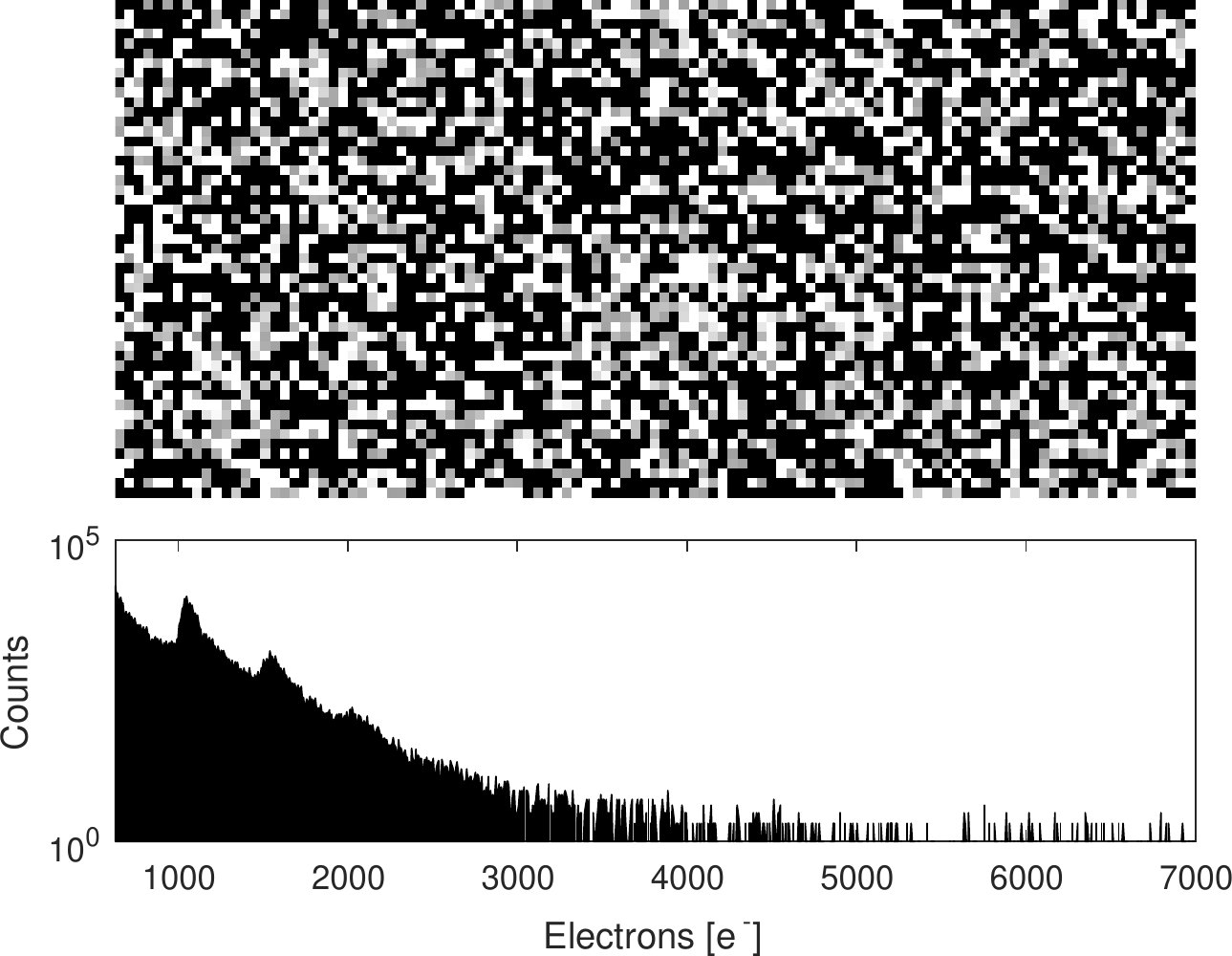}
\caption{Andor iKon-L (E2V CCD42-40).}
\end{subfigure}~~
\begin{subfigure}[b]{0.48\linewidth}
\includegraphics[width=\textwidth]{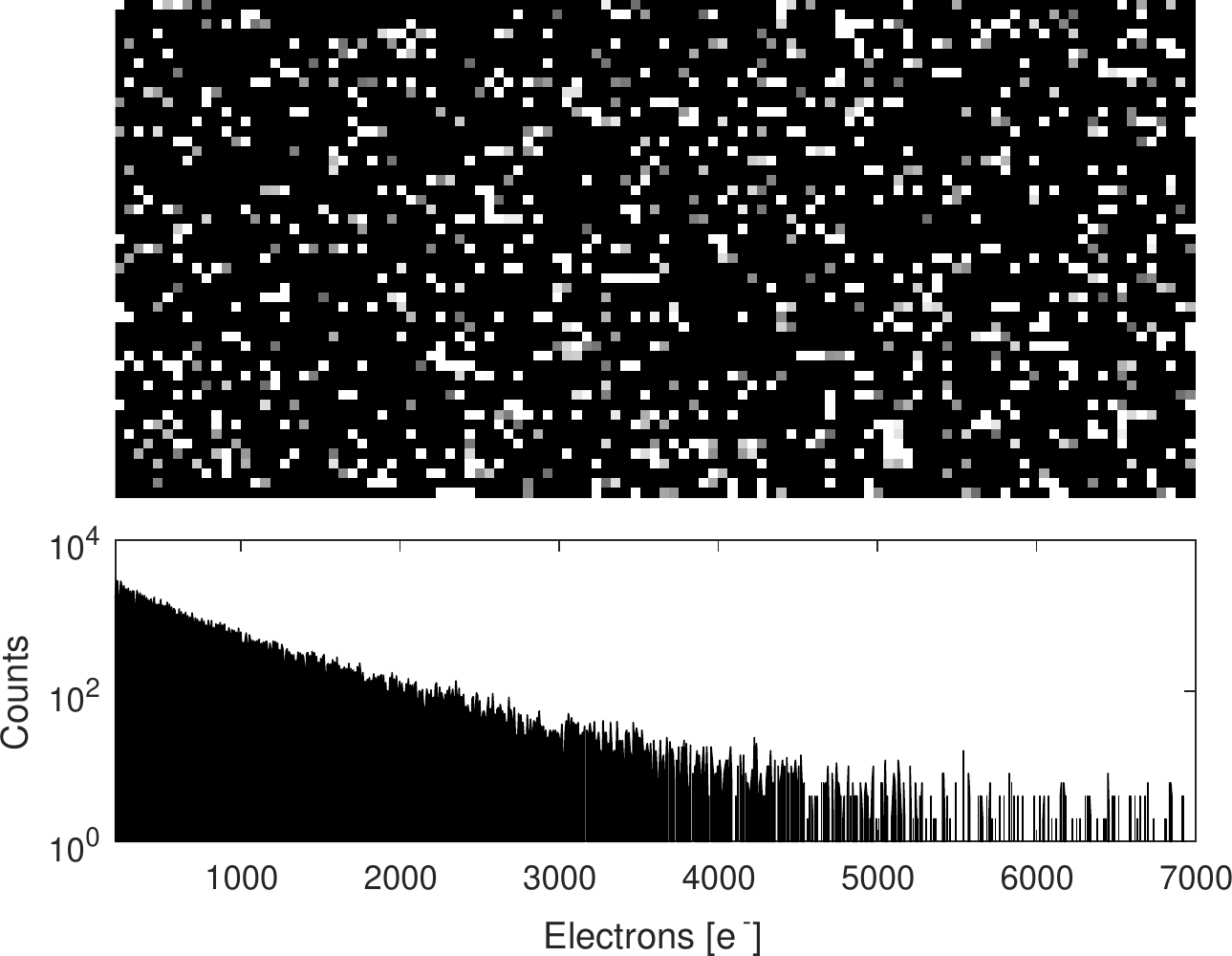}
\caption{Andor LucaS (TC247SPD).}
\end{subfigure}
\\\vspace{0.5cm}
\begin{subfigure}[b]{0.48\linewidth}
\includegraphics[width=\textwidth]{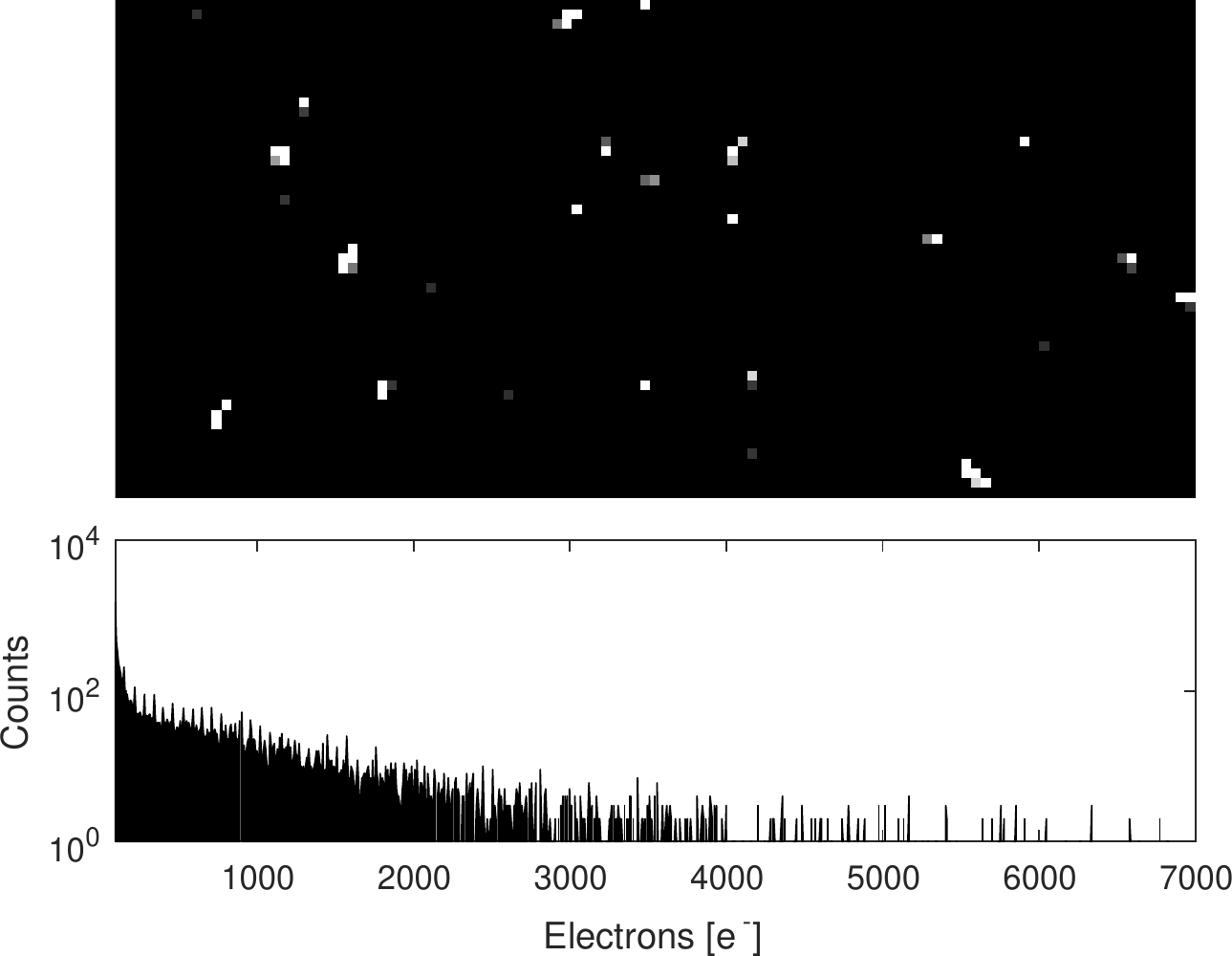};
\caption{Andor iKon-L (cosmic rays).}
\end{subfigure}~~
\begin{subfigure}[b]{0.48\linewidth}
\includegraphics[width=\textwidth]{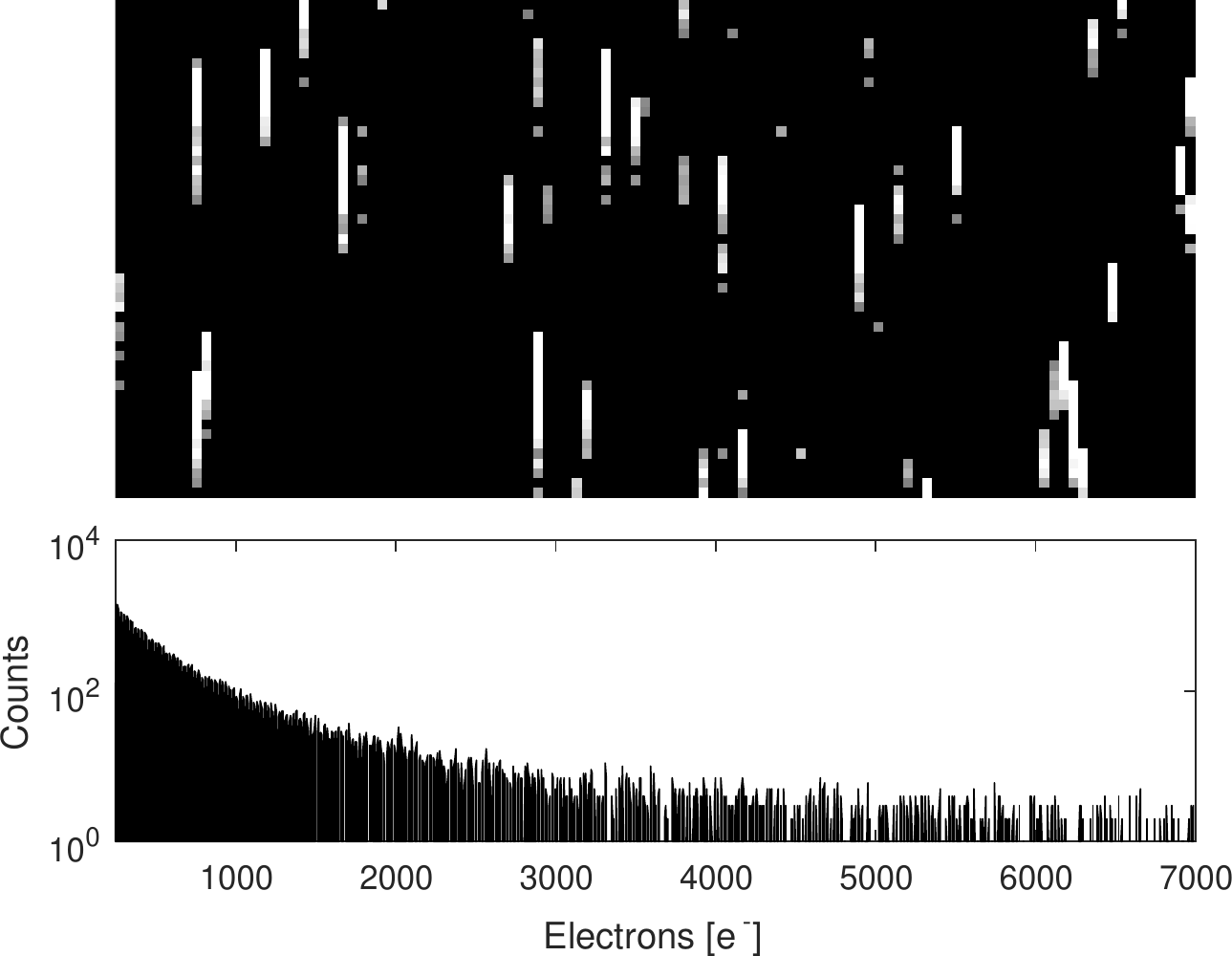}
\caption{BRITE Toronto (KAI 11000M).}
\end{subfigure}
\\\vspace{0.5cm}
\begin{subfigure}[b]{0.48\linewidth}
\includegraphics[width=\textwidth]{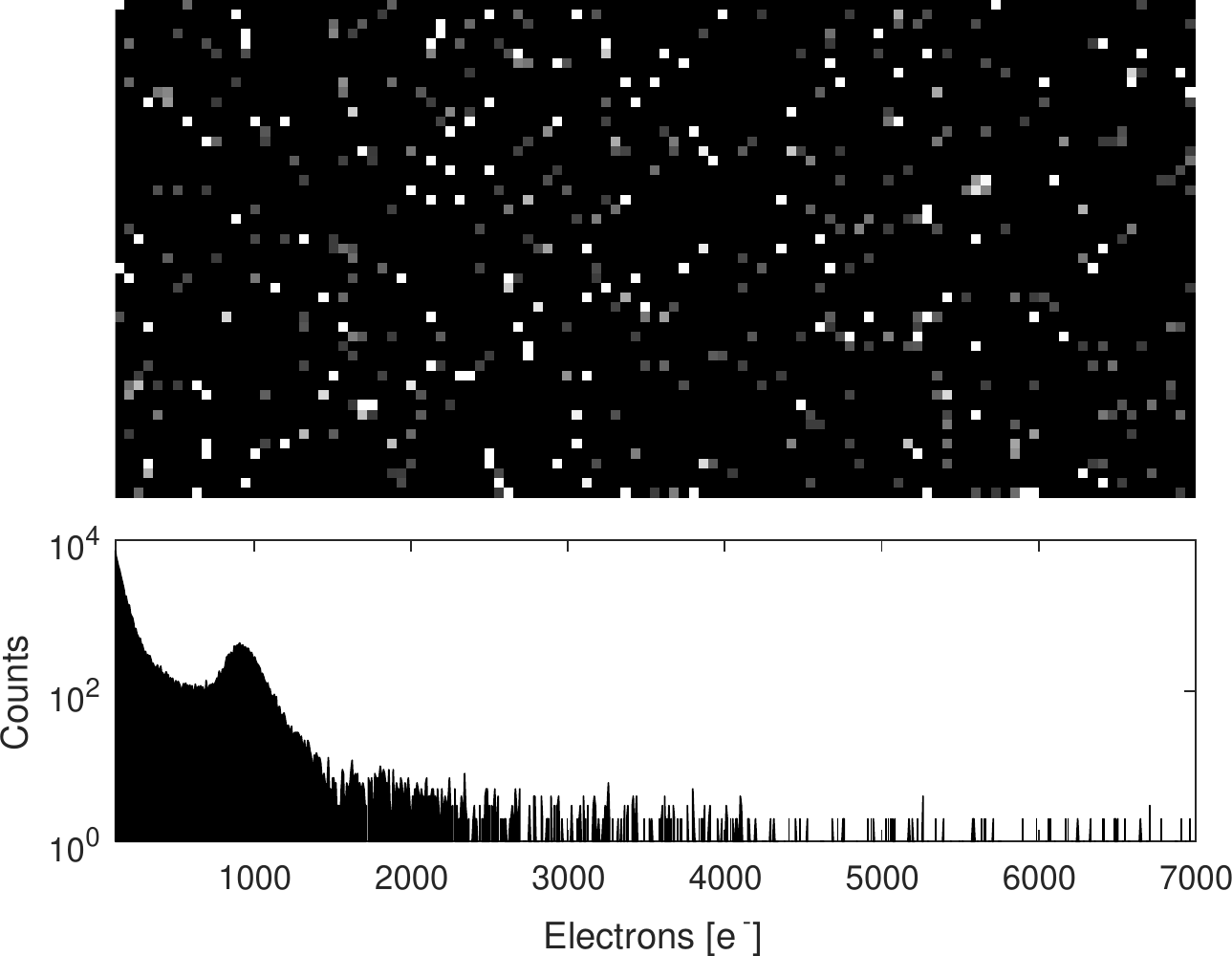}
\caption{SBIG 2000XM (KAI-2020M).}
\end{subfigure}~~
\begin{subfigure}[b]{0.48\linewidth}
\includegraphics[width=\textwidth]{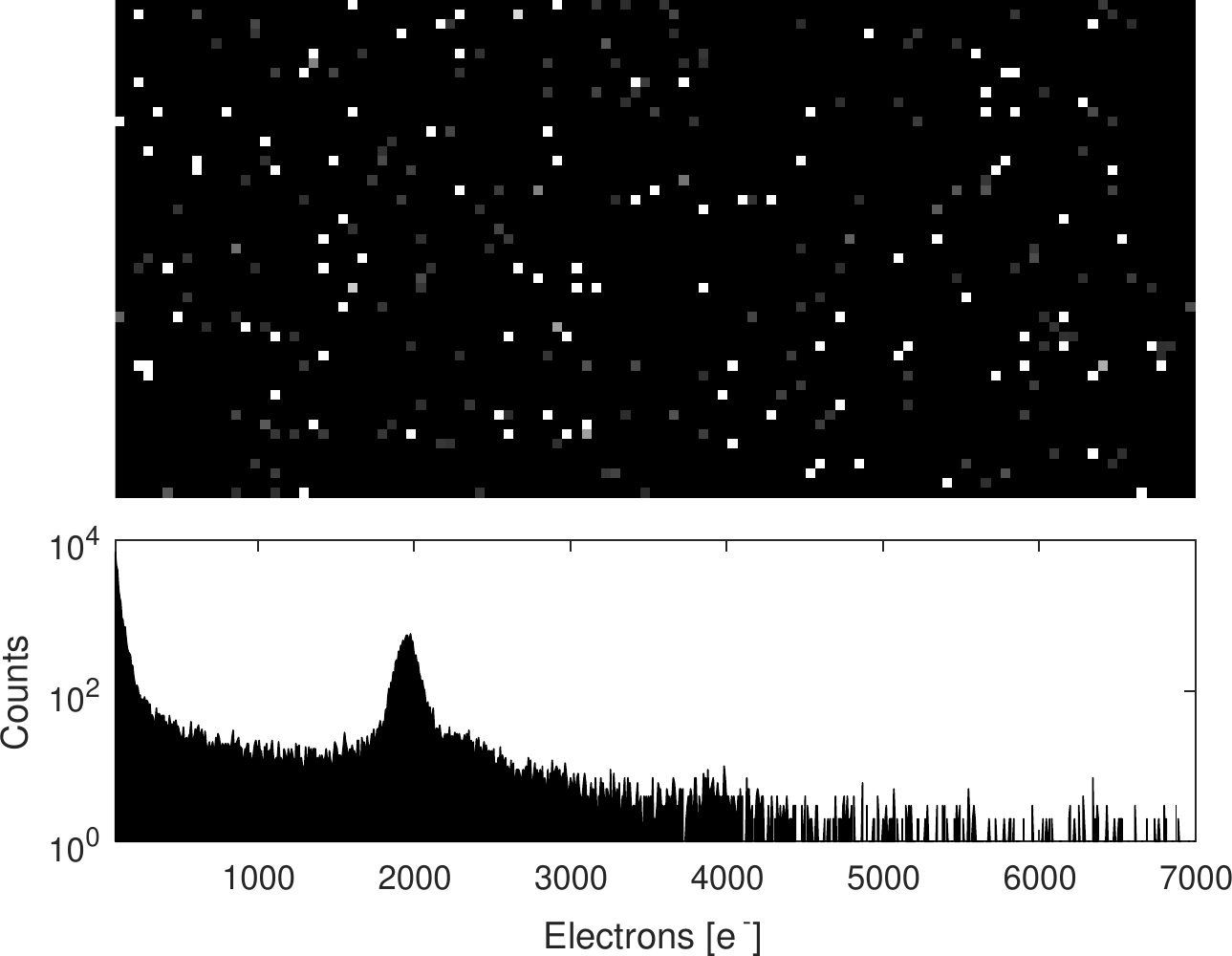}
\caption{SBIG ST-10XME (KAF-3200ME).}
\end{subfigure}

\caption{The parts (100$\times$50 pixels) of dark frames for the 6 employed cameras and their corresponding noise histograms. The dark frames were scaled so that mean intensity of an impulse equals 1000  [$e^-$]. The gray scale was set from 0 (black) to 1000 [$e^-$] (white).}
\label{allframes}
\end{figure*}

The description of employed CCD and EMCCD sensors is given in Tab. \ref{CameraTable1}. The frames utilized in our tests, with their corresponding noise distributions, are depicted in Fig. \ref{allframes}. All the templates used in our collection were calibrated by subtracting the high signal-to-noise bias frame (averaged over 9$\sim$41 frames, depending on camera) and multiplied by previously determined gain factor to expose the charge in electrons instead of digital units (ADUs - Analog to Digital Units). Since in the tests we needed to include only the impulsive noise pattern without the readout noise background, first we fitted the Gaussian distribution of counts resulting from the electronic noise ($\sigma_{e}$) and then we excluded all the pixels, with intensities below the threshold level of $5\sigma_{e}$. This assures that the probability of including a pixel not affected by impulsive noise is as low as $0.3\times10^{-6}$.The histograms of remaining impulsive pixels for the 6 employed cameras are presented in Fig. \ref{allframes}. In Tab. \ref{CameraTable2} we show basic statistics of extracted impulsive pixels, their median and mean charge.

For consistency of our database and for its future utilization for various purposes, the 1000$\times$1000 pixels regions from each dark frame were extracted. Only the frame obtained from EMCCD LucaS, due to its limited chip resolution (658$\times$496 pixels), had to be extended by replicating its lines and then columns (mirror copy), so that it fits the required size.

As it can be seen in the histograms (Fig. \ref{allframes}) and in the statistics of impulses (Tab. \ref{CameraTable2}), the noise varies noticeably from one sensor to another. Some CCDs have very strong multiple peaks (e.g. SBIG ST-10XME), while the others are characterized by smooth, continuous noise distribution (e.g. Andor LucaS). They also have various number of defective pixels and a wide range of generation rates. We would like to note, that the noise templates were selected from a much wider set of collected dark frames. However, they showed similar noise distributions, thus we assumed that the reduced set will sufficiently cover the range of impulsive noise scenarios, which may appear in astronomical cameras.

\begin{table*}
\centering
\caption{The overview of astronomical cameras utilized in the experiments. \vspace{0.2cm}}
\begin{tabular}{l | l | l | c | c | c}  
\hline
Camera & Sensor & Type & Exposure time & Temperature & Readout noise $\sigma_e$\\ \hline
Andor iKon L & E2V CCD42-40 & Back illuminated CCD & 1800 s & -10\degree C & $2.9~e^-$\vspace{0.5mm}\\
Andor Luca S EMCCD & Texas Instruments TC247SPD & EMCCD & 1000 s & 5\degree C &$15~e^-$\vspace{0.5mm}\vspace{0.5mm}\\
Andor iKon L (CR) & E2V CCD42-40 & Back illuminated CCD & 9$\times$3000 s & -90\degree C &$2.9~e^-$\vspace{0.5mm}\\ 
BRITE Toronto (BTr) & Kodak KAI-11002M & Interilne CCD & 1 s & 20\degree C &  20 e$^-$\vspace{0.5mm} \\
SBIG ST-10XME &  Kodak KAF-3200ME & Full frame CCD & 180 s & 0\degree C & 8.8 e$^-$\vspace{0.5mm}\\
SBIG 2000 & Kodak KAI-2001M & Interline CCD & 1000 s & 5\degree C & 13.5 e$^-$\vspace{0.5mm}\\ \hline

\end{tabular}
\label{CameraTable1}
\end{table*}

\begin{table}
\centering
\caption{Basic statistics of impulse noise in employed cameras.\vspace{0.2cm}}
\begin{tabular}{l |c | c | c }  
\hline
Camera & Impulsive pixels & Mean & Median \\ \hline\vspace{0.5mm}
Andor iKon L & 49\% & 844 e$^-$ & 848 e$^-$\\\vspace{0.5mm}
Andor Luca S EMCCD & 17.4\% & 356 e$^-$ &216 e$^-$\\\vspace{0.5mm}
Andor iKon L (CR) & 1.04\% & 63 e$^-$  &24 e$^-$ \\\vspace{0.5mm}
BRITE Toronto& 3.9\% & 409 e$^-$ & 196 e$^-$\\\vspace{0.5mm}
SBIG ST-10XME & 4.6\% & 503 e$^-$ & 74 e$^-$\\\vspace{0.5mm}
SBIG 2000 & 8.3\% & 585 e$^-$ & 111 e$^-$\\\hline

\end{tabular}
\label{CameraTable2}
\end{table}

\section{Experiments}

\subsection{Synthetic stellar profiles}
\begin{figure}
\centering
\begin{subfigure}[b]{0.49\linewidth}
\begin{tikzpicture}[scale=0.5,spy using outlines={circle,white,magnification=10,size=2.5cm, connect spies}]
\node {\includegraphics[width=\textwidth]{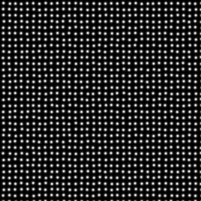}};
\spy on (-1.13,-0.95) in node [left] at (4,1.5);
\end{tikzpicture}
\caption{$\sigma_{\tiny{\textrm{PSF}}}=1$ [pix]\vspace{2mm}}
\end{subfigure}
\begin{subfigure}[b]{0.49\linewidth}
\begin{tikzpicture}[scale=0.5,spy using outlines={circle,gray,magnification=12,size=2.5cm, connect spies}]
\node {\includegraphics[width=\textwidth]{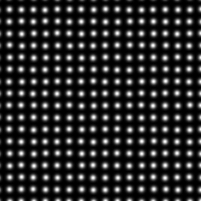}};
\spy on (-1.05,-1.08) in node [left] at (4,1.5);
\end{tikzpicture}
\caption{$\sigma_{\tiny{\textrm{PSF}}}=2$ [pix]\vspace{2mm}}
\end{subfigure}
\begin{subfigure}[b]{0.49\linewidth}
\begin{tikzpicture}[scale=0.5,spy using outlines={circle,gray,magnification=12,size=2.5cm, connect spies}]
\node {\includegraphics[width=\textwidth]{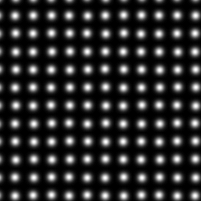}};
\spy on (-1.3,-1.2) in node [left] at (4,1.5);
\end{tikzpicture}
\caption{$\sigma_{\tiny{\textrm{PSF}}}=3$ [pix]\vspace{2mm}}
\end{subfigure}
\begin{subfigure}[b]{0.49\linewidth}
\begin{tikzpicture}[scale=0.5,spy using outlines={circle,gray,magnification=12,size=2.5cm, connect spies}]
\node {\includegraphics[width=\textwidth]{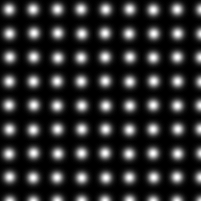}};
\spy on (-1.3,-1.08) in node [left] at (4,1.5);
\end{tikzpicture}
\caption{$\sigma_{\tiny{\textrm{PSF}}}=4$ [pix]\vspace{2mm}}
\end{subfigure}
\caption{Parts of reference simulated images for various PSF widths.}
\label{starsPSFs}
\end{figure}

In the first part of our experiments, we used synthetic star images corrupted by impulsive noise of various distribution and power. For this purpose, the four matrices of Gaussian stellar profiles (\cite{BadPixelPopowicz,gauss2}) were created, with various widths of point spread function (PSF) $\sigma_{\tiny{\textrm{PSF}}}=1,2,3,4$ pixels. The amplitude of the Gaussians was set permanently to 1000 [e$^-$]. The stars were arranged one next to another, so that their centroids were separated by $8~\sigma_{\tiny{\textrm{PSF}}}$. They were well separated and, simultaneously, they filled the whole image area ($1000\times10000$) to maximize the probability of disturbing a profile by an impulse. Only small 11-pixels margins were left at the frame boundaries to avoid filtering artifacts while using larger sliding windows (the margin width of 11 pixels, was determined from the largest filtering mask used in BDND method, where $r_2=10$). To include the effect of various sampling, the centroids were slightly shifted in $x$ and $y$ axis by $-0.5\sim0.5$ pixels (uniform distribution). This can be especially important for lower $\sigma_{\tiny{\textrm{PSF}}}$ where high PSF gradients appear. Exemplary cropped and zoomed parts from reference images are exposed in Fig. \ref{starsPSFs}.

To embed impulsive noise, the set of dark frames obtained from astronomical cameras was added to noiseless, reference images. Since the stellar profiles have a constant amplitude (1000 [e$^-$]), the varying signal to noise ratio (SNR) was achieved by using dark frames with scaled intensities of pixels. Therefore, the SNR in our experiments was defined as the ratio of a profile amplitude to the mean intensity of a single impulse. Thus, our SNR is defined on a per-pixel basis, not on the estimation of signal and noise in a whole stellar profile (or in whole image) as it is usually assumed. We believe that such a SNR definition allows for comfortable utilization of the comparison results. With the parameters easily accessible from a raw image (star amplitude, its ratio to the mean intensity of an impulse and a stellar width $\sigma_{\tiny{\textrm{PSF}}}$), one can easily find the most accurate denoising algorithm and its optimal parameter settings.

We evaluated the photometric and astrometric performance after applying various filtering algorithms. In the case of photometry, each stellar profile was measured using circular aperture of $3\sigma_{\tiny{\textrm{PSF}}}$ radius around a reference centroid position. For the astrometry, first the image was convolved with the Gaussian profile, which is a widely used step to enhance desired objects in noisy backgrounds (\cite{Masias,PopowiczBackground}). In the next step, the local maximum was found around reference centroid position, within the radius of $6\sigma_{\tiny{\textrm{PSF}}}$. Eventually, to obtain a fine, subpixel centroid position, the center of gravity was computed using the pixels within $3\sigma_{\tiny{\textrm{PSF}}}$ around the found maximum.

\begin{figure*}
\centering
\begin{subfigure}[b]{0.48\linewidth}
\includegraphics[width=\textwidth]{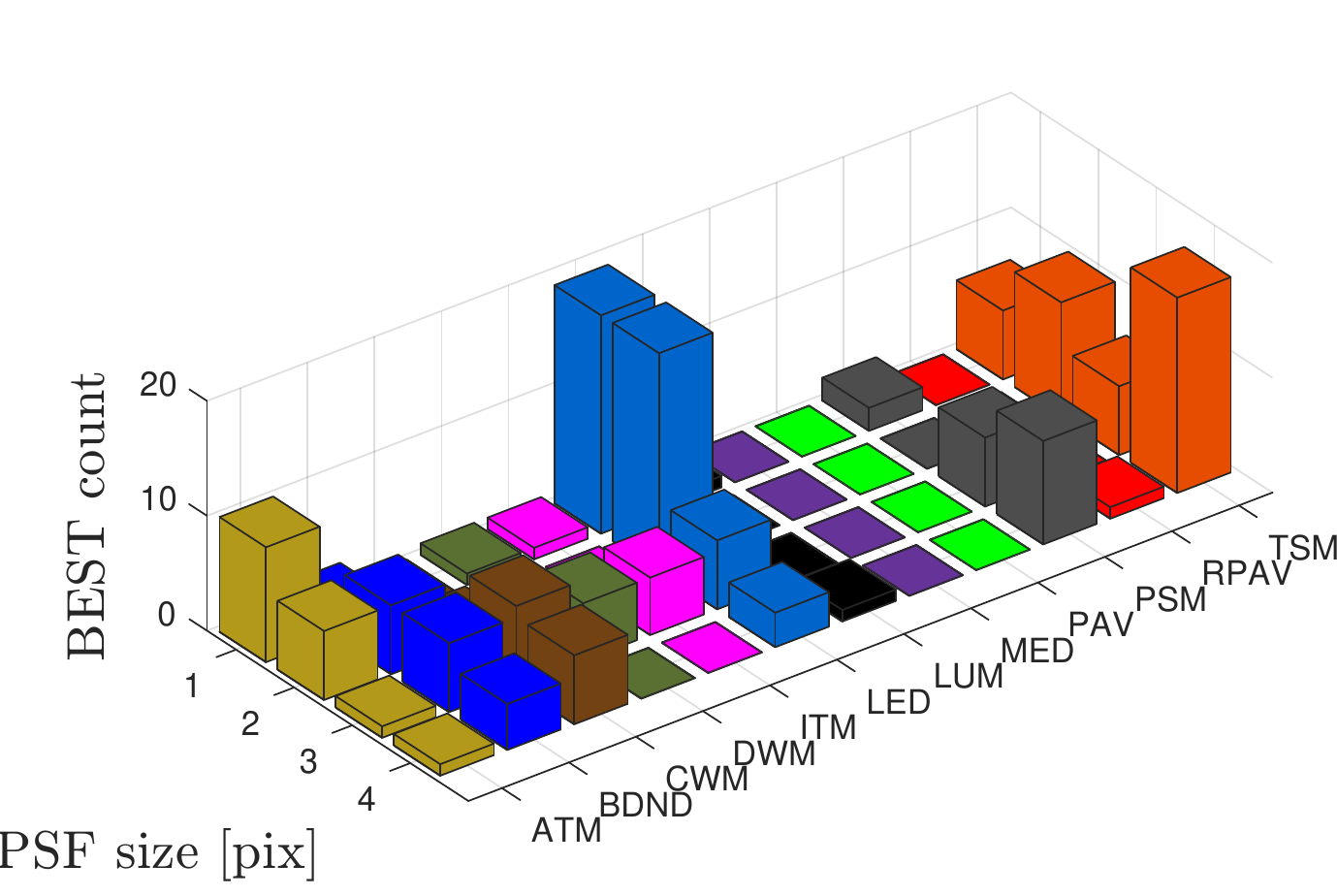}
\end{subfigure}~
\begin{subfigure}[b]{0.48\linewidth}
\includegraphics[width=\textwidth]{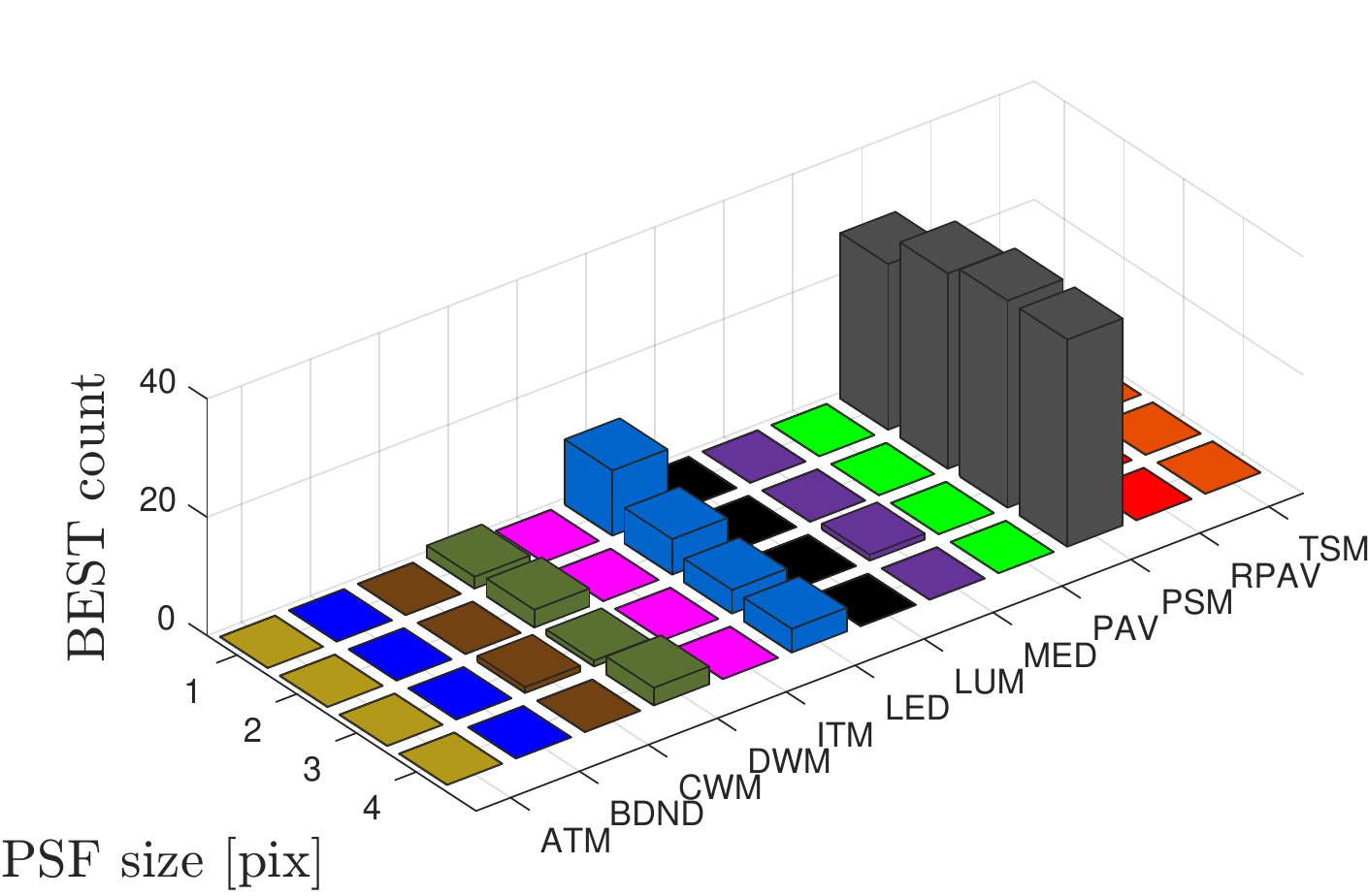}
\end{subfigure}
\\
\begin{subfigure}[b]{0.48\linewidth}
\includegraphics[width=\textwidth]{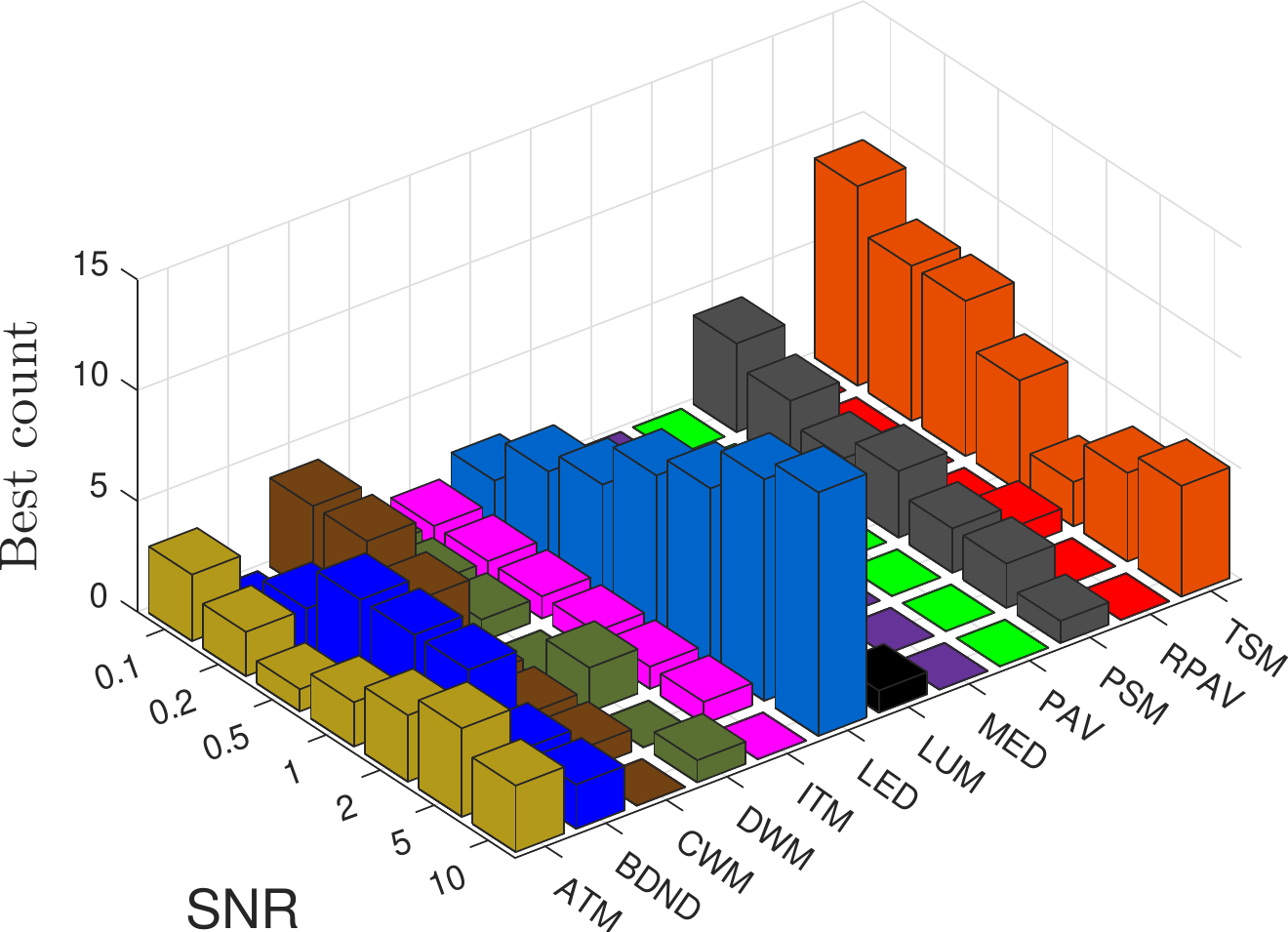}
\end{subfigure}~
\begin{subfigure}[b]{0.48\linewidth}
\includegraphics[width=\textwidth]{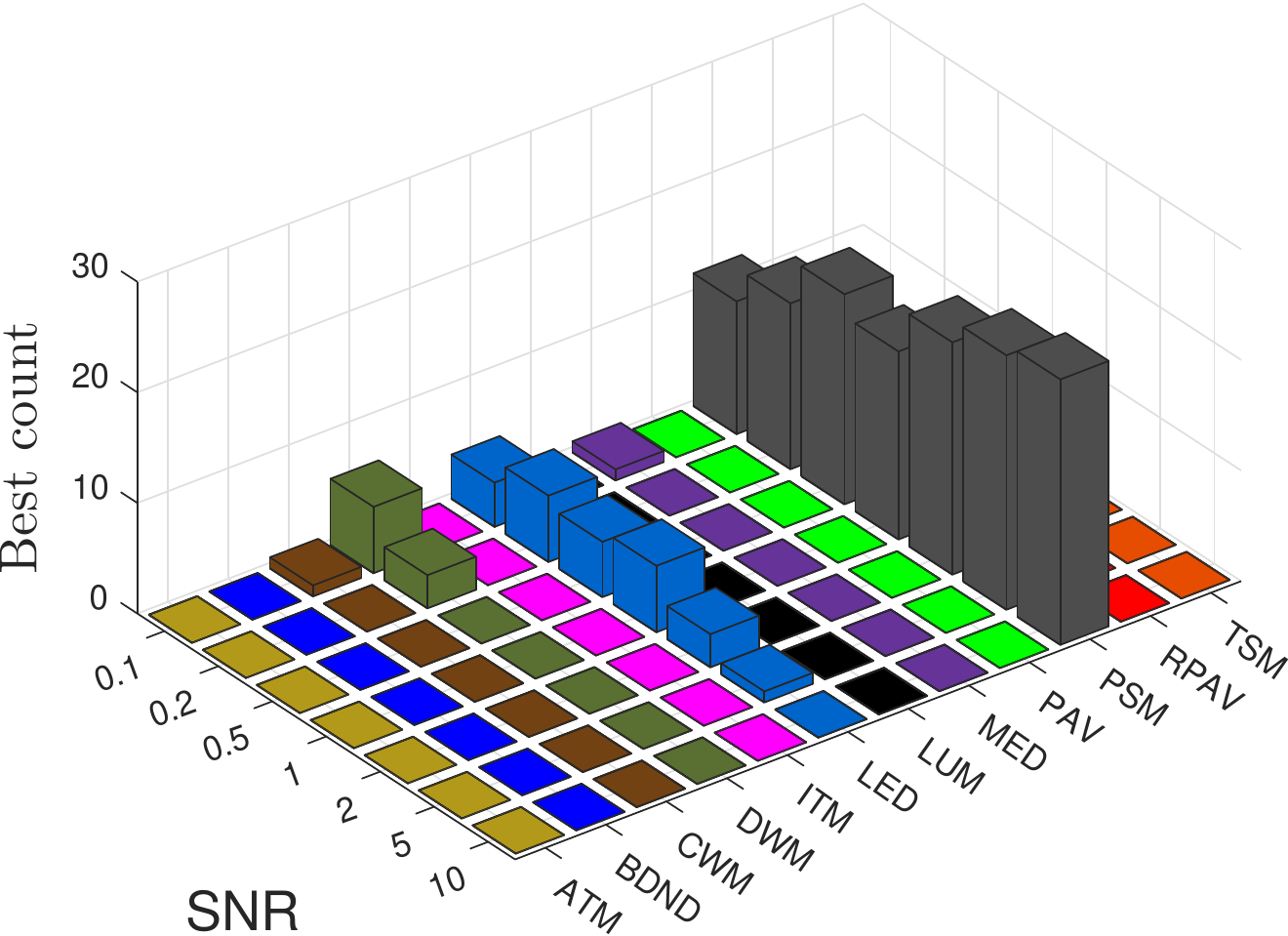}
\end{subfigure}
\\
\begin{subfigure}[b]{0.48\linewidth}
\includegraphics[width=\textwidth]{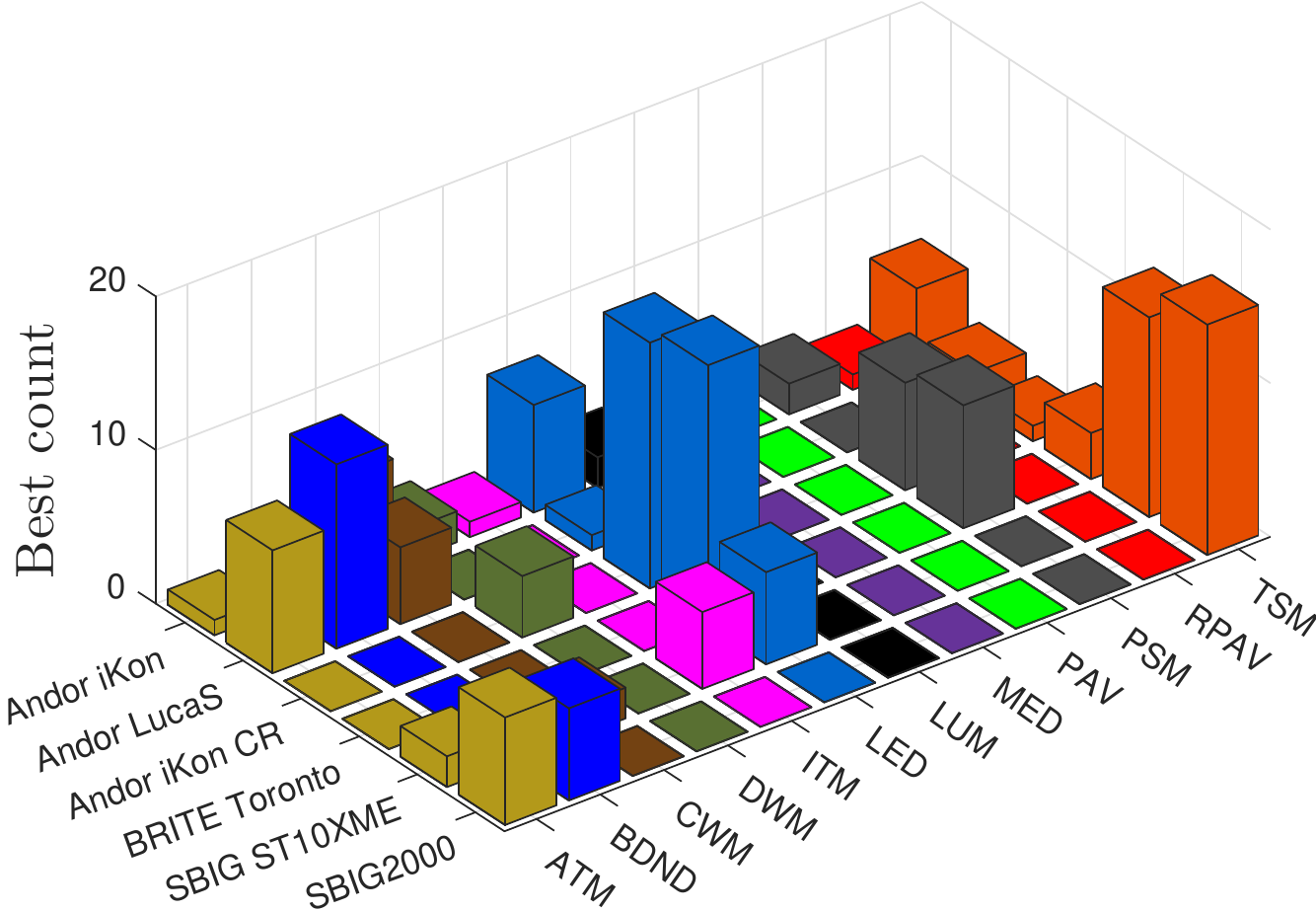}
\caption{Photometry}
\end{subfigure}~
\begin{subfigure}[b]{0.48\linewidth}
\includegraphics[width=\textwidth]{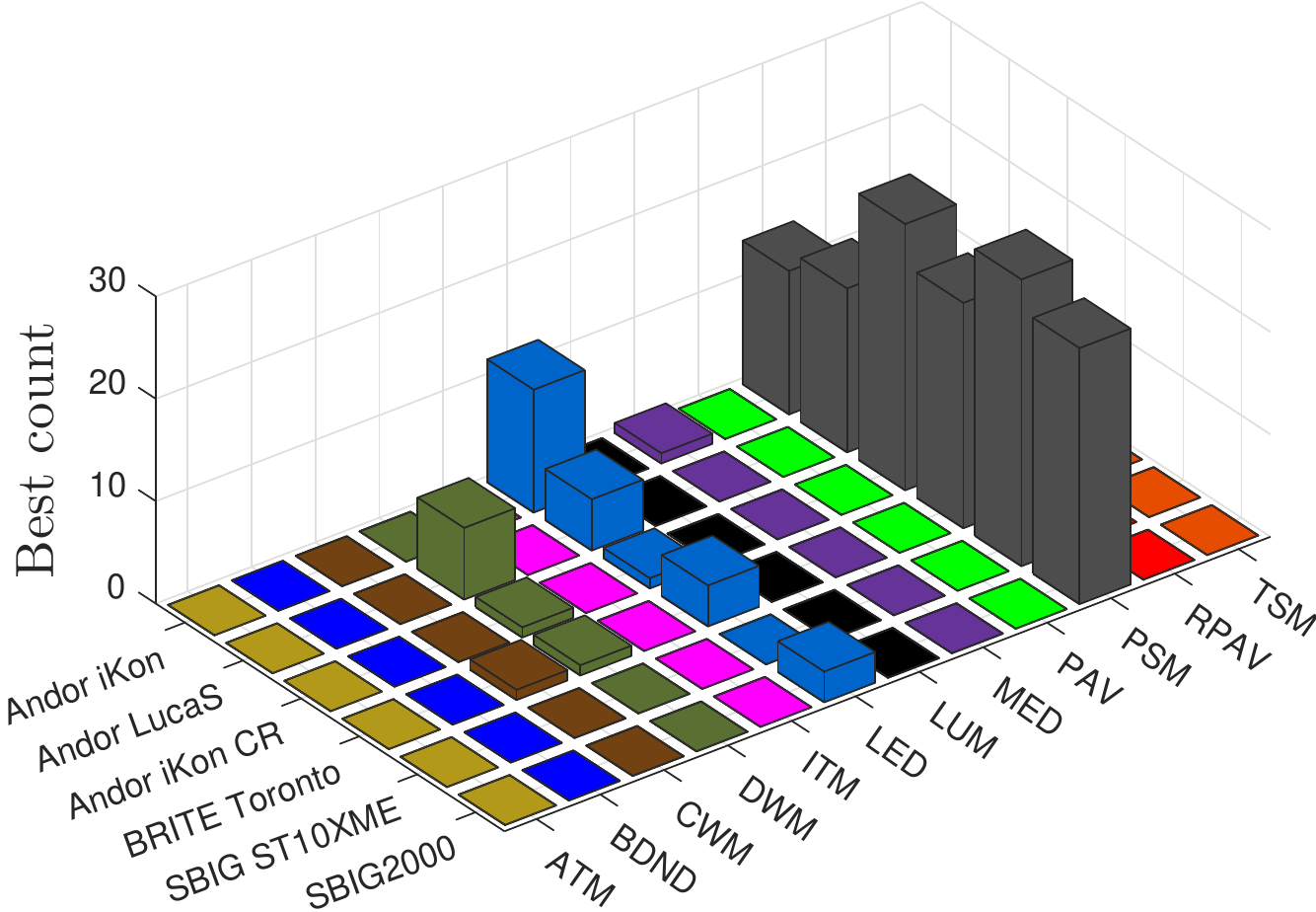}
\caption{Astrometry}
\end{subfigure}
\caption{Ranking of impulsive noise reduction methods based on photometric (a) and astrometric (b) evaluation on synthetic stellar profiles. The \emph{Best count} corresponds to the number of occurrences a method achieved the highest accuracy (the lowest photometric/astrometric error). }
\label{artificial_statisics}
\end{figure*}

The outcomes of photometry were compared with the reference values and the mean magnitude error ($E_m$, expressed in magnitudes) was determined. Similarly, the centroid estimations of stars were compared with the reference positions. The distance of estimated centroid to the true one was averaged over all stars in image array to obtain the mean centroid error ($E_a$, expressed in pixels). The most accurate methods, depending on the noise template, SNR ratio and PSF width, are listed in Tabs. \ref{photometric_accuracy} and \ref{astrometric_accuracy}, respectively for photometric and astrometric performance. The rankings of methods accuracy are displayed using bar plots in Fig. \ref{artificial_statisics}. The coloring employed in tables and in bar plots are consistent so that each method is assigned with a unique color.

\begin{table*}

\caption{Photometric accuracy of filtering methods evaluated on synthetic images. The most effective method with its optimal parameters and the corresponding magnitude error ($E_m$ [mag]) is provided in each Table cell. The SNR is defined as the ratio of stellar amplitude to the mean intensity of an impulse.}
\label{my-label}
\begin{tabular}{lllllllll}
     & \multicolumn{2}{c}{$\sigma_{\textrm{\tiny{PSF}}}$ = 1 pix.}& \multicolumn{2}{c}{$\sigma_{\textrm{\tiny{PSF}}}$ = 2 pix.} & \multicolumn{2}{c}{$\sigma_{\textrm{\tiny{PSF}}}$ = 3 pix.} & \multicolumn{2}{c}{$\sigma_{\textrm{\tiny{PSF}}}$ = 4 pix.} \\\cmidrule(lr){2-3} \cmidrule(lr){4-5} \cmidrule(lr){6-7} \cmidrule(lr){8-9} 
 SNR    & Method & $E_m$ & Method & $E_m$ & Method & $E_m$ & Method & $E_m$ \\\hline
\multicolumn{9}{c}{Andor iKon L (49\% defective pixels)}\\ \hline0.1& \color[rgb]{0.9,0.3,0}TSM (10/7) & \footnotesize{0.501}& \color[rgb]{0.9,0.3,0}TSM (50/6) & \footnotesize{0.318}& \color[rgb]{0.451,0.258,0.07}CWM (2/15) & \footnotesize{0.262}& \color[rgb]{0.451,0.258,0.07}CWM (2/15) & \footnotesize{0.246}\\
0.2& \color[rgb]{0,0.398,0.796}LED (0.25/20) & \footnotesize{0.47}& \color[rgb]{0.9,0.3,0}TSM (100/6) & \footnotesize{0.307}& \color[rgb]{0.451,0.258,0.07}CWM (2/15) & \footnotesize{0.254}& \color[rgb]{0.451,0.258,0.07}CWM (2/15) & \footnotesize{0.237}\\
0.5& \color[rgb]{0,0.398,0.796}LED (0.25/20) & \footnotesize{0.357}& \color[rgb]{0.9,0.3,0}TSM (100/6) & \footnotesize{0.281}& \color[rgb]{0.451,0.258,0.07}CWM (2/15) & \footnotesize{0.231}& \color[rgb]{0.451,0.258,0.07}CWM (2/15) & \footnotesize{0.216}\\
1.0& \color[rgb]{0,0.398,0.796}LED (0.25/20) & \footnotesize{0.294}& \color[rgb]{0.9,0.3,0}TSM (100/6) & \footnotesize{0.247}& \color[rgb]{0.451,0.258,0.07}CWM (2/15) & \footnotesize{0.203}& \color[rgb]{0,0,0}LUM (4/3) & \footnotesize{0.188}\\
2.0& \color[rgb]{0,0.398,0.796}LED (0.25/20) & \footnotesize{0.23}& \color[rgb]{0,0.398,0.796}LED (0.25/20) & \footnotesize{0.182}& \color[rgb]{0.35,0.44,0.1967}DWM (1/5000) & \footnotesize{0.164}& \color[rgb]{1,0,0}RPAV (2/2) & \footnotesize{0.15}\\
5.0& \color[rgb]{1,0,1}ITM (4/2) & \footnotesize{0.143}& \color[rgb]{0,0.398,0.796}LED (0.25/20) & \footnotesize{0.099}& \color[rgb]{0.3,0.3,0.3}PSM (2000/4) & \footnotesize{0.098}& \color[rgb]{0.7,0.6,0.1}ATM (4/4) & \footnotesize{0.092}\\
10.0& \color[rgb]{0,0,0}LUM (3/3) & \footnotesize{0.085}& \color[rgb]{0,0.398,0.796}LED (0.25/20) & \footnotesize{0.059}& \color[rgb]{0.35,0.44,0.1967}DWM (2/10000) & \footnotesize{0.055}& \color[rgb]{0.3,0.3,0.3}PSM (2000/4) & \footnotesize{0.053}\\
\hline
\multicolumn{9}{c}{Andor LucaS (17.4\% noised defective pixels)}\\ \hline0.1& \color[rgb]{0.7,0.6,0.1}ATM (4/1) & \footnotesize{0.25}& \color[rgb]{0.9,0.3,0}TSM (10000/7) & \footnotesize{0.299}& \color[rgb]{0.451,0.258,0.07}CWM (1/10) & \footnotesize{0.532}& \color[rgb]{0.451,0.258,0.07}CWM (1/10) & \footnotesize{0.304}\\
0.2& \color[rgb]{0.7,0.6,0.1}ATM (4/1) & \footnotesize{0.222}& \color[rgb]{0,0,1}BDND (4/5) & \footnotesize{0.273}& \color[rgb]{0.451,0.258,0.07}CWM (1/10) & \footnotesize{0.495}& \color[rgb]{0.451,0.258,0.07}CWM (1/10) & \footnotesize{0.281}\\
0.5& \color[rgb]{0.7,0.6,0.1}ATM (4/2) & \footnotesize{0.16}& \color[rgb]{0,0,1}BDND (4/2) & \footnotesize{0.208}& \color[rgb]{0,0,1}BDND (4/4) & \footnotesize{0.296}& \color[rgb]{0.451,0.258,0.07}CWM (1/10) & \footnotesize{0.229}\\
1.0& \color[rgb]{0.7,0.6,0.1}ATM (4/4) & \footnotesize{0.113}& \color[rgb]{0,0,1}BDND (3/2) & \footnotesize{0.162}& \color[rgb]{0,0,1}BDND (4/4) & \footnotesize{0.151}& \color[rgb]{0,0,1}BDND (4/4) & \footnotesize{0.126}\\
2.0& \color[rgb]{0.7,0.6,0.1}ATM (2/2) & \footnotesize{0.112}& \color[rgb]{0.7,0.6,0.1}ATM (4/1) & \footnotesize{0.089}& \color[rgb]{0,0,1}BDND (4/5) & \footnotesize{0.082}& \color[rgb]{0,0,1}BDND (4/7) & \footnotesize{0.069}\\
5.0& \color[rgb]{0.9,0.3,0}TSM (1000/7) & \footnotesize{0.069}& \color[rgb]{0.7,0.6,0.1}ATM (4/1) & \footnotesize{0.033}& \color[rgb]{0,0,1}BDND (4/3) & \footnotesize{0.051}& \color[rgb]{0,0,1}BDND (4/3) & \footnotesize{0.036}\\
10.0& \color[rgb]{0,0.398,0.796}LED (20/10) & \footnotesize{0.036}& \color[rgb]{0.7,0.6,0.1}ATM (4/2) & \footnotesize{0.025}& \color[rgb]{0,0,1}BDND (3/3) & \footnotesize{0.044}& \color[rgb]{0,0,1}BDND (3/9) & \footnotesize{0.023}\\
\hline
\multicolumn{9}{c}{Andor iKon CR (1.04\% defective pixels)}\\ \hline0.1& \color[rgb]{0.35,0.44,0.1967}DWM (4/5000) & \footnotesize{0.019}& \color[rgb]{0,0.398,0.796}LED (1/0.25) & \footnotesize{0.006}& \color[rgb]{0.3,0.3,0.3}PSM (1000/3) & \footnotesize{0.131}& \color[rgb]{0.3,0.3,0.3}PSM (10000/2) & \footnotesize{0.515}\\
0.2& \color[rgb]{0,0.398,0.796}LED (10/2) & \footnotesize{0.024}& \color[rgb]{0,0.398,0.796}LED (2/0.25) & \footnotesize{0.004}& \color[rgb]{0.35,0.44,0.1967}DWM (2/5000) & \footnotesize{0.069}& \color[rgb]{0.3,0.3,0.3}PSM (10000/2) & \footnotesize{0.321}\\
0.5& \color[rgb]{0,0.398,0.796}LED (0.25/5) & \footnotesize{0.028}& \color[rgb]{0,0.398,0.796}LED (2/0.25) & \footnotesize{0.008}& \color[rgb]{0.35,0.44,0.1967}DWM (2/5000) & \footnotesize{0.027}& \color[rgb]{0.3,0.3,0.3}PSM (10000/3) & \footnotesize{0.152}\\
1.0& \color[rgb]{0,0.398,0.796}LED (1/5) & \footnotesize{0.023}& \color[rgb]{0,0.398,0.796}LED (2/0.25) & \footnotesize{0.005}& \color[rgb]{0,0.398,0.796}LED (0.25/1) & \footnotesize{0.01}& \color[rgb]{0.3,0.3,0.3}PSM (10000/3) & \footnotesize{0.081}\\
2.0& \color[rgb]{0,0.398,0.796}LED (10/5) & \footnotesize{0.017}& \color[rgb]{0,0.398,0.796}LED (1/1) & \footnotesize{0.004}& \color[rgb]{0.35,0.44,0.1967}DWM (3/10000) & \footnotesize{0.006}& \color[rgb]{0.3,0.3,0.3}PSM (10000/3) & \footnotesize{0.041}\\
5.0& \color[rgb]{0,0.398,0.796}LED (10/5) & \footnotesize{0.016}& \color[rgb]{0,0.398,0.796}LED (1/1) & \footnotesize{0.004}& \color[rgb]{0,0.398,0.796}LED (5/0.25) & \footnotesize{0.001}& \color[rgb]{0.3,0.3,0.3}PSM (10000/4) & \footnotesize{0.016}\\
10.0& \color[rgb]{0.9,0.3,0}TSM (2000/7) & \footnotesize{0.015}& \color[rgb]{0,0.398,0.796}LED (2/1) & \footnotesize{0.004}& \color[rgb]{0,0.398,0.796}LED (5/0.25) & \footnotesize{0.002}& \color[rgb]{0,0.398,0.796}LED (0.25/1) & \footnotesize{0.003}\\
\hline
\multicolumn{9}{c}{BRITE Toronto (3.9\% defective pixels)}\\ \hline0.1& \color[rgb]{0.3,0.3,0.3}PSM (100/4) & \footnotesize{0.035}& \color[rgb]{0,0.398,0.796}LED (1/0.25) & \footnotesize{0.012}& \color[rgb]{0.3,0.3,0.3}PSM (500/4) & \footnotesize{0.256}& \color[rgb]{0.9,0.3,0}TSM (500/3) & \footnotesize{0.272}\\
0.2& \color[rgb]{0.3,0.3,0.3}PSM (200/4) & \footnotesize{0.036}& \color[rgb]{0,0.398,0.796}LED (2/0.25) & \footnotesize{0.01}& \color[rgb]{0.3,0.3,0.3}PSM (500/3) & \footnotesize{0.153}& \color[rgb]{0.9,0.3,0}TSM (500/3) & \footnotesize{0.167}\\
0.5& \color[rgb]{0,0.398,0.796}LED (10/2) & \footnotesize{0.047}& \color[rgb]{0,0.398,0.796}LED (2/0.25) & \footnotesize{0.008}& \color[rgb]{0.3,0.3,0.3}PSM (500/4) & \footnotesize{0.07}& \color[rgb]{0.9,0.3,0}TSM (500/3) & \footnotesize{0.078}\\
1.0& \color[rgb]{0,0.398,0.796}LED (2/2) & \footnotesize{0.049}& \color[rgb]{0,0.398,0.796}LED (5/0.25) & \footnotesize{0.009}& \color[rgb]{0.3,0.3,0.3}PSM (500/3) & \footnotesize{0.037}& \color[rgb]{0.3,0.3,0.3}PSM (500/2) & \footnotesize{0.041}\\
2.0& \color[rgb]{0,0.398,0.796}LED (5/2) & \footnotesize{0.042}& \color[rgb]{0,0.398,0.796}LED (5/0.25) & \footnotesize{0.008}& \color[rgb]{0,0.398,0.796}LED (0.25/1) & \footnotesize{0.017}& \color[rgb]{0.3,0.3,0.3}PSM (500/2) & \footnotesize{0.02}\\
5.0& \color[rgb]{0,0.398,0.796}LED (1/5) & \footnotesize{0.031}& \color[rgb]{0,0.398,0.796}LED (5/0.25) & \footnotesize{0.01}& \color[rgb]{0,0.398,0.796}LED (0.5/1) & \footnotesize{0.006}& \color[rgb]{0,0.398,0.796}LED (0.25/1) & \footnotesize{0.008}\\
10.0& \color[rgb]{0,0.398,0.796}LED (2/5) & \footnotesize{0.023}& \color[rgb]{0,0.398,0.796}LED (5/0.5) & \footnotesize{0.007}& \color[rgb]{0,0.398,0.796}LED (5/1) & \footnotesize{0.004}& \color[rgb]{0,0.398,0.796}LED (10/0.25) & \footnotesize{0.004}\\
\hline
\multicolumn{9}{c}{SBIG ST10XME (4.6\% defective pixels)}\\ \hline0.1& \color[rgb]{0.7,0.6,0.1}ATM (3/4) & \footnotesize{0.101}& \color[rgb]{0.9,0.3,0}TSM (1000/3) & \footnotesize{0.191}& \color[rgb]{1,0,1}ITM (1/1) & \footnotesize{0.196}& \color[rgb]{0.9,0.3,0}TSM (5000/7) & \footnotesize{0.192}\\
0.2& \color[rgb]{0.9,0.3,0}TSM (200/3) & \footnotesize{0.091}& \color[rgb]{0.7,0.6,0.1}ATM (4/3) & \footnotesize{0.102}& \color[rgb]{1,0,1}ITM (1/1) & \footnotesize{0.122}& \color[rgb]{0.9,0.3,0}TSM (5000/7) & \footnotesize{0.147}\\
0.5& \color[rgb]{0.9,0.3,0}TSM (500/2) & \footnotesize{0.048}& \color[rgb]{0.9,0.3,0}TSM (1000/2) & \footnotesize{0.045}& \color[rgb]{1,0,1}ITM (1/1) & \footnotesize{0.056}& \color[rgb]{0.9,0.3,0}TSM (5000/7) & \footnotesize{0.086}\\
1.0& \color[rgb]{0,0.398,0.796}LED (20/5) & \footnotesize{0.028}& \color[rgb]{0.9,0.3,0}TSM (1000/2) & \footnotesize{0.022}& \color[rgb]{1,0,1}ITM (1/1) & \footnotesize{0.028}& \color[rgb]{0.9,0.3,0}TSM (5000/7) & \footnotesize{0.051}\\
2.0& \color[rgb]{0,0.398,0.796}LED (5/5) & \footnotesize{0.02}& \color[rgb]{0.451,0.258,0.07}CWM (1/2) & \footnotesize{0.01}& \color[rgb]{1,0,1}ITM (2/1) & \footnotesize{0.014}& \color[rgb]{0.9,0.3,0}TSM (5000/7) & \footnotesize{0.028}\\
5.0& \color[rgb]{0,0.398,0.796}LED (10/5) & \footnotesize{0.02}& \color[rgb]{0,0.398,0.796}LED (20/1) & \footnotesize{0.004}& \color[rgb]{0.451,0.258,0.07}CWM (1/2) & \footnotesize{0.004}& \color[rgb]{0.9,0.3,0}TSM (5000/5) & \footnotesize{0.012}\\
10.0& \color[rgb]{0,0.398,0.796}LED (0.25/5) & \footnotesize{0.021}& \color[rgb]{0,0.398,0.796}LED (5/1) & \footnotesize{0.003}& \color[rgb]{0.9,0.3,0}TSM (20/3) & \footnotesize{0.002}& \color[rgb]{0.9,0.3,0}TSM (5000/5) & \footnotesize{0.006}\\
\hline
\multicolumn{9}{c}{SBIG2000 (8.3\% defective pixels)}\\ \hline0.1& \color[rgb]{0.7,0.6,0.1}ATM (4/1) & \footnotesize{0.935}& \color[rgb]{0.9,0.3,0}TSM (10000/7) & \footnotesize{0.65}& \color[rgb]{0.9,0.3,0}TSM (10000/7) & \footnotesize{0.491}& \color[rgb]{0.9,0.3,0}TSM (10000/7) & \footnotesize{0.287}\\
0.2& \color[rgb]{0,0,1}BDND (4/4) & \footnotesize{0.586}& \color[rgb]{0.9,0.3,0}TSM (10000/7) & \footnotesize{0.523}& \color[rgb]{0.9,0.3,0}TSM (10000/7) & \footnotesize{0.393}& \color[rgb]{0.9,0.3,0}TSM (10000/7) & \footnotesize{0.226}\\
0.5& \color[rgb]{0,0,1}BDND (4/4) & \footnotesize{0.231}& \color[rgb]{0,0,1}BDND (4/4) & \footnotesize{0.239}& \color[rgb]{0.9,0.3,0}TSM (10000/7) & \footnotesize{0.248}& \color[rgb]{0.9,0.3,0}TSM (10000/7) & \footnotesize{0.139}\\
1.0& \color[rgb]{0.7,0.6,0.1}ATM (4/1) & \footnotesize{0.076}& \color[rgb]{0,0,1}BDND (4/4) & \footnotesize{0.115}& \color[rgb]{0.9,0.3,0}TSM (10000/7) & \footnotesize{0.153}& \color[rgb]{0.9,0.3,0}TSM (10000/7) & \footnotesize{0.085}\\
2.0& \color[rgb]{0.7,0.6,0.1}ATM (4/4) & \footnotesize{0.063}& \color[rgb]{0,0,1}BDND (4/2) & \footnotesize{0.084}& \color[rgb]{0,0,1}BDND (4/4) & \footnotesize{0.07}& \color[rgb]{0.9,0.3,0}TSM (10000/7) & \footnotesize{0.048}\\
5.0& \color[rgb]{0.7,0.6,0.1}ATM (2/1) & \footnotesize{0.056}& \color[rgb]{0.7,0.6,0.1}ATM (4/1) & \footnotesize{0.025}& \color[rgb]{0.9,0.3,0}TSM (10000/7) & \footnotesize{0.037}& \color[rgb]{0.9,0.3,0}TSM (10000/7) & \footnotesize{0.02}\\
10.0& \color[rgb]{0.9,0.3,0}TSM (10000/3) & \footnotesize{0.033}& \color[rgb]{0.7,0.6,0.1}ATM (3/1) & \footnotesize{0.022}& \color[rgb]{0.7,0.6,0.1}ATM (4/1) & \footnotesize{0.017}& \color[rgb]{0.9,0.3,0}TSM (10000/7) & \footnotesize{0.01}\\
\hline
\label{photometric_accuracy}\end{tabular}
\end{table*}

\begin{table*}

\caption{Astrometric accuracy accuracy of filtering methods evaluated on synthetic images. The most effective method with its optimal parameters and the corresponding centroid error ($E_a$ [pix]) is provided in each Table cell. The SNR is defined as the ratio of stellar amplitude to the mean intensity of an impulse.}
\label{my-label}
\begin{tabular}{lllllllll}
     & \multicolumn{2}{c}{$\sigma_{\textrm{\tiny{PSF}}}$ = 1 pix.}& \multicolumn{2}{c}{$\sigma_{\textrm{\tiny{PSF}}}$ = 2 pix.} & \multicolumn{2}{c}{$\sigma_{\textrm{\tiny{PSF}}}$ = 3 pix.} & \multicolumn{2}{c}{$\sigma_{\textrm{\tiny{PSF}}}$ = 4 pix.} \\\cmidrule(lr){2-3} \cmidrule(lr){4-5} \cmidrule(lr){6-7} \cmidrule(lr){8-9} 
 SNR    & Method & $E_a$ & Method & $E_m$ & Method & $E_a$ & Method & $E_a$ \\\hline
\multicolumn{9}{c}{Andor iKon L (49\% defective pixels)}\\ \hline0.1& \color[rgb]{0.3,0.3,0.3}PSM (5000/2) & \footnotesize{2.879}& \color[rgb]{0,0.398,0.796}LED (0.5/0.25) & \footnotesize{4.994}& \color[rgb]{0.4,0.2,0.6}MED (4/0) & \footnotesize{6.514}& \color[rgb]{0,0.398,0.796}LED (0.5/0.25) & \footnotesize{8.557}\\
0.2& \color[rgb]{0.3,0.3,0.3}PSM (5000/3) & \footnotesize{2.71}& \color[rgb]{0,0.398,0.796}LED (0.25/0.25) & \footnotesize{3.992}& \color[rgb]{0,0.398,0.796}LED (0.5/0.25) & \footnotesize{4.609}& \color[rgb]{0,0.398,0.796}LED (0.25/0.25) & \footnotesize{4.792}\\
0.5& \color[rgb]{0,0.398,0.796}LED (0.25/20) & \footnotesize{2.169}& \color[rgb]{0,0.398,0.796}LED (0.25/0.25) & \footnotesize{1.823}& \color[rgb]{0,0.398,0.796}LED (0.25/0.25) & \footnotesize{1.443}& \color[rgb]{0.3,0.3,0.3}PSM (2000/1) & \footnotesize{1.505}\\
1.0& \color[rgb]{0.9,0.3,0}TSM (2000/3) & \footnotesize{1.356}& \color[rgb]{0,0.398,0.796}LED (0.25/0.25) & \footnotesize{0.779}& \color[rgb]{0,0.398,0.796}LED (0.5/0.25) & \footnotesize{0.817}& \color[rgb]{0,0.398,0.796}LED (0.5/1) & \footnotesize{0.86}\\
2.0& \color[rgb]{0,0.398,0.796}LED (0.25/10) & \footnotesize{0.759}& \color[rgb]{0.3,0.3,0.3}PSM (50/1) & \footnotesize{0.596}& \color[rgb]{0.3,0.3,0.3}PSM (100/2) & \footnotesize{0.605}& \color[rgb]{0.3,0.3,0.3}PSM (10/2) & \footnotesize{0.556}\\
5.0& \color[rgb]{0.3,0.3,0.3}PSM (500/4) & \footnotesize{0.566}& \color[rgb]{0.3,0.3,0.3}PSM (500/3) & \footnotesize{0.461}& \color[rgb]{0.3,0.3,0.3}PSM (50/2) & \footnotesize{0.425}& \color[rgb]{0.3,0.3,0.3}PSM (50/2) & \footnotesize{0.395}\\
10.0& \color[rgb]{0.3,0.3,0.3}PSM (500/4) & \footnotesize{0.562}& \color[rgb]{0.3,0.3,0.3}PSM (500/2) & \footnotesize{0.434}& \color[rgb]{0.3,0.3,0.3}PSM (100/2) & \footnotesize{0.437}& \color[rgb]{0.3,0.3,0.3}PSM (100/2) & \footnotesize{0.393}\\
\hline
\multicolumn{9}{c}{Andor LucaS (17.4\% noised defective pixels)}\\ \hline0.1& \color[rgb]{0.35,0.44,0.1967}DWM (4/5000) & \footnotesize{0.635}& \color[rgb]{0.35,0.44,0.1967}DWM (4/5000) & \footnotesize{0.524}& \color[rgb]{0.35,0.44,0.1967}DWM (4/2000) & \footnotesize{0.677}& \color[rgb]{0.35,0.44,0.1967}DWM (3/2000) & \footnotesize{0.713}\\
0.2& \color[rgb]{0.35,0.44,0.1967}DWM (4/10000) & \footnotesize{0.65}& \color[rgb]{0.35,0.44,0.1967}DWM (4/5000) & \footnotesize{0.523}& \color[rgb]{0,0.398,0.796}LED (0.5/0.25) & \footnotesize{0.654}& \color[rgb]{0.35,0.44,0.1967}DWM (4/2000) & \footnotesize{0.753}\\
0.5& \color[rgb]{0,0.398,0.796}LED (20/0.25) & \footnotesize{0.638}& \color[rgb]{0.3,0.3,0.3}PSM (100/4) & \footnotesize{0.491}& \color[rgb]{0.3,0.3,0.3}PSM (50/4) & \footnotesize{0.586}& \color[rgb]{0.3,0.3,0.3}PSM (20/4) & \footnotesize{0.649}\\
1.0& \color[rgb]{0,0.398,0.796}LED (0.25/2) & \footnotesize{0.563}& \color[rgb]{0.3,0.3,0.3}PSM (200/4) & \footnotesize{0.477}& \color[rgb]{0.3,0.3,0.3}PSM (50/4) & \footnotesize{0.5}& \color[rgb]{0.3,0.3,0.3}PSM (10/4) & \footnotesize{0.52}\\
2.0& \color[rgb]{0,0.398,0.796}LED (0.25/2) & \footnotesize{0.544}& \color[rgb]{0.3,0.3,0.3}PSM (200/3) & \footnotesize{0.444}& \color[rgb]{0.3,0.3,0.3}PSM (10/2) & \footnotesize{0.457}& \color[rgb]{0.3,0.3,0.3}PSM (10/3) & \footnotesize{0.426}\\
5.0& \color[rgb]{0,0.398,0.796}LED (0.25/2) & \footnotesize{0.517}& \color[rgb]{0.3,0.3,0.3}PSM (500/3) & \footnotesize{0.432}& \color[rgb]{0.3,0.3,0.3}PSM (20/2) & \footnotesize{0.429}& \color[rgb]{0.3,0.3,0.3}PSM (100/4) & \footnotesize{0.436}\\
10.0& \color[rgb]{0.3,0.3,0.3}PSM (2000/1) & \footnotesize{0.521}& \color[rgb]{0.3,0.3,0.3}PSM (2000/4) & \footnotesize{0.445}& \color[rgb]{0.3,0.3,0.3}PSM (100/2) & \footnotesize{0.46}& \color[rgb]{0.3,0.3,0.3}PSM (200/3) & \footnotesize{0.382}\\
\hline
\multicolumn{9}{c}{Andor iKon CR (1.04\% defective pixels)}\\ \hline0.1& \color[rgb]{0.3,0.3,0.3}PSM (20/1) & \footnotesize{0.519}& \color[rgb]{0,0.398,0.796}LED (0.25/0.5) & \footnotesize{0.472}& \color[rgb]{0.3,0.3,0.3}PSM (10/4) & \footnotesize{0.618}& \color[rgb]{0.35,0.44,0.1967}DWM (4/1000) & \footnotesize{0.746}\\
0.2& \color[rgb]{0.3,0.3,0.3}PSM (100/3) & \footnotesize{0.507}& \color[rgb]{0.3,0.3,0.3}PSM (50/3) & \footnotesize{0.474}& \color[rgb]{0.3,0.3,0.3}PSM (20/4) & \footnotesize{0.489}& \color[rgb]{0.3,0.3,0.3}PSM (10/4) & \footnotesize{0.635}\\
0.5& \color[rgb]{0.3,0.3,0.3}PSM (200/4) & \footnotesize{0.481}& \color[rgb]{0.3,0.3,0.3}PSM (100/4) & \footnotesize{0.459}& \color[rgb]{0.3,0.3,0.3}PSM (50/4) & \footnotesize{0.478}& \color[rgb]{0.3,0.3,0.3}PSM (20/3) & \footnotesize{0.472}\\
1.0& \color[rgb]{0.3,0.3,0.3}PSM (200/1) & \footnotesize{0.474}& \color[rgb]{0.3,0.3,0.3}PSM (200/4) & \footnotesize{0.45}& \color[rgb]{0.3,0.3,0.3}PSM (100/4) & \footnotesize{0.464}& \color[rgb]{0.3,0.3,0.3}PSM (10/2) & \footnotesize{0.477}\\
2.0& \color[rgb]{0.3,0.3,0.3}PSM (500/1) & \footnotesize{0.482}& \color[rgb]{0.3,0.3,0.3}PSM (500/3) & \footnotesize{0.438}& \color[rgb]{0.3,0.3,0.3}PSM (200/4) & \footnotesize{0.463}& \color[rgb]{0.3,0.3,0.3}PSM (100/4) & \footnotesize{0.46}\\
5.0& \color[rgb]{0.3,0.3,0.3}PSM (1000/1) & \footnotesize{0.467}& \color[rgb]{0.3,0.3,0.3}PSM (1000/4) & \footnotesize{0.424}& \color[rgb]{0.3,0.3,0.3}PSM (500/4) & \footnotesize{0.448}& \color[rgb]{0.3,0.3,0.3}PSM (200/3) & \footnotesize{0.417}\\
10.0& \color[rgb]{0.3,0.3,0.3}PSM (2000/1) & \footnotesize{0.463}& \color[rgb]{0.3,0.3,0.3}PSM (2000/4) & \footnotesize{0.43}& \color[rgb]{0.3,0.3,0.3}PSM (1000/4) & \footnotesize{0.446}& \color[rgb]{0.3,0.3,0.3}PSM (500/4) & \footnotesize{0.391}\\
\hline
\multicolumn{9}{c}{BRITE Toronto (3.9\% defective pixels)}\\ \hline0.1& \color[rgb]{0.3,0.3,0.3}PSM (50/4) & \footnotesize{0.559}& \color[rgb]{0.35,0.44,0.1967}DWM (4/2000) & \footnotesize{0.531}& \color[rgb]{0.451,0.258,0.07}CWM (4/2) & \footnotesize{0.693}& \color[rgb]{0,0.398,0.796}LED (20/0.25) & \footnotesize{0.643}\\
0.2& \color[rgb]{0.3,0.3,0.3}PSM (100/4) & \footnotesize{0.539}& \color[rgb]{0,0.398,0.796}LED (0.5/0.25) & \footnotesize{0.499}& \color[rgb]{0.3,0.3,0.3}PSM (20/4) & \footnotesize{0.584}& \color[rgb]{0.3,0.3,0.3}PSM (10/3) & \footnotesize{0.605}\\
0.5& \color[rgb]{0.3,0.3,0.3}PSM (200/4) & \footnotesize{0.529}& \color[rgb]{0.3,0.3,0.3}PSM (100/4) & \footnotesize{0.465}& \color[rgb]{0.3,0.3,0.3}PSM (50/4) & \footnotesize{0.484}& \color[rgb]{0.3,0.3,0.3}PSM (20/4) & \footnotesize{0.447}\\
1.0& \color[rgb]{0,0.398,0.796}LED (0.25/2) & \footnotesize{0.533}& \color[rgb]{0.3,0.3,0.3}PSM (200/4) & \footnotesize{0.439}& \color[rgb]{0.3,0.3,0.3}PSM (100/4) & \footnotesize{0.462}& \color[rgb]{0.3,0.3,0.3}PSM (50/4) & \footnotesize{0.435}\\
2.0& \color[rgb]{0,0.398,0.796}LED (0.25/2) & \footnotesize{0.519}& \color[rgb]{0.3,0.3,0.3}PSM (200/2) & \footnotesize{0.438}& \color[rgb]{0.3,0.3,0.3}PSM (200/4) & \footnotesize{0.445}& \color[rgb]{0.3,0.3,0.3}PSM (50/4) & \footnotesize{0.395}\\
5.0& \color[rgb]{0.3,0.3,0.3}PSM (1000/1) & \footnotesize{0.478}& \color[rgb]{0.3,0.3,0.3}PSM (1000/4) & \footnotesize{0.424}& \color[rgb]{0.3,0.3,0.3}PSM (500/4) & \footnotesize{0.461}& \color[rgb]{0.3,0.3,0.3}PSM (200/4) & \footnotesize{0.411}\\
10.0& \color[rgb]{0.3,0.3,0.3}PSM (2000/1) & \footnotesize{0.474}& \color[rgb]{0.3,0.3,0.3}PSM (2000/4) & \footnotesize{0.417}& \color[rgb]{0.3,0.3,0.3}PSM (1000/4) & \footnotesize{0.43}& \color[rgb]{0.3,0.3,0.3}PSM (500/4) & \footnotesize{0.406}\\
\hline
\multicolumn{9}{c}{SBIG ST10XME (4.6\% defective pixels)}\\ \hline0.1& \color[rgb]{0.3,0.3,0.3}PSM (20/1) & \footnotesize{0.49}& \color[rgb]{0.3,0.3,0.3}PSM (20/4) & \footnotesize{0.465}& \color[rgb]{0.3,0.3,0.3}PSM (10/4) & \footnotesize{0.517}& \color[rgb]{0.3,0.3,0.3}PSM (10/4) & \footnotesize{0.58}\\
0.2& \color[rgb]{0.3,0.3,0.3}PSM (50/1) & \footnotesize{0.509}& \color[rgb]{0.3,0.3,0.3}PSM (20/3) & \footnotesize{0.446}& \color[rgb]{0.3,0.3,0.3}PSM (20/4) & \footnotesize{0.508}& \color[rgb]{0.3,0.3,0.3}PSM (10/4) & \footnotesize{0.455}\\
0.5& \color[rgb]{0.3,0.3,0.3}PSM (100/1) & \footnotesize{0.498}& \color[rgb]{0.3,0.3,0.3}PSM (100/4) & \footnotesize{0.44}& \color[rgb]{0.3,0.3,0.3}PSM (20/3) & \footnotesize{0.469}& \color[rgb]{0.3,0.3,0.3}PSM (20/4) & \footnotesize{0.409}\\
1.0& \color[rgb]{0.3,0.3,0.3}PSM (200/1) & \footnotesize{0.492}& \color[rgb]{0.3,0.3,0.3}PSM (200/4) & \footnotesize{0.422}& \color[rgb]{0.3,0.3,0.3}PSM (100/4) & \footnotesize{0.463}& \color[rgb]{0.3,0.3,0.3}PSM (50/4) & \footnotesize{0.376}\\
2.0& \color[rgb]{0.3,0.3,0.3}PSM (500/1) & \footnotesize{0.498}& \color[rgb]{0.3,0.3,0.3}PSM (200/2) & \footnotesize{0.443}& \color[rgb]{0.3,0.3,0.3}PSM (200/4) & \footnotesize{0.464}& \color[rgb]{0.3,0.3,0.3}PSM (10/2) & \footnotesize{0.399}\\
5.0& \color[rgb]{0.3,0.3,0.3}PSM (1000/1) & \footnotesize{0.483}& \color[rgb]{0.3,0.3,0.3}PSM (1000/4) & \footnotesize{0.425}& \color[rgb]{0.3,0.3,0.3}PSM (500/4) & \footnotesize{0.446}& \color[rgb]{0.3,0.3,0.3}PSM (20/2) & \footnotesize{0.38}\\
10.0& \color[rgb]{0.3,0.3,0.3}PSM (2000/1) & \footnotesize{0.476}& \color[rgb]{0.3,0.3,0.3}PSM (2000/4) & \footnotesize{0.438}& \color[rgb]{0.3,0.3,0.3}PSM (1000/4) & \footnotesize{0.449}& \color[rgb]{0.3,0.3,0.3}PSM (50/2) & \footnotesize{0.422}\\
\hline
\multicolumn{9}{c}{SBIG2000 (8.3\% defective pixels)}\\ \hline0.1& \color[rgb]{0.3,0.3,0.3}PSM (50/4) & \footnotesize{0.515}& \color[rgb]{0.3,0.3,0.3}PSM (20/4) & \footnotesize{0.487}& \color[rgb]{0.3,0.3,0.3}PSM (10/3) & \footnotesize{0.562}& \color[rgb]{0.3,0.3,0.3}PSM (10/4) & \footnotesize{0.61}\\
0.2& \color[rgb]{0,0.398,0.796}LED (0.25/2) & \footnotesize{0.534}& \color[rgb]{0.3,0.3,0.3}PSM (20/3) & \footnotesize{0.483}& \color[rgb]{0.3,0.3,0.3}PSM (10/4) & \footnotesize{0.517}& \color[rgb]{0.3,0.3,0.3}PSM (10/4) & \footnotesize{0.545}\\
0.5& \color[rgb]{0,0.398,0.796}LED (0.25/2) & \footnotesize{0.517}& \color[rgb]{0.3,0.3,0.3}PSM (100/4) & \footnotesize{0.449}& \color[rgb]{0.3,0.3,0.3}PSM (20/3) & \footnotesize{0.459}& \color[rgb]{0.3,0.3,0.3}PSM (10/4) & \footnotesize{0.465}\\
1.0& \color[rgb]{0,0.398,0.796}LED (0.25/2) & \footnotesize{0.518}& \color[rgb]{0.3,0.3,0.3}PSM (200/4) & \footnotesize{0.447}& \color[rgb]{0.3,0.3,0.3}PSM (10/2) & \footnotesize{0.499}& \color[rgb]{0.3,0.3,0.3}PSM (20/3) & \footnotesize{0.446}\\
2.0& \color[rgb]{0.3,0.3,0.3}PSM (500/1) & \footnotesize{0.517}& \color[rgb]{0.3,0.3,0.3}PSM (200/3) & \footnotesize{0.446}& \color[rgb]{0.3,0.3,0.3}PSM (200/4) & \footnotesize{0.459}& \color[rgb]{0.3,0.3,0.3}PSM (100/4) & \footnotesize{0.413}\\
5.0& \color[rgb]{0.3,0.3,0.3}PSM (1000/1) & \footnotesize{0.492}& \color[rgb]{0.3,0.3,0.3}PSM (1000/4) & \footnotesize{0.42}& \color[rgb]{0.3,0.3,0.3}PSM (500/4) & \footnotesize{0.457}& \color[rgb]{0.3,0.3,0.3}PSM (200/4) & \footnotesize{0.468}\\
10.0& \color[rgb]{0.3,0.3,0.3}PSM (2000/1) & \footnotesize{0.474}& \color[rgb]{0.3,0.3,0.3}PSM (2000/4) & \footnotesize{0.42}& \color[rgb]{0.3,0.3,0.3}PSM (1000/4) & \footnotesize{0.445}& \color[rgb]{0.3,0.3,0.3}PSM (500/4) & \footnotesize{0.424}\\
\hline
\label{astrometric_accuracy}\end{tabular}
\end{table*}

\subsection{Real images}
\begin{figure*}
\centering
\begin{subfigure}[b]{0.3\linewidth}
\includegraphics[width=\textwidth]{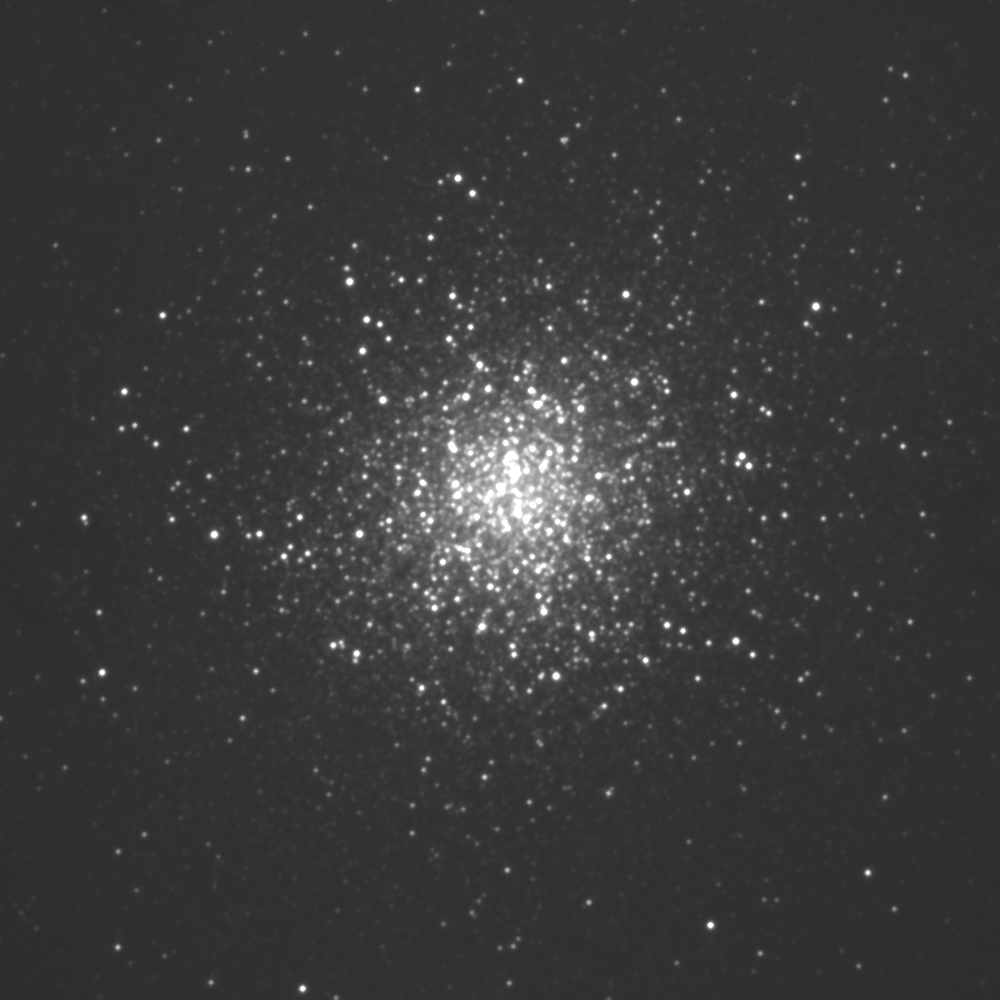}
\end{subfigure}~
\begin{subfigure}[b]{0.3\linewidth}
\includegraphics[width=\textwidth]{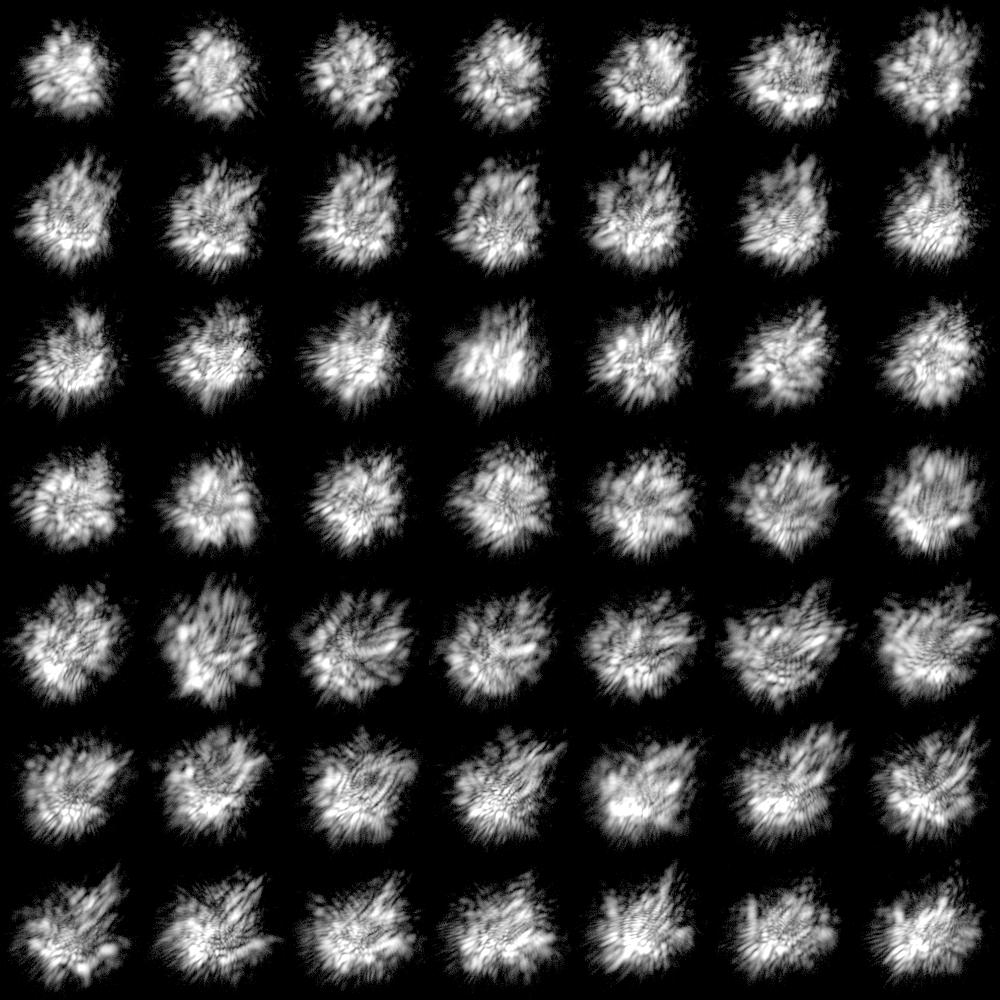}
\end{subfigure}~
\begin{subfigure}[b]{0.3\linewidth}
\includegraphics[width=\textwidth]{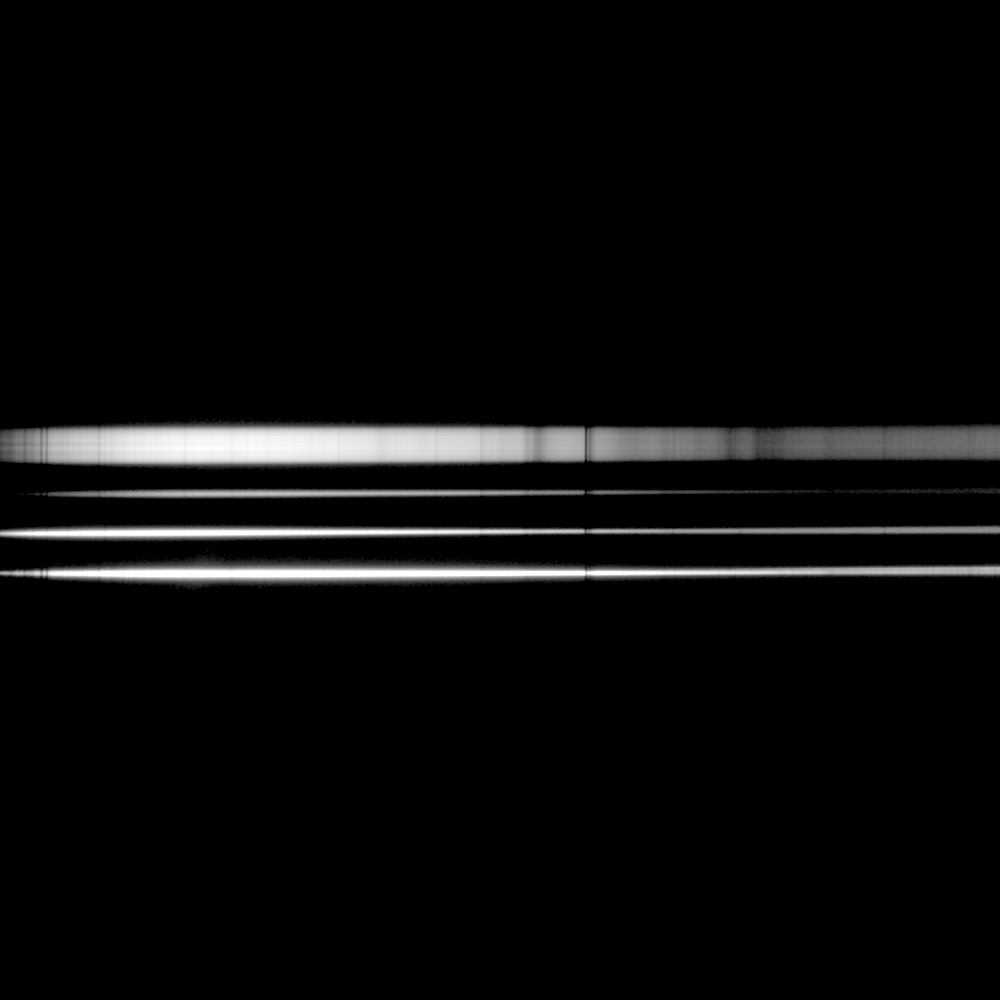}
\end{subfigure}
\\\vspace{0.18cm}
\begin{subfigure}[b]{0.3\linewidth}
\includegraphics[width=\textwidth]{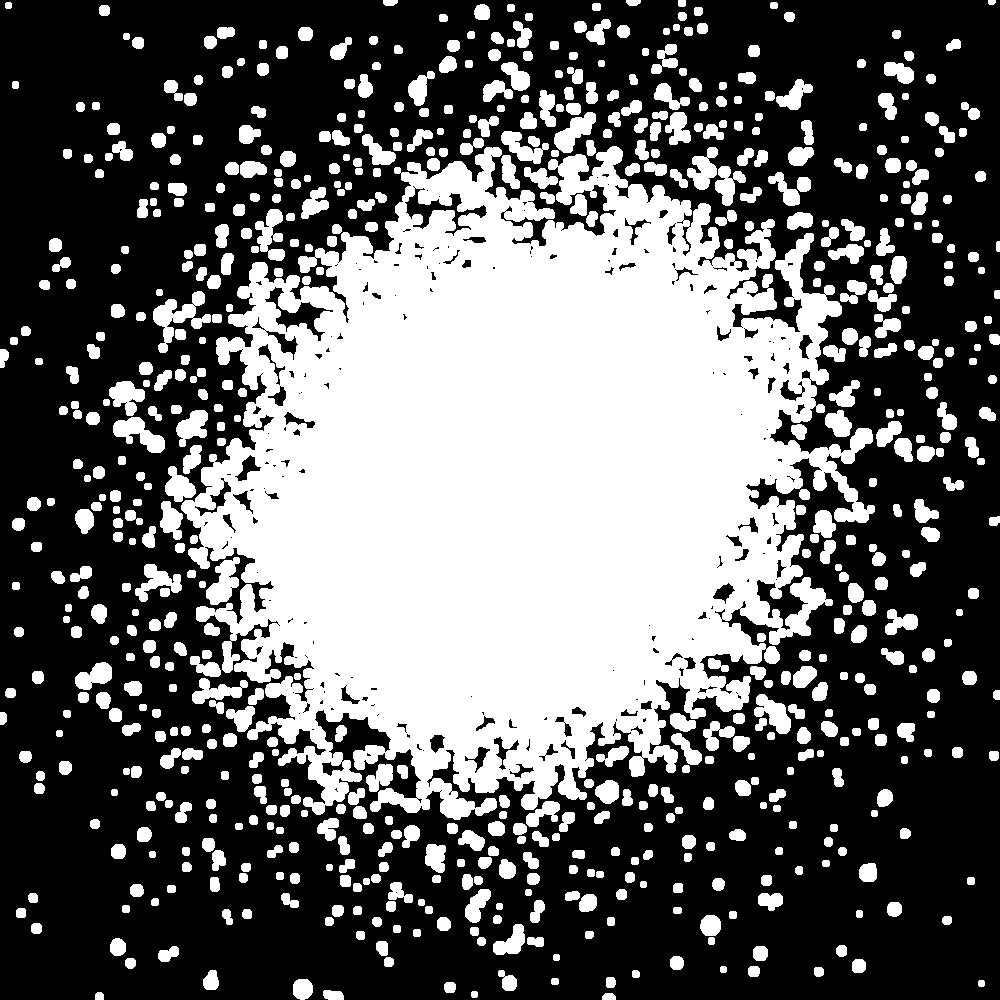}
\caption{M13 globular cluster.\\~}
\end{subfigure}~
\begin{subfigure}[b]{0.3\linewidth}
\includegraphics[width=\textwidth]{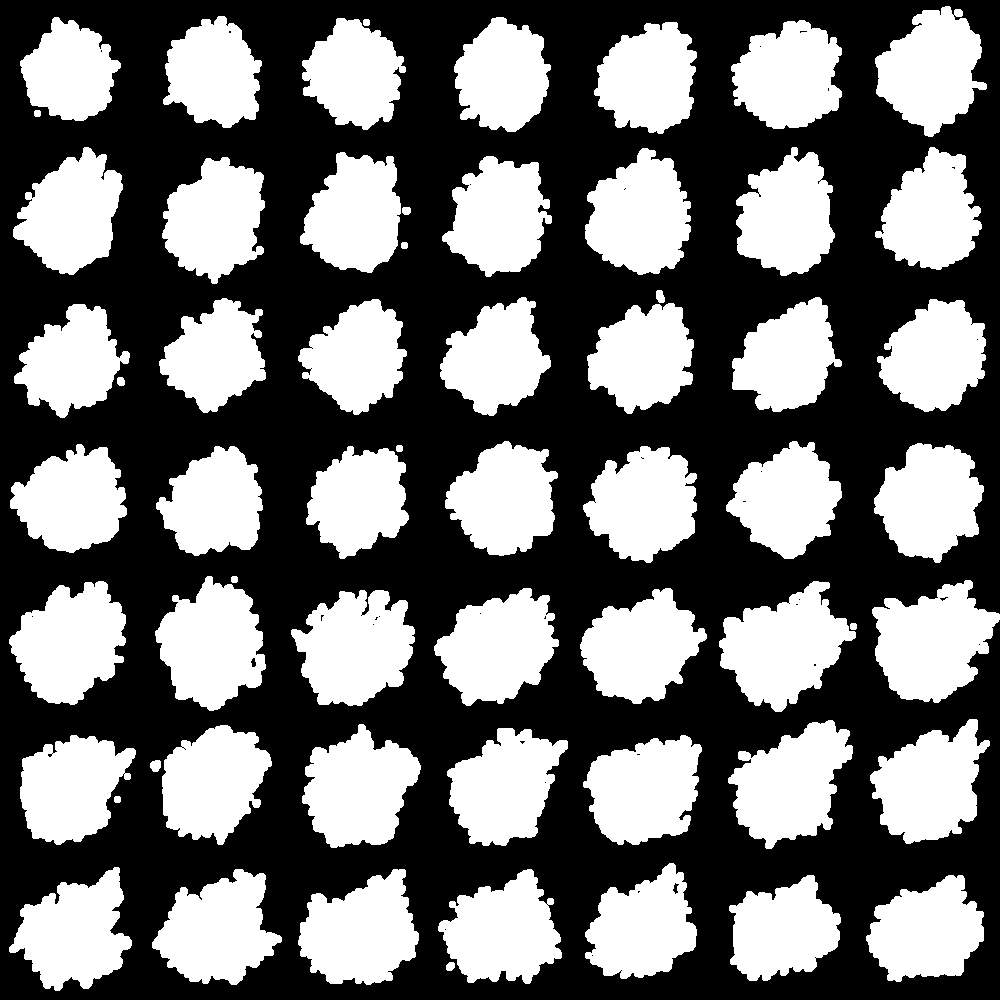}
\caption{Speckle series of Arcturus.\\~}
\end{subfigure}~
\begin{subfigure}[b]{0.3\linewidth}
\includegraphics[width=\textwidth]{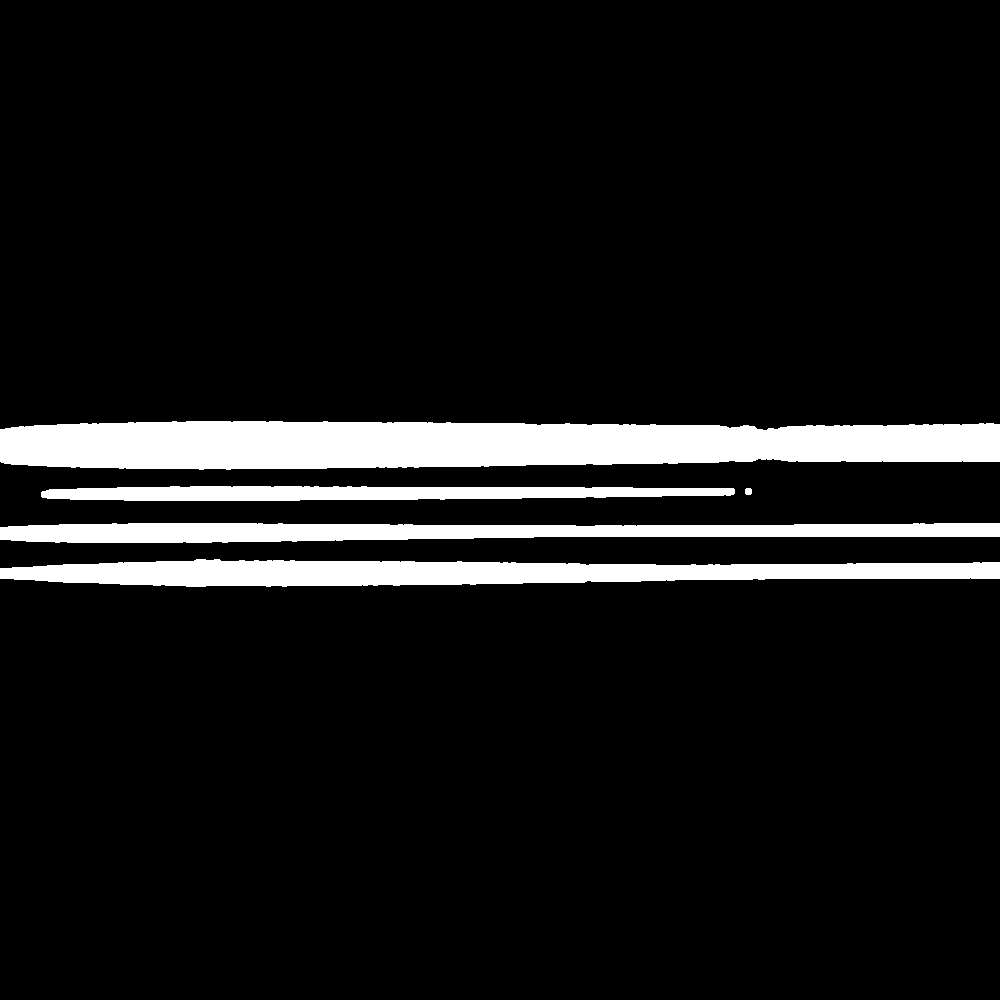}
\caption{Spectra of Jupiter, Ganymede, Spica and Arcturus (top to bottom).}
\end{subfigure}
\caption{Reference real images (above) and their corresponding object masks (below) obtained in various imaging techniques: (a) traditional imaging, (b) speckle imaging and (c) spectroscopy. The gray scale in images is logarithmic to visualize both dim and bright objects.}
\label{real_images}
\end{figure*}

Samples of real images were obtained by 30 cm Ritchey-Chretien (2.4 m focal length, by GSO) with ATIK 11000M (full-frame CCD) and LucaS (CCD/EMCCD) cameras, the equipment held by Institute of Automatic Control, Silesian University of Technology. The observations were performed in Kamieniec (Poland) school observatory held by Polish Amateur Astronomical Society of Gliwice. Our representative set of astronomical images includes: (1) Messier 13 globular cluster (direct imaging), (2) speckle series of Arcturus (high-speed speckle imaging) and (3) low-resolution spectra of Jupiter, Ganymede moon, Arcturus and Spica (spectroscopy). For the spectroscopy, DADOS self-guiding, slit spectrograph was used. For each image, the mask of the objects of interest was created by thresholding original frame and then dilating the binary outcome by 3 pixels. The threshold was set to: MEDIAN($I$) + 0.3$\cdot$MAD($I$), where $I$ stands for pixels intensities in an image and MAD is a median absolute deviation. The original frames with their objects masks are presented in Fig. \ref{real_images}.   

An example of direct imaging - M13 cluster - was the outcome of 3 h exposure, (50 unfiltered exposures, 3 minutes each one). Although our full-frame camera was cooled down to -48\degree C (ambient -10\degree C), the master dark frame, constructed from 15 frames, was subtracted from each frame, prior to the median averaging, to reduce very few residual impulses. The high number of exposures and relatively deep cooling, reduced significantly the readout noise in the final frame, so that it can be considered as nearly noiseless reference. The M13 globular cluster was selected due to high number of objects within the field of view. The spread of PSF of a stellar profile was $\sigma_{\tiny{\textrm{PSF}}}=1.5$ pix, which corresponds to moderate seeing conditions FWHM=2.3 arcsec.

The speckle series of Arcturus was obtained using high-quality Barlow lenses (10 m focal length) and employing LucaS camera running in CCD mode, which assured that the acquired images were not affected by CIC noise. Additionally, due to the very short exposure times (10 ms), the dark current was not present. The high intensity of observed object allowed for the assumption that the readout noise is also negligible when compared with Poisson statistics of counts. The final reference frame was constructed from 49 speckle patterns arranged in 7$\times$7 array, as presented in Fig. \ref{real_images}b. For the employed focal length (10 m) and the pixel size in LucaS (10 $\mu$m) the Airy disk diameter is approximately 4 pixels, which taking into account possible optical aberrations, results in $\sigma_{\tiny{\textrm{PSF}}}$ close to 1 pixel.

Four samples of spectroscopic data were obtained employing DADOS low-resolution spectrograph. The spectra of observed objects were registered by full-frame camera, while LucaS was used as a guider. Due to the high intensity of selected objects (Spica, Arcturus, Jupiter and Ganymede) the exposures times could be relatively short (1$\sim$4 sec.). For each object we registered 50 images and then the frames were median averaged. Similarly to the examples of traditional and speckle imaging, the reference spectra could be considered as nearly noiseless, due to the averaging and the high intensity of observed objects. The four spectra were arranged vertically, one below another, in a reference image, as depicted in Fig. \ref{real_images}c. The selected objects were of high (Jupiter) and low (the others) angular size, which manifests itself in various widths of spectral strips.

The reference frames were initially scaled, so that the mean intensity of the object's brightest pixel equals to 1000 [e$^-$]. Then, the noise templates were added to create a set of corrupted images. Due to the variety of objects in real images (speckle, spectrum stripes, stars), the SNR had to be defined slightly different than for the synthetic images. Therefore, it was defined as the ratio of average intensity of an object pixel (i.e. a pixel included in object mask) to the mean intensity of an impulse. Finally, the noisy frames were processed by filtering algorithms and the final accuracy was calculated using simple root-mean-square (RMS) formula:
\begin{equation}
E_{\textrm{\tiny{RMS}}}  = \sqrt{\frac{1}{Z} \sum_{p\in O}{(I_p'-I_p)^2}}
\end{equation}
where $E_{\textrm{\tiny{RMS}}}$ stands for root mean square error, $O$ is the set of $Z$ pixels which belong to the object mask (i.e. the white pixels in binary masks presented in lower row of Fig. \ref{real_images}), $I_p$ and $I_p'$ are respectively the intensity of reference and denoised pixel at position $p$.

\begin{figure*}
\centering
\begin{subfigure}[b]{0.48\linewidth}
\includegraphics[width=\textwidth]{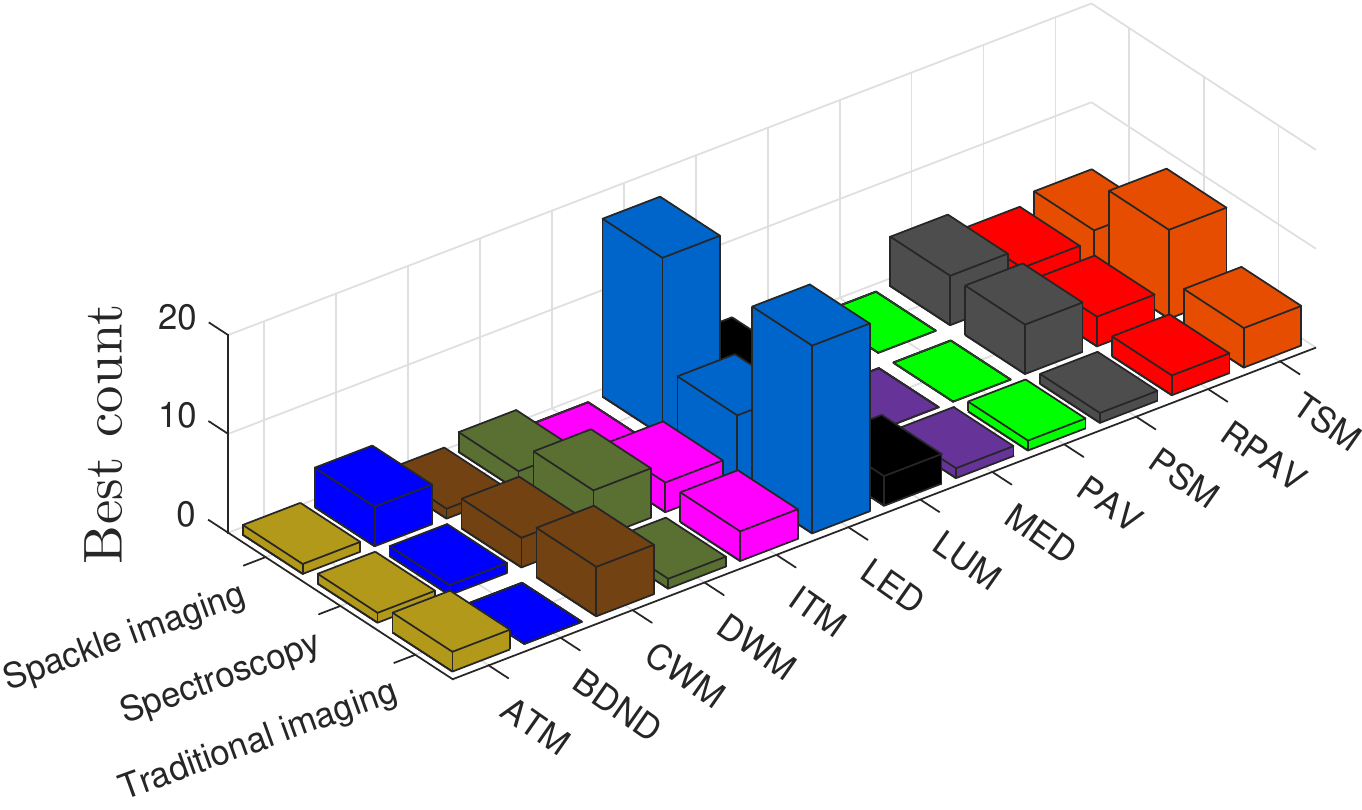}
\end{subfigure}~
\begin{subfigure}[b]{0.48\linewidth}
\includegraphics[width=\textwidth]{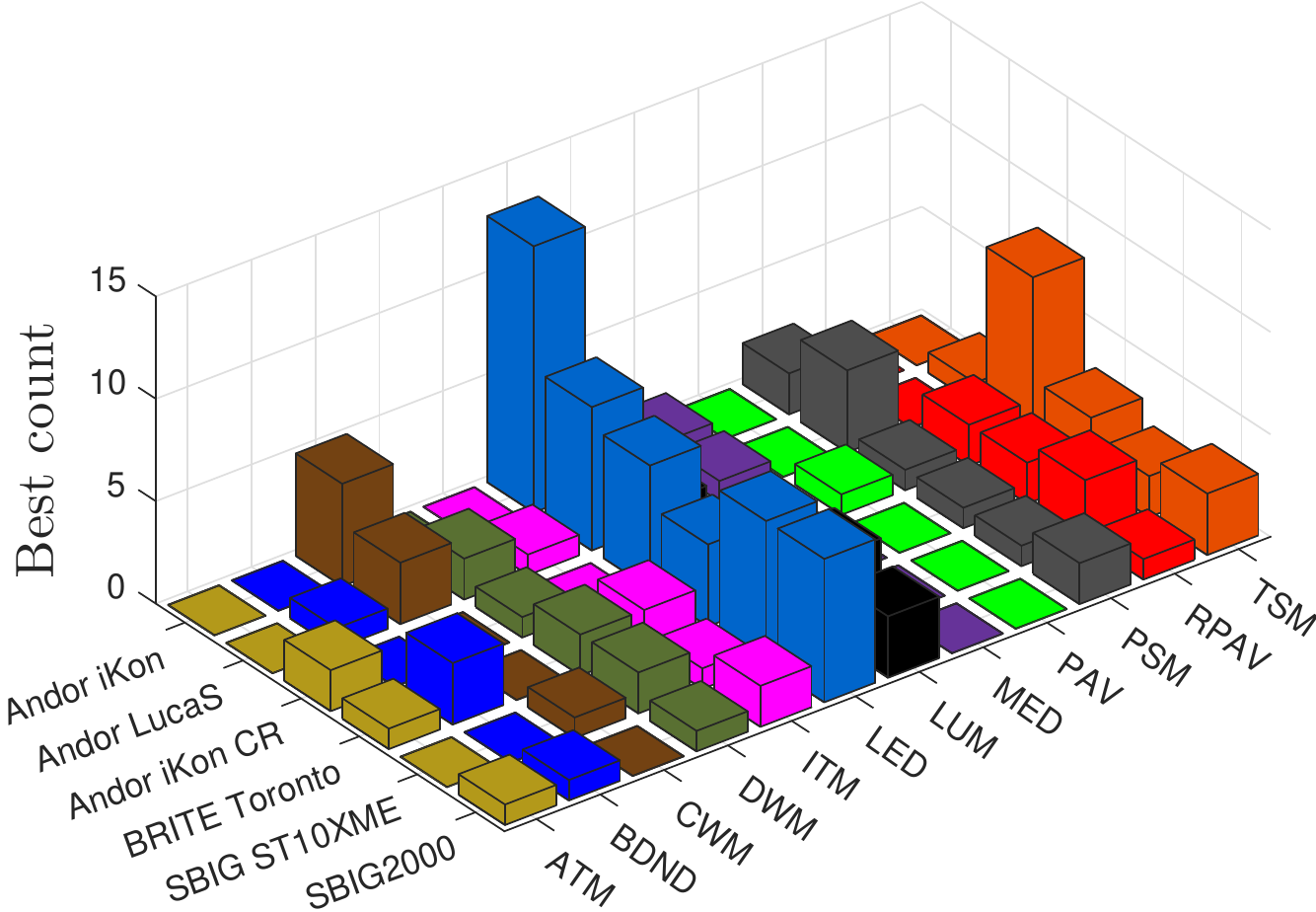}
\end{subfigure}~
\\
\begin{subfigure}[b]{0.48\linewidth}
\includegraphics[width=\textwidth]{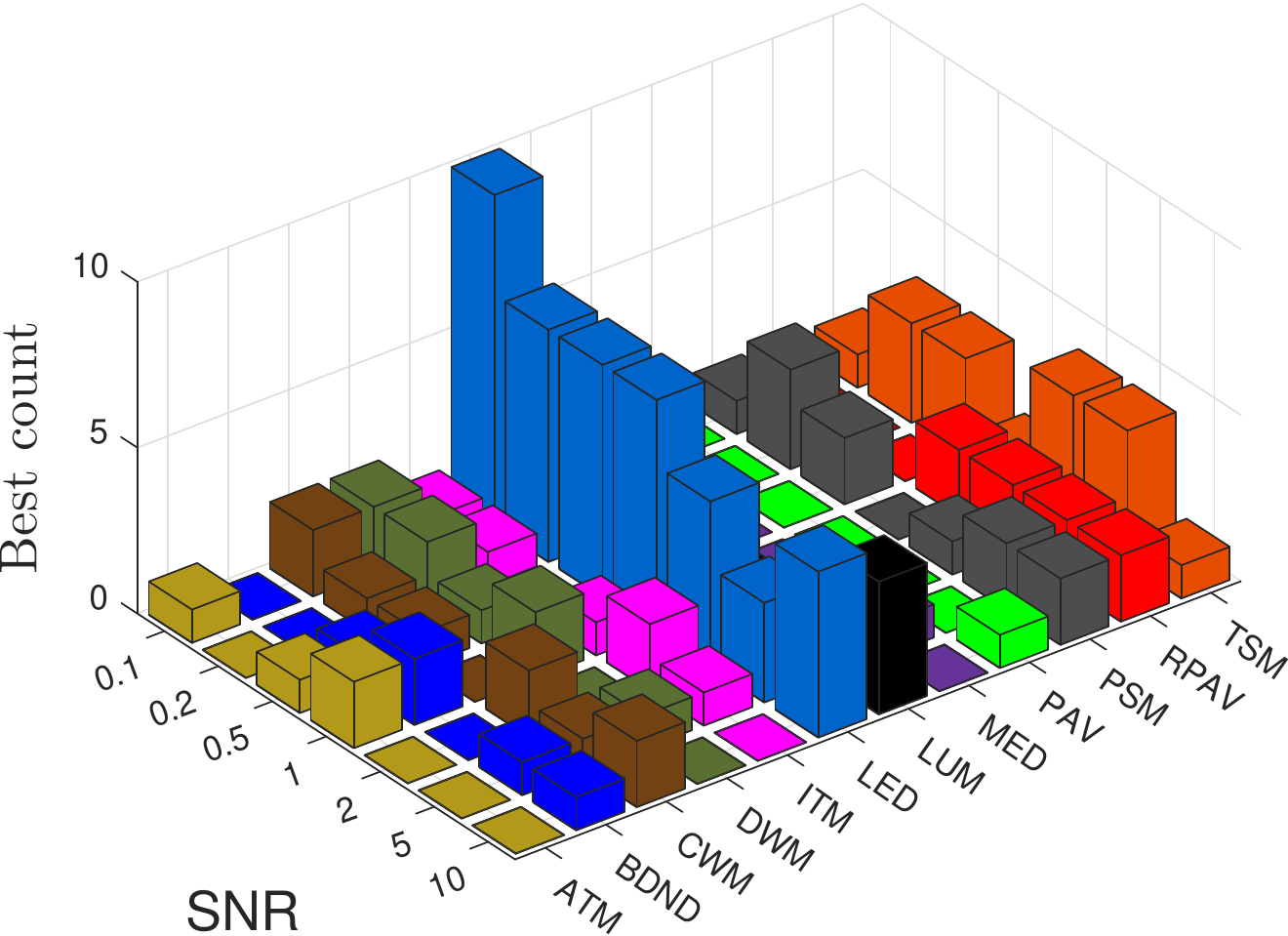}
\end{subfigure}~
\caption{Ranking of impulsive noise reduction methods based on the evaluation on real astronomical images. The \emph{Best count} corresponds to the number of occurrences a method achieved the highest accuracy (the lowest RMS error). }
\label{real_statisics}
\end{figure*}

Similarly to the results of synthetic data, the outcomes of our comparison are presented collectively in Tab. \ref{individual_accuracy} and by bar plots in Fig. \ref{real_statisics}. The same coloring scheme was applied for better distinguishability of the methods.

\begin{table*}

\caption{Photometric accuracy of filtering methods evaluated on real images. The most effective method with its optimal parameters and the corresponding RMS error ($E_{\textrm{\tiny{RMS}}}$ [ADU]) is provided in each Table cell. The SNR is defined as the ratio of mean intensity of object pixel to the mean intensity of an impulse.}
\label{my-label}
\begin{tabular}{lllllll}
     & \multicolumn{2}{c}{Spackle imaging}& \multicolumn{2}{c}{Spectroscopy} & \multicolumn{2}{c}{Traditional imaging} \\\cmidrule(lr){2-3} \cmidrule(lr){4-5} \cmidrule(lr){6-7}  
 SNR    & Method & $E_{\textrm{\tiny{RMS}}}$ & Method & $E_{\textrm{\tiny{RMS}}}$ & Method & $E_{\textrm{\tiny{RMS}}}$  \\\hline
\multicolumn{7}{c}{Andor iKon L (49\% defective pixels)}\\ \hline0.1& \color[rgb]{0,0.398,0.796}LED (0.25/0.25) & \footnotesize{1032.938}& \color[rgb]{0,0.398,0.796}LED (0.25/0.25) & \footnotesize{1168.898}& \color[rgb]{0,0.398,0.796}LED (0.25/0.25) & \footnotesize{849.539}\\
0.2& \color[rgb]{0,0.398,0.796}LED (0.25/0.25) & \footnotesize{512.426}& \color[rgb]{0,0.398,0.796}LED (0.25/0.25) & \footnotesize{578.715}& \color[rgb]{0,0.398,0.796}LED (0.25/0.25) & \footnotesize{409.203}\\
0.5& \color[rgb]{0,0.398,0.796}LED (0.25/0.25) & \footnotesize{196.569}& \color[rgb]{0,0.398,0.796}LED (0.25/0.25) & \footnotesize{219.568}& \color[rgb]{0,0.398,0.796}LED (0.25/0.25) & \footnotesize{134.959}\\
1.0& \color[rgb]{0,0.398,0.796}LED (0.25/0.25) & \footnotesize{83.913}& \color[rgb]{0,0.398,0.796}LED (0.25/0.25) & \footnotesize{86.287}& \color[rgb]{0,0.398,0.796}LED (0.25/0.25) & \footnotesize{33.973}\\
2.0& \color[rgb]{0,0.398,0.796}LED (0.25/0.25) & \footnotesize{15.861}& \color[rgb]{0.451,0.258,0.07}CWM (4/5) & \footnotesize{4.405}& \color[rgb]{0.451,0.258,0.07}CWM (4/5) & \footnotesize{0.819}\\
5.0& \color[rgb]{0.4,0.2,0.6}MED (4/0) & \footnotesize{1.141}& \color[rgb]{0.3,0.3,0.3}PSM (200/4) & \footnotesize{2.71}& \color[rgb]{0.451,0.258,0.07}CWM (3/15) & \footnotesize{2.862}\\
10.0& \color[rgb]{0.451,0.258,0.07}CWM (3/10) & \footnotesize{3.152}& \color[rgb]{0.451,0.258,0.07}CWM (2/2) & \footnotesize{0.458}& \color[rgb]{0.3,0.3,0.3}PSM (50/2) & \footnotesize{0.104}\\
\hline
\multicolumn{7}{c}{Andor LucaS (17.4\% noised defective pixels)}\\ \hline0.1& \color[rgb]{0,0.398,0.796}LED (10/0.25) & \footnotesize{2.095}& \color[rgb]{0,0.398,0.796}LED (10/0.5) & \footnotesize{2.17}& \color[rgb]{0.451,0.258,0.07}CWM (2/1) & \footnotesize{1.129}\\
0.2& \color[rgb]{0,0.398,0.796}LED (10/0.5) & \footnotesize{0.324}& \color[rgb]{0.3,0.3,0.3}PSM (1000/2) & \footnotesize{0.182}& \color[rgb]{0.451,0.258,0.07}CWM (3/15) & \footnotesize{2.516}\\
0.5& \color[rgb]{0,0.398,0.796}LED (2/2) & \footnotesize{0.058}& \color[rgb]{0.35,0.44,0.1967}DWM (3/5000) & \footnotesize{0.172}& \color[rgb]{0.451,0.258,0.07}CWM (2/5) & \footnotesize{0.558}\\
1.0& \color[rgb]{0.35,0.44,0.1967}DWM (3/1000) & \footnotesize{0.78}& \color[rgb]{1,0,1}ITM (3/2) & \footnotesize{0.127}& \color[rgb]{0,0.398,0.796}LED (0.5/1) & \footnotesize{0.359}\\
2.0& \color[rgb]{0,0.398,0.796}LED (10/0.5) & \footnotesize{1.097}& \color[rgb]{0.3,0.3,0.3}PSM (10/3) & \footnotesize{0.451}& \color[rgb]{0.4,0.2,0.6}MED (1/0) & \footnotesize{0.002}\\
5.0& \color[rgb]{0.3,0.3,0.3}PSM (100/4) & \footnotesize{0.022}& \color[rgb]{0.9,0.3,0}TSM (50/2) & \footnotesize{0.159}& \color[rgb]{0,0,0}LUM (4/3) & \footnotesize{0.335}\\
10.0& \color[rgb]{0,0,1}BDND (2/3) & \footnotesize{0.029}& \color[rgb]{0.3,0.3,0.3}PSM (500/1) & \footnotesize{0.068}& \color[rgb]{0,0.398,0.796}LED (5/1) & \footnotesize{0.308}\\
\hline
\multicolumn{7}{c}{Andor iKon CR (1.04\% defective pixels)}\\ \hline0.1& \color[rgb]{0.3,0.3,0.3}PSM (2000/2) & \footnotesize{0.246}& \color[rgb]{0.7,0.6,0.1}ATM (3/2) & \footnotesize{0.014}& \color[rgb]{0.9,0.3,0}TSM (500/2) & \footnotesize{0.301}\\
0.2& \color[rgb]{0.9,0.3,0}TSM (200/3) & \footnotesize{0.236}& \color[rgb]{0.9,0.3,0}TSM (50/4) & \footnotesize{0.171}& \color[rgb]{0,0.398,0.796}LED (1/2) & \footnotesize{0.087}\\
0.5& \color[rgb]{0.7,0.6,0.1}ATM (1/1) & \footnotesize{0.008}& \color[rgb]{0,0.398,0.796}LED (2/1) & \footnotesize{0.054}& \color[rgb]{0.9,0.3,0}TSM (2000/1) & \footnotesize{0.098}\\
1.0& \color[rgb]{1,0,0}RPAV (1/3) & \footnotesize{0.006}& \color[rgb]{0,0.398,0.796}LED (2/1) & \footnotesize{0.125}& \color[rgb]{0,0.398,0.796}LED (0.5/2) & \footnotesize{0.013}\\
2.0& \color[rgb]{0.9,0.3,0}TSM (1000/1) & \footnotesize{0.079}& \color[rgb]{0.9,0.3,0}TSM (20/7) & \footnotesize{0.076}& \color[rgb]{0,0.398,0.796}LED (5/2) & \footnotesize{0.006}\\
5.0& \color[rgb]{0.9,0.3,0}TSM (1000/2) & \footnotesize{0.045}& \color[rgb]{0.35,0.44,0.1967}DWM (2/10000) & \footnotesize{0.109}& \color[rgb]{0,0.398,0.796}LED (2/5) & \footnotesize{0.006}\\
10.0& \color[rgb]{1,0,0}RPAV (1/1) & \footnotesize{0.055}& \color[rgb]{0.9,0.3,0}TSM (200/6) & \footnotesize{0.002}& \color[rgb]{0,1,0}PAV (2/0) & \footnotesize{0.007}\\
\hline
\multicolumn{7}{c}{BRITE Toronto (3.9\% defective pixels)}\\ \hline0.1& \color[rgb]{0,0.398,0.796}LED (20/0.5) & \footnotesize{0.185}& \color[rgb]{0.35,0.44,0.1967}DWM (2/1000) & \footnotesize{0.621}& \color[rgb]{1,0,1}ITM (2/3) & \footnotesize{0.879}\\
0.2& \color[rgb]{0,0.398,0.796}LED (20/0.5) & \footnotesize{0.034}& \color[rgb]{0.35,0.44,0.1967}DWM (3/10000) & \footnotesize{0.051}& \color[rgb]{1,0,1}ITM (3/1) & \footnotesize{0.029}\\
0.5& \color[rgb]{0.3,0.3,0.3}PSM (50/3) & \footnotesize{0.33}& \color[rgb]{0,0,1}BDND (1/5) & \footnotesize{0.051}& \color[rgb]{0,0,0}LUM (4/4) & \footnotesize{0.127}\\
1.0& \color[rgb]{0,0,1}BDND (1/3) & \footnotesize{0.123}& \color[rgb]{0.9,0.3,0}TSM (20/5) & \footnotesize{0.016}& \color[rgb]{0.7,0.6,0.1}ATM (3/1) & \footnotesize{0.415}\\
2.0& \color[rgb]{0,0.398,0.796}LED (2/20) & \footnotesize{0.037}& \color[rgb]{0.9,0.3,0}TSM (50/5) & \footnotesize{0.093}& \color[rgb]{1,0,0}RPAV (3/3) & \footnotesize{0.127}\\
5.0& \color[rgb]{0,0,1}BDND (4/5) & \footnotesize{0.048}& \color[rgb]{0.9,0.3,0}TSM (100/5) & \footnotesize{0.002}& \color[rgb]{1,0,0}RPAV (2/1) & \footnotesize{0.031}\\
10.0& \color[rgb]{0,0,0}LUM (1/3) & \footnotesize{0.007}& \color[rgb]{0,0,0}LUM (1/2) & \footnotesize{0.007}& \color[rgb]{0,0.398,0.796}LED (1/2) & \footnotesize{0.002}\\
\hline
\multicolumn{7}{c}{SBIG ST10XME (4.6\% defective pixels)}\\ \hline0.1& \color[rgb]{0,0.398,0.796}LED (20/2) & \footnotesize{0.038}& \color[rgb]{0.451,0.258,0.07}CWM (2/10) & \footnotesize{0.159}& \color[rgb]{0,0.398,0.796}LED (5/2) & \footnotesize{0.008}\\
0.2& \color[rgb]{0.3,0.3,0.3}PSM (20/2) & \footnotesize{0.049}& \color[rgb]{0,0,0}LUM (4/4) & \footnotesize{0.423}& \color[rgb]{0.35,0.44,0.1967}DWM (3/100000) & \footnotesize{0.052}\\
0.5& \color[rgb]{0.9,0.3,0}TSM (50/3) & \footnotesize{0.04}& \color[rgb]{0,0,0}LUM (1/3) & \footnotesize{0.146}& \color[rgb]{0,0.398,0.796}LED (20/1) & \footnotesize{0.024}\\
1.0& \color[rgb]{0.35,0.44,0.1967}DWM (4/100000) & \footnotesize{0.029}& \color[rgb]{1,0,0}RPAV (4/4) & \footnotesize{0.034}& \color[rgb]{0,0.398,0.796}LED (1/2) & \footnotesize{0.025}\\
2.0& \color[rgb]{1,0,0}RPAV (1/4) & \footnotesize{0.025}& \color[rgb]{1,0,1}ITM (1/3) & \footnotesize{0.083}& \color[rgb]{0,0,0}LUM (3/2) & \footnotesize{0.101}\\
5.0& \color[rgb]{0,0.398,0.796}LED (20/0.25) & \footnotesize{0.118}& \color[rgb]{0.9,0.3,0}TSM (200/5) & \footnotesize{0.037}& \color[rgb]{0,0.398,0.796}LED (20/1) & \footnotesize{0.005}\\
10.0& \color[rgb]{0,0,0}LUM (2/2) & \footnotesize{0.003}& \color[rgb]{1,0,0}RPAV (4/1) & \footnotesize{0.001}& \color[rgb]{0,0.398,0.796}LED (0.5/2) & \footnotesize{0.036}\\
\hline
\multicolumn{7}{c}{SBIG2000 (8.3\% defective pixels)}\\ \hline0.1& \color[rgb]{0,0.398,0.796}LED (2/5) & \footnotesize{0.04}& \color[rgb]{0.35,0.44,0.1967}DWM (1/500) & \footnotesize{0.34}& \color[rgb]{0,0.398,0.796}LED (20/1) & \footnotesize{1.015}\\
0.2& \color[rgb]{0.3,0.3,0.3}PSM (200/1) & \footnotesize{0.057}& \color[rgb]{0.9,0.3,0}TSM (100/1) & \footnotesize{0.073}& \color[rgb]{0,0.398,0.796}LED (20/1) & \footnotesize{0.589}\\
0.5& \color[rgb]{0,0.398,0.796}LED (10/2) & \footnotesize{0.176}& \color[rgb]{0.3,0.3,0.3}PSM (5000/4) & \footnotesize{0.037}& \color[rgb]{0.9,0.3,0}TSM (200/2) & \footnotesize{0.242}\\
1.0& \color[rgb]{0,0,1}BDND (2/9) & \footnotesize{0.011}& \color[rgb]{0,0,0}LUM (2/3) & \footnotesize{0.047}& \color[rgb]{0.7,0.6,0.1}ATM (2/1) & \footnotesize{0.053}\\
2.0& \color[rgb]{0,0.398,0.796}LED (0.5/20) & \footnotesize{0.089}& \color[rgb]{1,0,1}ITM (2/1) & \footnotesize{0.04}& \color[rgb]{0.9,0.3,0}TSM (200/3) & \footnotesize{0.021}\\
5.0& \color[rgb]{0,0,0}LUM (2/3) & \footnotesize{0.178}& \color[rgb]{1,0,0}RPAV (4/2) & \footnotesize{0.095}& \color[rgb]{1,0,1}ITM (1/3) & \footnotesize{0.084}\\
10.0& \color[rgb]{0,0.398,0.796}LED (10/5) & \footnotesize{0.007}& \color[rgb]{0,0,0}LUM (1/2) & \footnotesize{0.004}& \color[rgb]{0,0.398,0.796}LED (2/2) & \footnotesize{0.031}\\
\hline
\label{individual_accuracy}\end{tabular}
\end{table*}

\section{Discussion}
\subsection{Synthetic stellar profiles - photometry}
The experiments conducted on images of synthetic stars revealed that the highest photometric accuracy was achieved by various methods depending on $\sigma_{\tiny{\textrm{PSF}}}$, noise type and its SNR. This finding confirms that there is no single algorithm which can deal with all noise scenarios. Nevertheless, the methods which proved to be most accurate, are LED and TSM. However, their predominance over the others is not so prevalent.

For cosmic-ray-like noise (Andor iKon CR and BRITE Torronto) the LED is a definite winner, which is not surprising since the method was constructed to deal with such complex artifacts. The superiority of LED is also more evident for images with the highest SNR values, where the impulses are less visible. On the other hand, for the lower SNR regime, the methods are more comparable and the TSM provides frequently the best outcomes. 

One should also notice that for very narrow stellar profiles ($\sigma_{\tiny{\textrm{PSF}}}$=1 and 2) the LED is often the best solution. It can be explained by its ability to suppress the filtering strength for small symmetrical objects, while the other methods treat such stars as a noise. On the other hand, for much wider profiles ($\sigma_{\tiny{\textrm{PSF}}}$=4) this advantage is not present and thus simpler methods, like TSM, PSM or even CWM, perform better.

\subsection{Synthetic stellar profiles - astrometry}
In contrast to the results of photometric evaluation, the astrometry revealed a clear superiority of the PSM method. This technique provided the lowest errors even for very noisy images, achieving centroid accuracy well below 1 pixel. There were some images, for which LED algorithm was better, which is consistent with its good filtering capabilities proved in photometric evaluation. The best results provided by LED appear for $\sigma_{\tiny{\textrm{PSF}}}$=1, which was also observed in photometric data. However, the PSM filter outperformed LED even for such narrow profiles.

It is worth to have a deeper look at the two winning methods. The PSM and LED share exactly the same basic schema of intensity replacement in faulty pixels. In both methods these pixels are first detected and then, having the mask of bad pixels defined, their intensities are exchanged with the median intensity of not corrupted neighbors. The algorithms, however, differ in the way of impulse detection: while in LED filter the approach is quite complicated, the PSM algorithm utilizes a simple comparison of pixel intensity with a median intensity of its neighbors, (a threshold is applied). 

The fact that PSM did not provide similarly high photometric performance indicates that for proper centroiding, the preservation of pixel intensities is not a vital factor. Slight over-smoothing, leading to relatively high photometric errors, can be better than retaining small impulses which can significantly bias a centroid. The predominance of PSM may be associated with its adjustable and simple detection threshold, which allows for fine tuning of smoothing strength to achieve the optimal astrometric results.

\subsection{Real images}
Since the quality of filtering of real images was assessed using RMS measure, it was expected that the results will be, at least to some degree, consistent with the ones of photometric evaluation on synthetic stars. Indeed, a similar spread of best methods can be observed. However, the LED algorithm seems to have the best average performance again. 

Its predominance is evident for very high density of noise (Andor iKon L, 49\% defective pixels), which was also observed in previous experiments for smaller PSF sizes (P$\sigma_{\tiny{\textrm{PSF}}}$=1 or 2, in photometry) and for lower SNRs (SNR=0.1$\sim$1, in astrometry). This can be explained by the ability of this method to reduce extended clusters of hot pixels (frequently observed in high-density noise), which are not handled properly by other methods. The results also proved the high efficiency of LED for smaller profiles, which were present in real examples of traditional and speckle imaging ($\sigma_{\tiny{\textrm{PSF}}}=1$ and $1.5$ respectively for speckles and M13). 

The removal of cosmic rays in traditional imaging, was the main goal of LED, thus it provides very good results for this noise/image combination (Andor iKon CR + Traditional imaging). However, the LED is significantly less efficient for spectroscopic observations, which was expected, as the algorithm recognizes gradients in spectral bar as cosmic ray impacts and filters them out. For this type of images, the outcomes of all the methods are more uniform, with TSM having slightly better performance. The average performance of TSM method should be also considered as it appeared to be the second best method. The high accuracy of TSM is consistent with the results of photometric evaluation on synthetic stars. Its high precision can be observed e.g. in spectroscopic images corrupted by cosmic rays.

\subsection{Conclusions}
The conducted experiments allowed for drawing several conclusions and hints for application of the most suitable algorithm for a given image-noise scenario. The LED algorithm, developed originally to deal with cosmic rays while preserving symmetrical objects, appeared to be very efficient solution for denoising the astronomical images. However, there are some ranges of sizes of stellar profiles ($\sigma_{\tiny{\textrm{PSF}}}=4$ or higher) for which other methods have to be applied. Importantly, it should be done for spectroscopic images, which are of definitely different type and therefore should be treated with care. Among others, the TSM algorithm gives good promise for reliable photometric outcomes.

In contrast to the photometry, where various methods were able to provide the best results, for astrometry the PSM method significantly outperformed the other approaches. It provided the most accurate outcomes for a wide range of $\sigma_{\tiny{\textrm{PSF}}}$ and SNR values. Assuming proper optimization of its tunable threshold, one should in the first place consider this denoising technique, while detecting and localizing stellar profiles buried in noise. However, when stellar profiles are narrow or noise density is high, the LED algorithm may be also employed.

\section{Summary}
The impulsive noise in astronomical images does not appear only in the form of well-defined hot pixels easily compensable with the dark frames. There are also several sources of non-stationary (in time and in space) impulsive noise, which can significantly reduce the quality of astronomical data. They include: the nonlinearity of dark current, its random telegraph signals in proton-radiated sensors working in space, clock-induced noise in EMCCDs and cosmic ray hits. For such types of  disturbances, there is no possibility to define the position of defective pixels or to compensate for the intensity offsets employing calibration frames. Therefore, the image filtering algorithms have to be utilized to properly interpolate over faulty pixels.

In this article we presented the details of all mentioned noise sources and performed an extensive evaluation of the efficiency of various filtering schema. Twelve widely-used algorithms were implemented and their accuracy was checked on astronomical images affected by real impulsive noise. As the noise templates, we acquired the dark frames from image sensors working on ground and in space. To perform its best, each method was adjusted by optimizing the values of up to two tunable parameters. Both synthetic stellar profiles and real astronomical images were utilized. For the latter ones, we employed 30 cm Ritchey-Chretien telescope installed in Kamieniec observatory (Poland), and created high-fidelity reference frames as examples of three imaging techniques used in astronomy: speckle observations, traditional imaging and spectrography. 

The evaluations confirmed that there is no single method, which provides the best photometric performance significantly in all tested cases. The Laplacian Edge Detector, widely used in astronomical pipelines, proved its high performance, however its superiority was limited to low width of stellar profiles and it performed slightly worse for very low signal-to-noise ratios. For such cases, other methods, especially Tri-State Median Filter, should be considered. For astrometric evaluations, the Progressive Switched Median filter performed the best, almost for all tested images. Its predominance was connected with its adjustable parameters, which allowed for the best fitting to the requirements of astrometry.

The conclusions drawn in this article can be treated as guidelines while creating pipelines for astronomical image processing. The collection of methods taken for comparisons can be used as a reference set for evaluations of novel denoising algorithms. The presented noise templates with an extended description of non-stationary noise creation, show a range of practical problems encountered not only in astronomical images. We showed that the simplified noise models (like salt and pepper) present in the rich literature and used as a reference in numerous evaluations, are far from reality. We hope, that the developers of new filtering algorithms, will consider using such realistic noise frames as they will be made available on demand by the authors.

\section*{Acknowledgments}
We would like to thank anonymous referee for a very detailed review and for numerous valuable comments which improved significantly the paper.
Adam Popowicz was supported by Polish National Science Center, grant no. 2013/11/N/ST6/03051: Novel Methods of Impulsive Noise Reduction in Astronomical Images. Bogdan Smolka acknowledges Polish National Science Center, grant: DEC-2012/05/B/ST6/03428. Aleksander R. Kurek acknowledges Polish National Science Center grant no. 2012/07/B/ST9/04425 and Polish Ministry of Science and Higher Education grant no. MNSW:3-388/M/2016. Valeri Orlov acknowledges the support from Direcci{\'o}n General de Asuntos del Personal Acad{\'e}mico (UNAM, M{\'e}xico) under the projects IN102514.
The research was performed using the infrastructure supported by POIG.02.03.01-24-099/13 grant: GeCONiI - Upper Silesian Center for Computational Science and Engineering.

\bibliography{ms}

\end{document}